\documentclass[reprint,amsmath,amssymb,amsfonts,aps,prb,superscriptaddress]{revtex4-1}
\usepackage{graphicx}
\usepackage{bm}
\usepackage{epsfig}
\usepackage{epstopdf}
\usepackage{xcolor}
\usepackage{hyperref}
\hypersetup{colorlinks=true,allcolors=blue}
\usepackage{natbib}

\begin{document}
\title{Entanglement Properties of
 Disordered Quantum Spin Chains with Long-Range Antiferromagnetic Interactions}

\author{Y. Mohdeb}
\email[]{y.mohdeb@jacobs-university.de}
\affiliation{Department of Physics and Earth Sciences, Jacobs University Bremen, Bremen 28759, Germany}  

\author{J. Vahedi}
\email[]{j.vahediaghmashhadi@tu-braunschweig.de}
\affiliation{Department of Physics and  Earth Sciences, Jacobs University Bremen, Bremen  28759, Germany}  
\affiliation{ Technische Univ. Braunschweig, Institut f. Mathematische Physik, Mendelssohnstr. 3, 38106 Braunschweig, Germany}  
\affiliation{Department of Physics, Sari Branch, Islamic Azad University, Sari 48164-194, Iran}

\author{N. Moure}
\email[]{mouregom@usc.edu}
\affiliation{Department of Physics and Astronomy  
University of Southern California, Los Angeles, CA 90089-0484}

\author{A. Roshani}
\email[]{roshani@usc.edu}
\affiliation{Department of Physics and Astronomy  
University of Southern California, Los Angeles, CA 90089-0484}

\author{Hyun-Yong Lee}
\email[]{hyunyong@korea.ac.kr}
\affiliation{Department of Applied Physics, Graduate School, Korea University, Sejong 30019, Korea}
\affiliation{Division of Display and Semiconductor Physics, Korea University, Sejong 30019, Korea}
\affiliation{Interdisciplinary Program in E$\cdot$ICT-Culture-Sports Convergence, Korea University, Sejong 30019, Korea}   

\author{R. N. Bhatt}
\email[]{ravin@princeton.edu}
\thanks{***(2019-2020)}
\affiliation{Department of Electrical Engineering, Princeton University, Princeton, New Jersey 08544, USA}
\affiliation{School of Natural Sciences, Institute for Advanced Study $***$, Princeton, New Jersey 08540, USA}

\author{Stefan Kettemann}
\email[]{s.kettemann@jacobs-university.de}
\affiliation{Department of  Physics and Earth Sciences, Jacobs University
  Bremen, Bremen 28759, Germany}  
\affiliation{Division of Advanced Materials Science, Pohang University of Science and Technology (POSTECH), Pohang 790-784, South Korea}

\author{Stephan Haas}
\email[]{shaas@usc.edu}
\affiliation{Department of Physics and Astronomy University of Southern California, Los Angeles, CA 90089-0484}
 \affiliation{Department of Physics and Earth Sciences, Jacobs University
  Bremen, Bremen 28759, Germany} 
\date{\today}

\begin{abstract}
Entanglement measures are useful tools in characterizing otherwise unknown quantum phases and indicating transitions between them. Here we examine the concurrence and entanglement entropy in quantum spin chains with random long-range couplings,  spatially decaying with a power-law exponent $\alpha$.
 Using  the strong disorder renormalization group (SDRG) technique, we find  by analytical solution of the master equation 
  a strong disorder fixed point, characterized by a  fixed point distribution  of the couplings with a finite dynamical exponent, which describes the system consistently in the regime 
  $\alpha > 1/2$. 
 A numerical implementation of the SDRG method yields  a power law spatial decay of the average concurrence, which is also confirmed by exact numerical diagonalization. 
However, we find that the lowest-order SDRG approach is not sufficient to obtain the  typical value of the concurrence. 
We therefore implement a correction scheme which allows us to obtain the leading order corrections to the random singlet state. This approach yields a power-law spatial decay  of the typical value of the   concurrence, which we derive both by a numerical implementation of the corrections and by  analytics.
Next, using numerical SDRG, the   entanglement entropy (EE) is found  to be logarithmically enhanced for all $\alpha$, corresponding to a critical behavior with an  effective central charge $c = {\rm ln} 2$, independent of $\alpha$. This is confirmed by an analytical derivation. 
Using numerical exact diagonalization  (ED), we  confirm the 
logarithmic enhancement of the EE and a weak dependence on $\alpha$. For a wide range of distances $l$, the EE fits a critical behavior with a  central charge  
close to $c=1$, which is the same as for  the clean Haldane-Shastry model with a power-la-decaying interaction with 
$\alpha =2$. Consistent with this observation, we  find  using  ED that the concurrence shows power law decay, albeit with smaller power exponents than obtained by  SDRG. 
We also present results obtained with DMRG and  find agreement with ED for sufficiently small $\alpha < 2 $, whereas for larger $\alpha$ DMRG tends to underestimate the   entanglement entropy and finds a faster decaying concurrence. 
\end{abstract} 

\maketitle

\section{Introduction}
Long-range interactions in disordered quantum many-body systems arise in a variety of physical contexts. For example, randomly placed magnetic impurities in doped semiconductors are known to interact with each other via exchange couplings that depend on their separation distance \cite{anderson58,mott76}. While in insulating regimes these interactions typically decay exponentially, $J(r)\propto \exp(-r/\xi)$, in the metallic regime they are mediated via the RKKY mechanism, and therefore decay with a power law, $J(r)\propto r^{-d} $, where $d$ is the dimension of the host system. Power-law random long-range interactions have also been found to occur in quantum glasses, where tunneling ions form local two-level systems which are interacting by dipole-dipole and elastic interactions \cite{yu}. More recently, tunable long-range Heisenberg interactions have been demonstrated in quantum simulators based on trapped ions \cite{grass}. 

 There have been many theoretical studies of quantum spin chains with random {\em short-range} interactions. For example, it is known that bond disorder drives spin-$1/2$ Heisenberg chains with short-ranged antiferromagnetic interactions into an  infinite-randomness fixed point (IRFP) \cite{bhattlee,fisher2,monthus},
  characterized by 
  a critical   ground state, i.e.  a product state of singlets
  which are formed at random distances.  The entanglement entropy of such critical spin chains  scales  logarithmically, with a central charge $c$ of the corresponding conformal field theory\cite{calabrese}. As the IRFP is a critical point,  the entanglement entropy is  logarithmically enhanced. The corresponding 
   central charge $\tilde{c}$  is, however,  smaller by a factor 
  $\ln 2$ than the central charge of the corresponding clean spin chain. 
  This has been derived using  the strong disorder renormalization group (SDRG) method \cite{refael-entropy,hoyos}.

More recently, the SDRG has also been applied  to disordered quantum spin systems with {\em long-range} interactions, where it has been shown to  lead to a self-consistent description of its thermodynamic properties  in terms of a novel strong disorder fixed point \cite{ours,ourPRB}.
However,  quantum information theoretical measures, such as the entanglement entropy and   the   concurrence have not yet been studied in such systems. Here, we analyze these quantities  for  disordered quantum spin-$1/2$-chains with  long-range power-law couplings, decaying with a power-law exponent $\alpha$, using the SDRG, exact diagonalization (ED), and density matrix renormalization group (DMRG).
  There have been indications that such system undergo a phase transition  at a critical decay exponent $\alpha_c$ from a localized regime for $\alpha > \alpha_c$, to a delocalized regime
 for $\alpha < \alpha_c$\cite{ours}, similar to the delocalization transition of disordered fermions with long-range hoppings \cite{mirlin96,cuevas04}. Thus, a logarithmic enhancement of the average  entanglement entropy is expected at some value of $\alpha$.
Moreover, since  the  spin-1/2 Haldane-Shastry model, a Heisenberg model  with power-law  long-range interactions, is known to be critical for $\alpha=2$ with a conformal charge $c=1$\cite{Haldane,Rao}, and as
  it was suggested in Ref. \onlinecite{refael-entropy} that disordered spin chains are critical when the clean spin chain is critical with a central charge $\tilde{c}$, smaller by  a factor 
   $\ln 2$   than the central charge of the clean  critical spin chain,
    one could expect a logarithmic enhancement of the entanglement entropy 
     at $\alpha =2$ with $\tilde{c} = \ln 2$ in the presence of disorder.
   This motivates us to examine the entanglement  properties of disordered
    quantum spin systems with long-range power-law couplings  as a function of  the power exponent $\alpha$.

Here,  we    focus on  the  bond disordered XX-spin  chain with  long-range couplings,  defined by  the Hamiltonian
\begin{equation}\label{H}
H=\sum_{i<j}J_{ij}\left(S_{i}^{x}\,S_{j}^{x}+S_{i}^{y}\,S_{j}^{y}\right), 
\end{equation}
describing $N$ interacting $S=1/2$ spins that are randomly placed at positions ${\bf r}_i$ on a  lattice of length $L$ and lattice spacing  $a$, with density $n_0 = N/L = 1/l_0$, where $l_0$ is the average distance between  them. 
The couplings between all pairs of sites $i,j,$ are taken to be antiferromagnetic and long-ranged, decaying with a power law, 
\begin{equation} \label{jcutoff}
J_{ij} = J_0\left|({\bf r}_i-{\bf r}_j)/a\right |^{-\alpha}.
\end{equation}
  The random placement of the spins on the lattice sites, excluding double occupation, results in an initial (bare) distribution of the separation distance $l$ between pairs of spins, 
  $p^0(l)$, yielding an  initial distribution of couplings $J$, $P^0(J)$. For example, for $N=2$ it is 
  given  exactly  by $p^0(l) = (1-l/L) 2/(L-1)$, where we set $a=1$, which yields  an initial distribution of couplings $J$,
  \begin{equation} \label{p0}
  P^0(J) =  \frac{2}{(L-2) \alpha J_0} \left( (\frac{J_0}{J})^{1 +1/\alpha} - \frac{1}{L}(\frac{J_0}{J})^{1 +2/\alpha}\right), 
  \end{equation}
  for couplings  restricted to the interval 
  $ J_0  ((L-1)/a)^{-{\alpha}}  <J<J_0$.
It is important to note that in the case of long-range interactions, this model does not simply map onto an effective fermionic tight-binding model with long-range hoppings, because the phase factors arising in the Jordan-Wigner transformation are   non-trivial.

\section{The Strong  Disorder Renormalization Group}
Here we describe how to  apply the SDRG to this model,   with the aim to evaluate  the concurrence and the entanglement entropy \cite{monthus3}. 

\begin{figure}
  \includegraphics[width=0.4\textwidth]{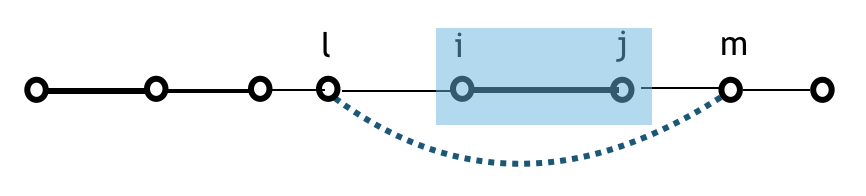}
\caption{Decimation  of the strongest-coupled 
 pair $i,j$ (thick line)  generates effective couplings between other spins  $l,m$ (dotted line). }
\label{rgnnl}
\end{figure}
Choosing
           the pair with  the largest coupling $(i,j)$,
            which 
         forms a singlet (see Fig. \ref{rgnnl}), we      
               take the expectation value of the Hamiltonian 
        in that particular singlet state within second-order perturbation 
             in  couplings with
              all other spins\cite{westerberg}. This yields the  
             long-range renormalization rule
             for the  couplings between  spins $(l,m)$ in the XX model, \cite{ours,ourPRB}
              \begin{equation} \label{jeff}
               (J^{x}_{lm})' =   J_{lm}^x - \frac{(J^x_{li}-J^x_{lj})(J^x_{im}-J^x_{jm})}{J^x_{ij}}.
              \end{equation}
             For the Heisenberg model, this rule  differs by a numerical prefactor $1/2$\cite{ours,ourPRB}.    Previously it was found that in long-range coupled disordered quantum spin chains there is
a strong disorder fixed point\cite{ourPRB}.  This was realized by  inspecting the evolution of the width of the distribution of couplings $J$ with the RG flow. 
 In the short-range case, i.e. at the infinite randomness fixed point (IRFP), this distribution gets wider at every RG step,
 with width $W=(\langle \ln (J/\Omega_0)^2 \rangle -  \langle \ln (J/\Omega_0) \rangle^2)^{1/2} =\ln(\Omega_0/\Omega) = \Gamma_{\Omega},$ increasing monotonically as  the RG scale $\Omega$ is lowered. In contrast, for long-range couplings with 
 finite $\alpha$ the width $W$  saturates and   converges to  $W = \Gamma$, with $\Gamma = 2 \alpha$ in the XX-limit\cite{ourPRB}.
  The convergence to a finite dynamical exponent $\Gamma$ characterizes the new
   strong disorder fixed point (SDFP). 
  For large number of spins $N,$  and in the limit of a small RG scale $\Omega$, 
            the resulting 
     distribution function of renormalized couplings $J$ at RG scale $\Omega$
     was found to converge 
      to\cite{monthus},
      \begin{equation} \label{pjsd}
      P(J,\Omega) =\frac{1}{\Omega \Gamma_{\Omega}} \left( \frac{\Omega}{J}\right)^{1-1/\Gamma_{\Omega}}.
      \end{equation}
  At the IRFP, $\Gamma_{\Omega}$ increases monotonically as 
  $\Gamma_{\Omega}= \ln \Omega_0/\Omega$,  
  when $\Omega$, the largest energy at this renormalization step, is lowered. Here 
   $\Omega_0$ is the initially largest energy in the SC.
 
  At the  SDFP, however,   $\Gamma_{\Omega}$ is found to  converge to an asymptotic finite value $\Gamma_{\Omega} \rightarrow  2 \alpha$\cite{ourPRB}, yielding 
   a narrower distribution with finite width $\Gamma$. We note that this distribution is less divergent as $J\rightarrow 0$ than the initial distribution Eq. (\ref{p0}).
  In order to check  whether Eq. (\ref{pjsd}) is indeed 
  a fixed point distribution of the RG with 
  RG rule Eq. (\ref{jeff}),  we first present  
 an analytical derivation. The  master equation governing the renormalization process is
  a differential equation for the distribution function $P(J, \Omega)$,   given by 
 \begin{equation}\label{Meq0}
\begin{aligned}
-    \frac{\partial P(J,\Omega)}{\partial \Omega}= & P(\Omega, \Omega)\int_{}\left(\prod_{i=1}^{5} dJ_i P(J_i,\Omega)\right)\\ & \times
     \delta\left(J-J_5+\frac{(J_1-J_2)(J_3-J_4)}{\Omega}\right).
     \end{aligned}
 \end{equation}
 
 \begin{figure}
 \begin{center}
  \includegraphics[width=0.35\textwidth]{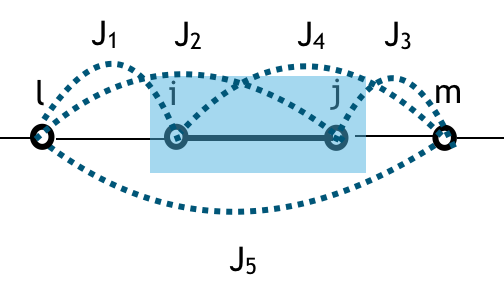}
  \vspace{-0.6cm}
\caption{ Definition of  couplings between removed spins $i,j$ and other spins  $l,m$ (dotted lines). }
\label{rgdef}
\end{center}
\end{figure}
The numbering of the couplings is defined in  Fig. \ref{rgdef}.
  In order to proceed, let us note  that the renormalization correction to the bare coupling $J_5$, given by $\delta J = (J_1-J_2)(J_3-J_4)/\Omega$, is always  smaller than $J_5$,  $\delta J < J_5$, as can be checked numerically \cite{ourPRB}. 
  Performing first the integral over $J_5$, we can expand the distribution function  of $J_5 = J+\delta J$ in $\delta J$,
  finding to all orders in $\delta J$
   \begin{equation}\label{Meq}
\begin{aligned}
  &  -\frac{\partial P(J,\Omega)}{\partial \Omega}   =
  \nonumber \\ &
    P(\Omega, \Omega)  \sum_{n=0}^{\infty} \frac{1}{n!} \langle \left(\frac{(J_1-J_2)(J_3-J_4)}{\Omega}\right)^n \rangle
    \partial_J^n P(J,\Omega),
     \end{aligned}
 \end{equation}
 where $\langle ... \rangle$ denotes the  averaging with the distribution functions $P(J_i,\Omega),$  for $i=1,2,3,4.$
 Thus, we find
   \begin{equation}\label{Meq2}
\begin{aligned}
   - \frac{ \partial}{\partial \Omega}  P (J, \Omega)  =P(\Omega, \Omega)  
   \left(1 +  \sum_{n =2}^{\infty} \frac{A_n}{n !} \Omega^n \partial_J^n \right) P(J, \Omega),
     \end{aligned}
 \end{equation}
 where 
 \begin{equation} \label{and}
 A_n =  \langle \left(\frac{(J_1-J_2)(J_3-J_4)}{\Omega^2}\right)^n \rangle.
 \end{equation}
 We see that the strong disorder scaling ansatz Eq. (\ref{pjsd}) is a solution of Eq. (\ref{Meq2})
 for $\Omega \rightarrow 0$ with $\mu = 1/\Gamma = const.$,
  when neglecting all renormalization corrections given by the terms 
  $n \ge 2$. 
  \begin{figure}
 \begin{center}
  \includegraphics[width=0.5\textwidth]{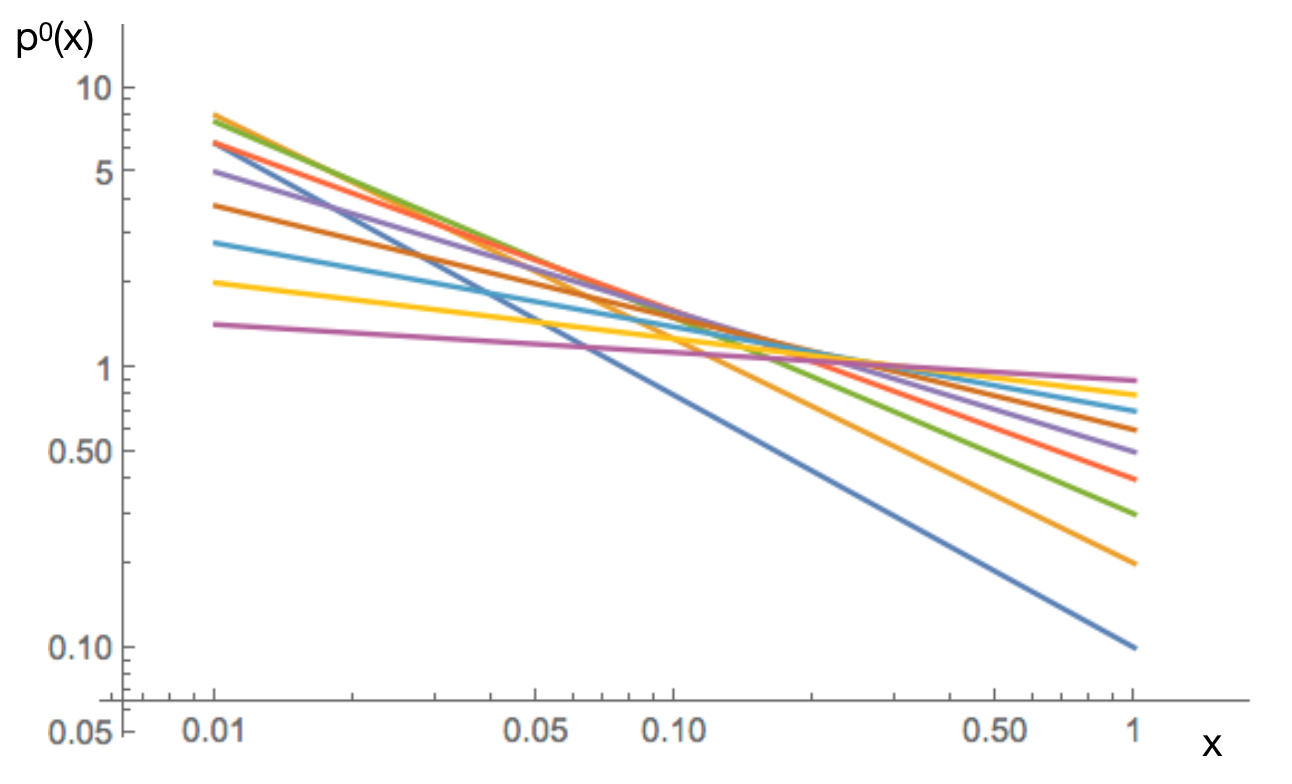}
  \vspace{-0.cm}
\caption{Probability density function $p^0(x= J/\Omega)$ on a double logarithmic scale for $\mu =  0.1$ (blue) to $ \mu = 0.9$ (magenta) in steps of $0.1$. }
\label{p0ofx}
\end{center}
\end{figure}
   \begin{figure}
 \begin{center}
  \includegraphics[width=0.5\textwidth]{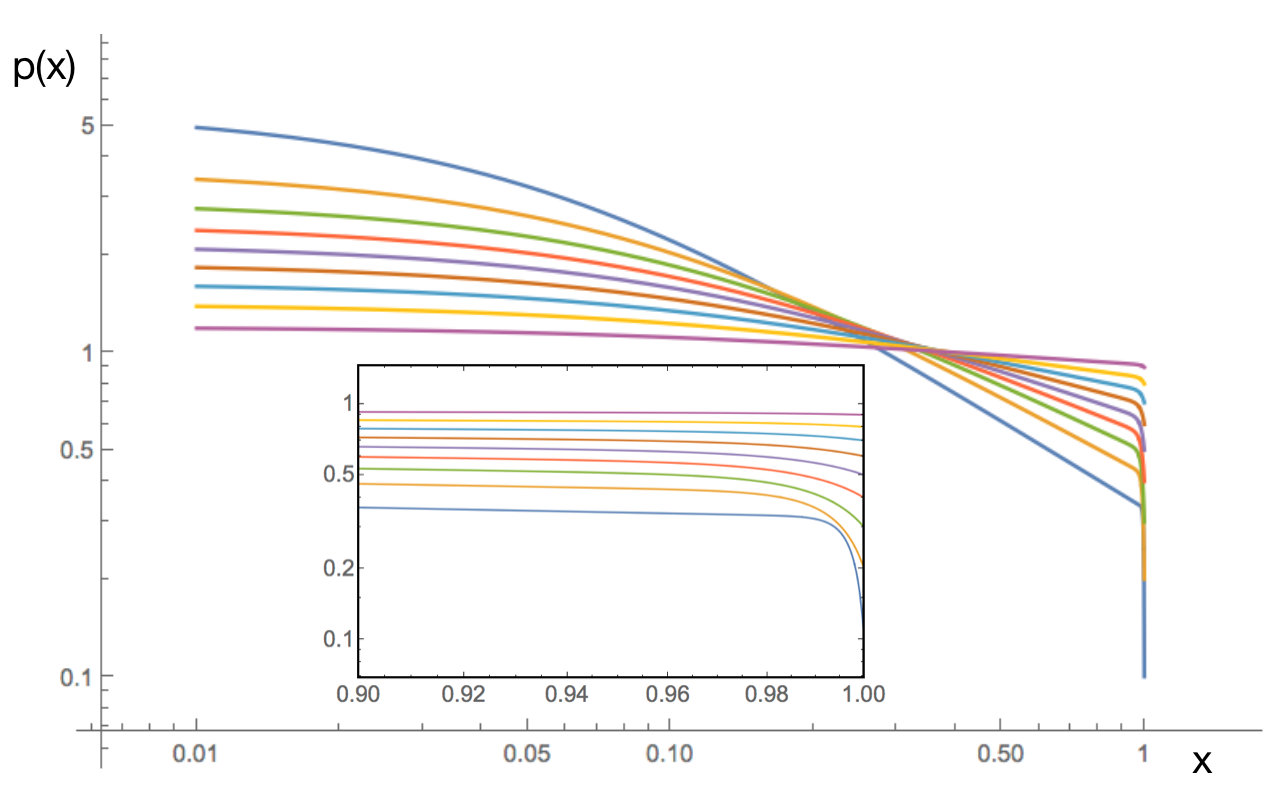}
  \vspace{-0.cm}
\caption{Probability density function $p(x= J/\Omega)$ on a double logarithmic scale for $\mu =  0.1$ (blue) to $ \mu = 0.9$ (magenta) in steps of $0.1$. Inset: 
 Magnification showing the exponential decay to $\mu$ as 
  $x = J/\Omega \rightarrow 1$.}
\label{pofx}
\end{center}
\end{figure}
  To solve Eq. (\ref{Meq2}), including the correction terms, 
  we make the ansatz 
  \begin{equation} \label{Ansatz}
  P(J,\Omega)= \frac{1}{\Omega} p(x=J/\Omega),
  \end{equation}
where  the probability density function  $p(x)$ is  to be derived with the condition that $p(1)=\mu$, so that $p(x) = p^0(x) =\mu x^{\mu-1}$ corresponds to strong disorder scaling  Eq. (\ref{pjsd}) with $\mu = const.$. 
 Inserting this ansatz, we find the following ordinary differential equation for $p(x)$,
 
 \begin{equation}\label{Meqfinal}
 \begin{aligned}
   (1-\mu) p(x) +x \frac{ \partial}{\partial x}  p (x)  = 
    \sum_{n =2}^{\infty} \frac{A_n}{n !}  \partial_x^n  p(x),
     \end{aligned}
    \end{equation}
    with the condition $p(1)=\mu$.
    
Choosing an iterative approach, we can  first approximate 
the coefficients $A_n$, by setting $p(x) = p^0(x) =\mu x^{\mu-1}$ in Eq. (\ref{Ansatz}), yielding
\begin{equation}\label{An}
\begin{aligned}
 A_n \approx \frac{\mu^4 \Gamma^2(\mu)  \Gamma^2(n+1)}{(\frac{n}{2}+\mu)^2  \Gamma^2(\mu+n+1)}.
     \end{aligned}
 \end{equation}
We note that $A_n \sim 1/n^2$ for $0< \mu \ll 1$.
For $n=2$,  this simplifies  with $\Gamma(z+1) = z \Gamma(z)$ to $A_2 = 4 \mu^2/((\mu+2)^2 (\mu+1)^4) $, and we can solve  to second order $n=2$ the resulting  2nd order ordinary differential equation  exactly in terms of products of Hermite polynomials and Hypergeometric functions, as plotted in Fig. \ref{pofx}. For comparison, we plot $p^0(x) = \mu x^{\mu-1}$ in Fig. \ref{p0ofx}. We see that $p(x)$ has the same power law 
$p(x) \sim  x^{\mu-1}$ dependence for intermediate $0 \ll x = J/\Omega <1$  as $p^0(x)$, but converges for $x \rightarrow 0$
to a constant which depends on $\mu$ approximately as $p ( x \rightarrow 0) \approx 1/\mu^{3/4}$. 
  We note that for small $0< \mu \ll 1$, $A_n \sim 4 \mu^2/n^2,$
  so that higher order terms decay fast with $n$, and the pdf $p(x)$ obtained for $n=2$ is a good approximation for small $\mu < 1$.
  
 An alternative way 
 to solve the master equation, is to 
 make use of a step which is commonly used to calculate the generating function of a distribution: we
multiply both sides  of Eq. (\ref{Meq}) by $z^J$, where $z$ is a real number, 
and then integrate both sides  over $J$. Thereby, one can generate all moments of the distribution by partial derivatives in $z$ and thus the entire distribution function.  Since we are interested in the fixed point solution for large systems, 
we set the lower integration limit to 0. The upper integration limit is provided by $\Omega$. We next insert the ansatz
for the distribution of 
renormalized couplings  
Eq. (\ref{pjsd}), and find a  differential equation  for  the exponent $\mu(\Omega) = 1/\Gamma_{\Omega}  $ (a detailed derivation is provided in  Appendix A),
\begin{eqnarray}
    \label{diffeq}
 && \sum_{n=1}^{\infty}\frac{-\mu'(\Omega)(\Omega t)^{n}}{(n-1)!(\mu+n)^2}\nonumber \\
 && =
   \sum_{k=1,k'=0}^{\infty}\frac{\mu^6(\Omega)\Gamma^2(\mu)(2k)!(\Omega t)^{2k+k'}}{(k+\mu)^2\Gamma^2(\mu+2k+1)k'!(\mu+k')\Omega}.
    \end{eqnarray}
Here,  $\Gamma(x)$ is the standard Gamma function.
Since we are  searching for the fixed point solution of the master equation, we keep only  the leading order in $\Omega$ of Eq. (\ref{diffeq}). This yields  for $\Omega\rightarrow 0$ the necessary condition 
\begin{equation}\label{LOdiffeq}
  \mu'(\Omega)=0 .
\end{equation}
Thus, the solution approaches for 
 $\Omega \rightarrow 0$ a fixed point with vanishing slope of the dynamical exponent. 

 The fixed point value 
 $\Gamma(\Omega=0)= 1/\mu(\Omega=0)  = \Gamma_0$ is  finite, but cannot be  determined  by this calculation alone. To this end we need to use a scaling argument, as outlined below.  
  The distribution of lengths $l$ of singlets at renormalization scale $\Omega$ can also  be derived from a master equation, as was done 
  in Ref. \cite{fisher,hoyos} for the IRFP.
    However, noting that the strong disorder fixed point distribution 
    $P(J,\Omega)$ was obtained above to lowest order in a Taylor expansion 
     in the difference between the renormalized and the bare coupling 
     $J^0(l) = J_0 (l/a)^{-\alpha}$, it is clear that there is a strong correlation between the distribution of lengths and bare couplings. 
     Therefore, we can make to lowest order the ansatz, 
     \begin{equation}
         P(J, \Omega, l) = P(J, \Omega) \delta (J - J_0 (l/a)^{-\alpha})
         |\frac{d J_0 (l/a)^{-\alpha}}{d l}|.
     \end{equation}
     
     We note that in each RG step, a fraction $dn/n(\Omega)$ 
     of the remaining spins $n(\Omega)$ at renormalization energy
     $\Omega$ are taken away. Since this is due to the formation of a singlet with coupling $J = \Omega$, this fraction should be equal to $2P (J = \Omega, \Omega) d \Omega  $, leading to the differential equation
     \begin{equation}
     \frac{d n}{ d \Omega } = 2 P(J= \Omega, \Omega) n (\Omega). 
\end{equation}
     Since at the SDFP  $P(J= \Omega, \Omega) = \mu/\Omega$, 
  the density   of not yet decimated singlets at the RG scale $\Omega  $  is 
  given by
\begin{equation}
\label{nomlr}
n (\Omega) =n_0 \left(\frac{\Omega}{\Omega_0}\right)^{2\mu},
\end{equation}
where $n_0 = N/L = 1/l_0$ is the initial density of spins. 
 Defining  $l_{\Omega}$ as the average distance between spins at RG scale $\Omega$, 
 we have $n(\Omega) = 1/l_{\Omega} $. 
    Thus, it follows from Eq. (\ref{nomlr}) that the RG energy scale is related to the length $l_{\Omega}$ via $\Omega \sim l_{\Omega}^{-\frac{1}{2 \mu}} $.
     As we discussed above, the strong disorder fixed point distribution is dominated by the bare couplings, which scale with distance $l$ as $J^0 \sim l^{-\alpha}$.
      In order for the energy-length scaling to be consistent at all RG scales $\Omega$, it follows necessarily that 
      \begin{equation}
      2 \mu = 1/\alpha.
    \end{equation}
    As we have shown above by solution of the master equation that the strong disorder  fixed point distribution is a solution for $\mu <1,$ we can conclude that it is expected to give 
     consistent  results for $\alpha > 1/2.$

We are now ready to evaluate the distribution of singlet lengths  by integrating 
over the renormalization energy $\Omega$

\begin{equation} 
P_s(l)= 2 \int_{\Omega^*}^{\Omega_0} d\Omega l_0 n (\Omega) P(J=\Omega,l;\Omega), 
\end{equation}
where we have used the fact that $l_0 n(\Omega)$ is the fraction of not yet decimated spins at $\Omega$, and the factor $2$ comes from 
 the normalization condition for $P_s(l)$, and $\Omega^* = J_0 ((L-1)/a)^{-\alpha}$ is the smallest possible energy scale.
 Thus,  for $L\rightarrow \infty$ we find 
\begin{eqnarray} 
P_s(l)&=& \int_{0}^{\Omega_0} \frac{d\Omega} {\Omega} \left(\frac{\Omega}{\Omega_0}\right)^{2\mu}
\delta(\Omega- J_0 \frac{a^{\alpha}}{l^{\alpha}}  ) J_0 \frac{a^{\alpha}}{l^{\alpha+1}} \\  \nonumber
&&=  \frac{a}{l^2}.
\label{singletLR}
\end{eqnarray}

\begin{figure}[h!]
\centering
\includegraphics[width=0.45\textwidth]{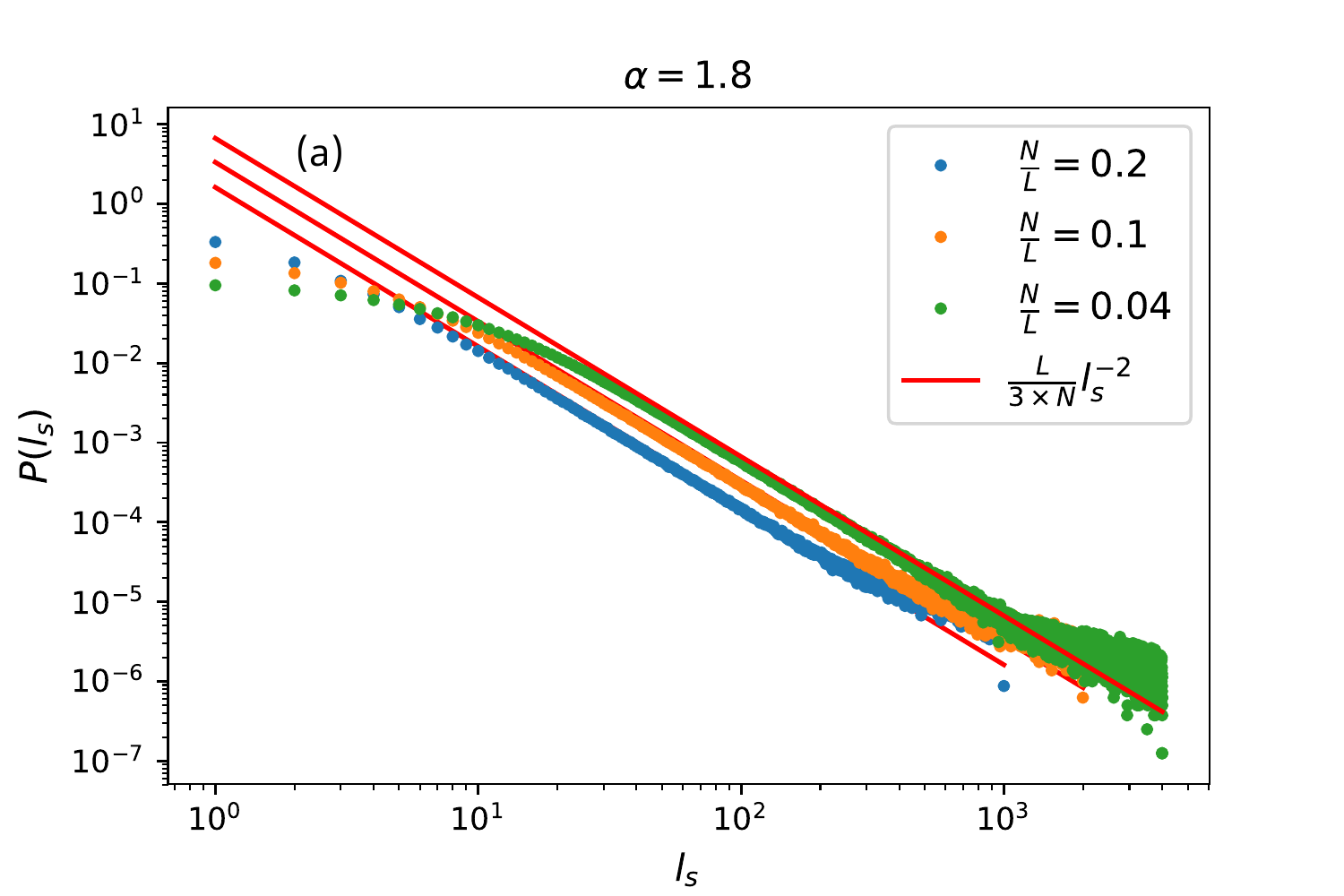}
\includegraphics[width=0.45\textwidth]{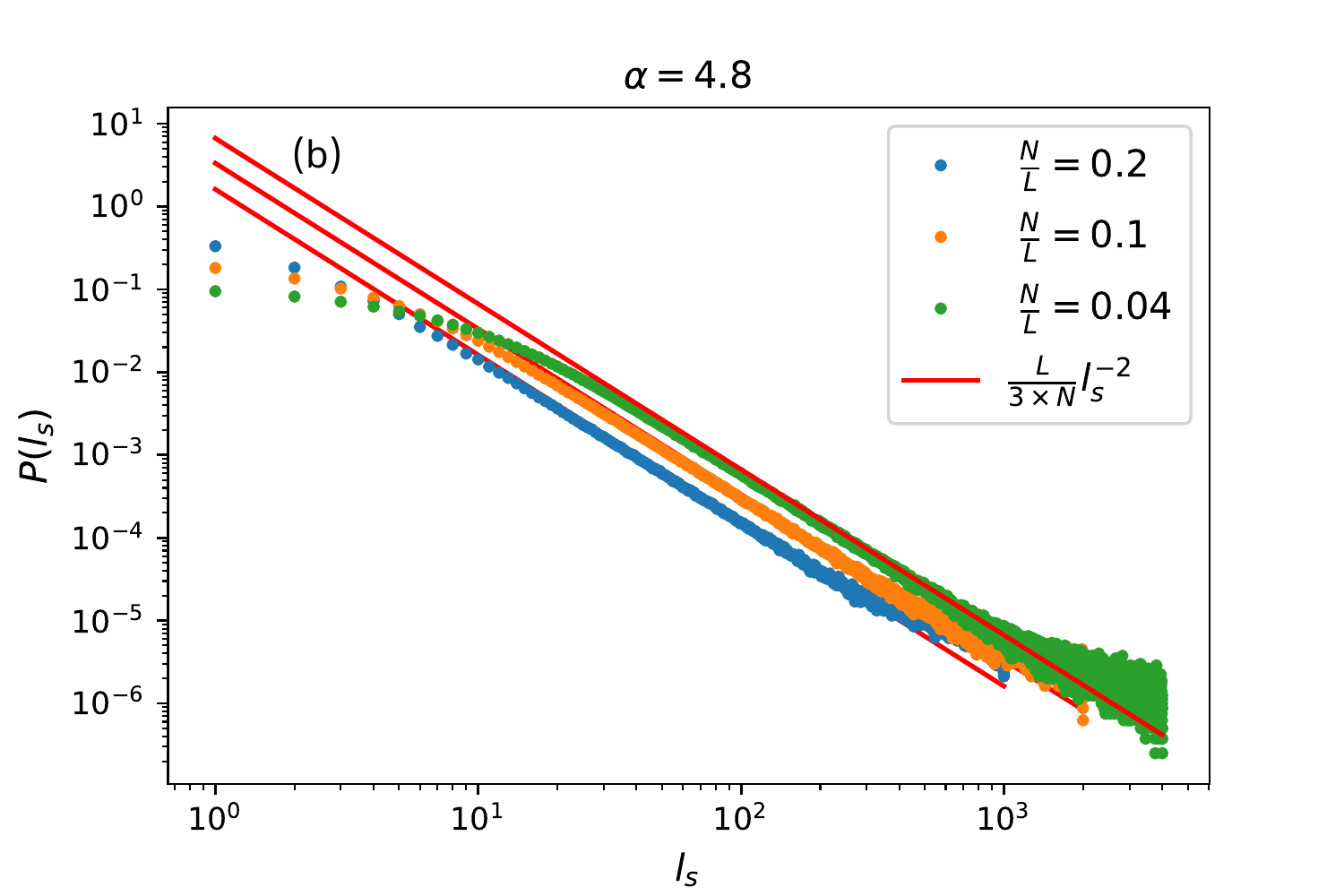}
\caption{Probability distribution of singlet lengths   $P(l_s)$, where  $l_s$ is the 
real physical distance, on a logarithmic scale, obtained via SDRG for $N=200$ and for various filling factors $N/L=0.2,0.1, 0.04$. The data were generated for $10^5$ disorder realizations with  fixed  $\alpha=1.8$ (a) and $\alpha=4.8$ (b). Red dashed lines: function 
$\frac{L}{3 N} l_{s}^{-2}$,  fitting 
the intermediate $a \ll l_s \ll  L$  scaling regime. }
\label{RealSLPDF}
\end{figure}

Here we used $\mu = 1/(2 \alpha)$ and the fact
that $\Omega_0 = J_0 $ is the largest energy
 scale. So far,  we have 
 assumed that $l$ can take any value within the interval $[a,L].$
   Taking into account that the distance can in fact 
   only take discrete values $l_i = i a,$
the properly normalized probability mass function $p_i$ for $L/a \rightarrow \infty$  would be
 $   p_i = 6/(\pi^2 i^2)$.
 However, as the renormalization of the couplings will become more important as the RG scale $\Omega$ is lowered, corresponding to intermediate and long length scales $l$, let us check the actual distribution in that regime, by numerical implementation of the SDRG. In Fig. \ref{RealSLPDF} we show  results for $N=200$ spins, $\alpha =1.8$ and  $\alpha =4.8$ and various 
  filling factors $\frac{N}{L}=0.2,0.1, 0.04$. We find that the prefactor scales with the inverse density, the average spacing between the spins $l_0 = L/N$ as
  \begin{equation} \label{plsnum}
  P(l_s) \approx \frac{L}{3 N} l_{s}^{-2}, ~{\rm for} ~a \ll l_s \ll L,
  \end{equation}
  for both $\alpha =1.8$ and  $\alpha =4.8$ in the intermediate scaling regime. We find a drop off at small $l_s < l_0 = L/N$, and 
   a slower decay for large $l_s \rightarrow L$. Note that the numerical
   $P(l_s)$ is properly normalized, when summing over all $l_s \in [a,L-a]$. 

\section{Concurrence}\label{sec:concurrence}
The entanglement between any two spins of a chain can be quantified by the concurrence between them \cite{wootters}. When 
 the spins  in  the chain are in a pure state,  $|\psi\rangle$, such as the RS state,   the concurrence between spins $m$ and $n$ is given by the correlation function
\begin{equation}\label{Conc_mn}
C_{mn}=|\langle\psi|\sigma_m^y\sigma_n^y|\psi\rangle|,
\end{equation}
which is the absolute value of the overlap between the original state and the state  obtained after spins $m$ and $n$ have been flipped. 
As reviewed above, 
the SDRG procedure  yields  a random singlet state (RSS) as an approximation of the ground state of the system, even when long-range couplings are present. The RSS is known to  become asymptotically exact for short-range models, characterized by the infinite randomness fixed point (IRFP) \cite{bhattlee}. This RSS is a product state that can be written in the form
\begin{equation}\label{RSS}
|\psi_{RS}\rangle=\bigotimes_{\{i,j\}\in RS}|0_{ij}\rangle,
\end{equation}
where $|0_{ij}\rangle=\left(|\uparrow_i\,\downarrow_{j}\rangle-|\downarrow_i\,\uparrow_j\rangle\right)/\sqrt{2}$ is the singlet state between spins enumerated by  $i$ and $j$, and the direct product extends over all singlets forming the RSS. From Eq. (\ref{RSS}) it becomes apparent that when the system is in the RSS, the concurrence between the two spins $i,j$ is given by
\begin{equation}\label{CmnRSS}
C_{ij}=\begin{cases}
				1 &\text{if $i$ and $j$ form a singlet in the RSS},\\
				0 &\text{otherwise}.
			\end{cases}
\end{equation}
Thus,  as the RSS disregards any but the strongest couplings,  it fails to account for corrections by residual weak couplings. 
In order to include  the finite amount of entanglement  which prevails  between spins that do not form a singlet in the RSS, and in turn weakens the entanglement between  the spins that do form a singlet during the SDRG procedure, we need to find a consistent strategy to include these corrections in the SDRG scheme.\par 
Before proceeding any further,  let us first  review  the known results for the scaling behavior of the mean and typical concurrence at the IRFP, i.e. the fixed point for short-range coupled random spin chains, which will serve as a baseline to compare with results obtained in the SDFP. 

\section{Mean and typical Concurrence at the IRFP}
As Fisher noted in Ref. \cite{fisher2}, the mean correlation function between spins  at long distances $l$ 
 is dominated by rare events.  Typically, two distant spins enumerated by  indices $n_1$ and $n_2=n_1+n$ will not form a singlet and will therefore  be very weakly correlated. However, in the rare event that they do form a singlet in the RSS, they will be strongly correlated and therefore will dominate the mean  correlation function, and thereby the concurrence,  at large distances. As a result, the mean correlation function must be proportional to the ratio of singlets formed at index  distance $n$, $P(n)$, which is related to the distribution function of real distances $l$, $P(l)$, which we have 
 reviewed above.  At the IRFP, $P(l) \sim a/ l^2$,
  and we see that the mean correlations decay faster when disorder is introduced: in the clean case they decay more slowly as $l^{-1/2}$.   Hoyos \textit{et. al.} explicitly noted that for chains with open boundary conditions $C(n)=0$ for even index distance $n$, which yields the more accurate result at the IRFP \cite{hoyos},\footnote{In Ref. \cite{hoyos} the spin  correlation function  instead of the concurrence is calculated, explaining  the difference in  by a factor $S(S+1) = 3/4$.},
\begin{equation}\label{meanC_RSS-hoyos}
\bar{C}(n) \sim \dfrac{1}{n^2}\times\begin{cases} 1 &\text{if}\ n \ \text{is odd},\\
0 &\text{if}\ n \ \text{is even}.
	\end{cases}					  
\end{equation}
 It was also noted in Ref. \cite{hoyos} that there are additional terms in 
 $\bar{C}(n)$ that decay faster for large $n$, the next leading term decaying as  
 $1/n^{5/2}$.
While  rare events dominate the mean concurrence, we expect a different behavior for the typical concurrence which  is  proportional to the typical value of the coupling between two spins $J$ at the RG scale $\Omega_l$ at which they are decimated. This value can be calculated using the full  IRFP distribution of $J$. Thereby, one finds  \cite{fisher2,hoyos},
\begin{equation}\label{typC_RSS}
C_{typ}(l)\sim e^{-k\sqrt{l}},
\end{equation}
where it was used that at the IRFP the distance between spins is related to the RG scale $\Omega_l$ by 
$l\sim \ln^2(\Omega_0/\Omega_l)$ to obtain the scaling behavior  as an extended exponential, with $k$ being a non-universal constant of order unity \cite{fisher2}.
 Thus, at the IRFP the typical value of the concurrence decays  exponentially  fast with distance, whereas its mean decays  with a power law. In 
 the next section we  introduce a 
 general approach to include  corrections to the RS state, and we will, in particular, investigate how the typical concurrence decays with $l$ at the SDFP. 

\section{Corrections to the Random Singlet State}
 To incorporate the effects of  the  couplings which are neglected in the RS state   we define  an effective Hamiltonian as the sum of the Hamiltonion $\tilde{H}_0$ with all  effective, renormalized couplings taken into account in the SDRG, and a perturbation term $\tilde{H}'$, which includes all couplings which are neglected in the SDRG, 
\begin{equation}\label{Hforcorrections}
\tilde{H}=\underbrace{\sum_{\{ij\}\in RS}\tilde{J}_{ij} S_i\,S_j}_{\tilde{H}_0}+\underbrace{\sum_{\{ij\}\notin RS}\tilde{J}_{ij}S_i\,S_j}_{\tilde{H}'}.
\end{equation}
Now, we can perform  perturbation theory in the term $\tilde{H}'$ to obtain  
the  ground state of the disordered XX chain  in  first order  of $\tilde{H}'$ as, 
\begin{equation}\label{Psi_perturbation}
| \psi \rangle=| \psi_{RS} \rangle + \sum_{\beta}\dfrac{\langle \psi_\beta|\tilde{H}'| \psi_{RS} \rangle}{E_{RS}-E_\beta}| \psi_\beta \rangle,
\end{equation}
where $E_{RS}$ is the ground state energy of the RS state $| \psi_{RS} \rangle$, and
the sum runs over all excited states of $\tilde{H}_0$, $| \psi_\beta \rangle$, as labeled by the index $\beta$, 
with eigen energies $E_{\beta}$. The excited states can be obtained by   combinations of    triplet  states, as 
  obtained by  excitations of  the  singlets  in  the  RS state. 
  \begin{figure}
\begin{center}
\includegraphics[scale=0.5]{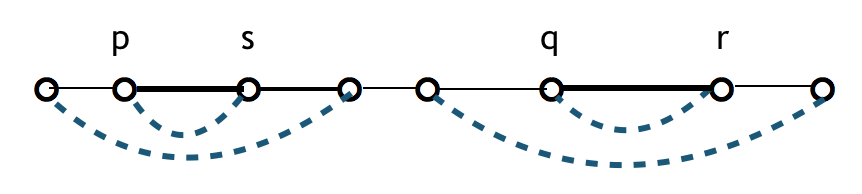}
\includegraphics[scale=0.5]{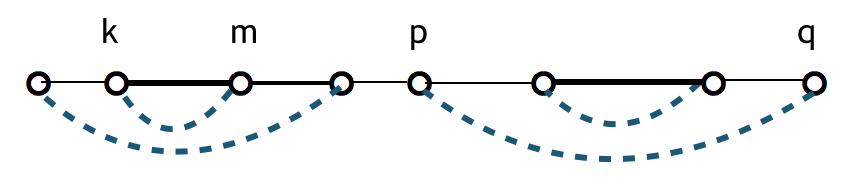}
\end{center}
\vspace*{-12pt}
\caption{ Full lines indicate strongest  bare couplings. Some weaker couplings are indicated by thinner lines. Dashed lines connect  spins that form singlets in the random singlet state. Top: Spins  $\{p,q\}$ of Eq. (\ref{CorrectedC}) do not form a singlet with each other. Bottom:  $\{pq\}$ of Eq. (\ref{CorrectedC}) do  form a singlet. }\label{fig:pq_cases}
\end{figure}
We note the useful relation
\begin{equation}\label{H'onRS}
\begin{split}
\tilde{H}' |\psi_{RS}\rangle & =   \dfrac{1}{4}\!\!\sum_{\{nl\}\neq\{mk\}}\!\!\!\!\!\!(J_{nm} ~ + J_{lk} ~ - J_{nk}~ - J_{ml})\times \\ &
(| +_{nl} \rangle | -_{mk} \rangle \!+\!| -_{nl} \rangle | +_{mk} \rangle)\!\!\!\!\bigotimes_{\substack{\{ij\}\neq \{nl\}\\ \phantom{\{ij\}}\neq\{mk\}}}\!\!\!\! |0_{ij} \rangle,
\end{split}
\end{equation}
where $|\pm_{nl}\rangle\equiv|(S=1,M=\pm1)\rangle$ are two of the triplet states
 formed of the spins  $\{l, n\}$ when  the spins  $l$ and $n$ form a singlet in the RSS.
 Thus, the double sum and the direct product run over all singlet pairs in the RSS, with the exceptions specified under the summation and direct product signs\footnote{This notation will be used throughout to simplify the long expressions involved.}. From this result, it becomes apparent that the only excited states that contribute to the sum in Eq. (\ref{Psi_perturbation}) are of the form
\begin{equation}\label{Psi_beta}
|\psi_\beta\rangle=|\pm_{nl}\rangle|\mp_{mk}\rangle\bigotimes_{\substack{\{ij\}\neq \{nl\}\\ \phantom{\{ij\}}\neq\{mk\}}}\!\!\!\! |0_{ij} \rangle,
\end{equation}
whose energy difference  to the RS state is given by  $E_\beta- E_{RS}=(J_{nl}+J_{mk})/2$. With this in mind, Eq. (\ref{Psi_perturbation}) transforms into the final form of the  ground state  with corrections,
 \begin{equation}\label{Psi_corrected}
\begin{split}
 |\psi\rangle =& c_{\psi}|\psi_{RS}\rangle  
 		-\frac{c_{\psi}}{2}\!\sum_{\{nl\}\neq\{mk\}}\!\!\!\!\!\! 
      \dfrac{J_{nm}+ J_{lk} -J_{nk} -J_{ml}}{J_{nl} + J_{mk}} \times \\ &( | +_{nl} \rangle | -_{mk} \rangle +| -_{nl} \rangle | +_{mk} \rangle
 )\!\!\!\!\!\!\bigotimes_{\substack{\{ij\}\neq \{nl\}\\ \phantom{\{ij\}}\neq\{mk\}}}\!\!\!\!\!\! |0_{ij}\rangle,
\end{split}
\end{equation}
with a normalization constant $c_{\psi}$ given by
\begin{equation}\label{c}
c_{\psi}=\left(1+\dfrac{1}{2}\sum_{\{nl\}\neq\{mk\}}
      \left(\dfrac{J_{nm}+ J_{lk} -J_{nk} -J_{ml}}{J_{nl} + J_{mk}}\right)^2\right)^{-1/2}.
\end{equation}

Now, we are all set, and we can  use the perturbed ground state, given by Eq. (\ref{Psi_corrected}), to derive the concurrence
between the spins with indices $p$ and $q$.
We find the conditional expression 
\begin{equation}\label{CorrectedC}
\begin{split}
C_{pq}^{NS}=&
			c_{\psi}^2\left|\dfrac{J_{pq}+J_{rs}-J_{pr}-J_{qs}}{J_{ps}+J_{qr}}\right| \text{if} \ \{pq\}\notin RS ,
			\\ C_{pq}^{S}=&
			1-\dfrac{c_{\psi}^2}{2}\sum\limits_{\{mk\}\neq\{pq\}}\left(\dfrac{J_{pm}+J_{qk}-J_{pk}-J_{mq}}{J_{pq}+J_{mk}}\right)^2 \\ & \text{if} \ \{pq\}\in RS, 
			\end{split}
\end{equation}
where $C_{pq}^{NS}$ is the concurrence between spins $p$ and $q$ if they do not form a singlet in the RS state, 
and $C_{pq}^{S}$ is the one, if they do. 
The indexes $r$ and $s$ in the first line  correspond to the spins that form a singlet in the RS state with spins $q$ and $p$, respectively, as shown in Fig. \ref{fig:pq_cases}.\par

Eq. (\ref{CorrectedC}) has all the properties expected for the concurrence between two spins in a chain with long-range couplings. It does  give a non-zero value for all pairs that do not form a singlet in the RS state, and it  gives a concurrence smaller than 1 for spins that do. \footnote{Note that with the given definition of $c$, the concurrence for $\{pq\}\in RS$ is always positive.} 

\begin{figure}[htb]
\begin{center}
\includegraphics[width=0.35\textwidth]{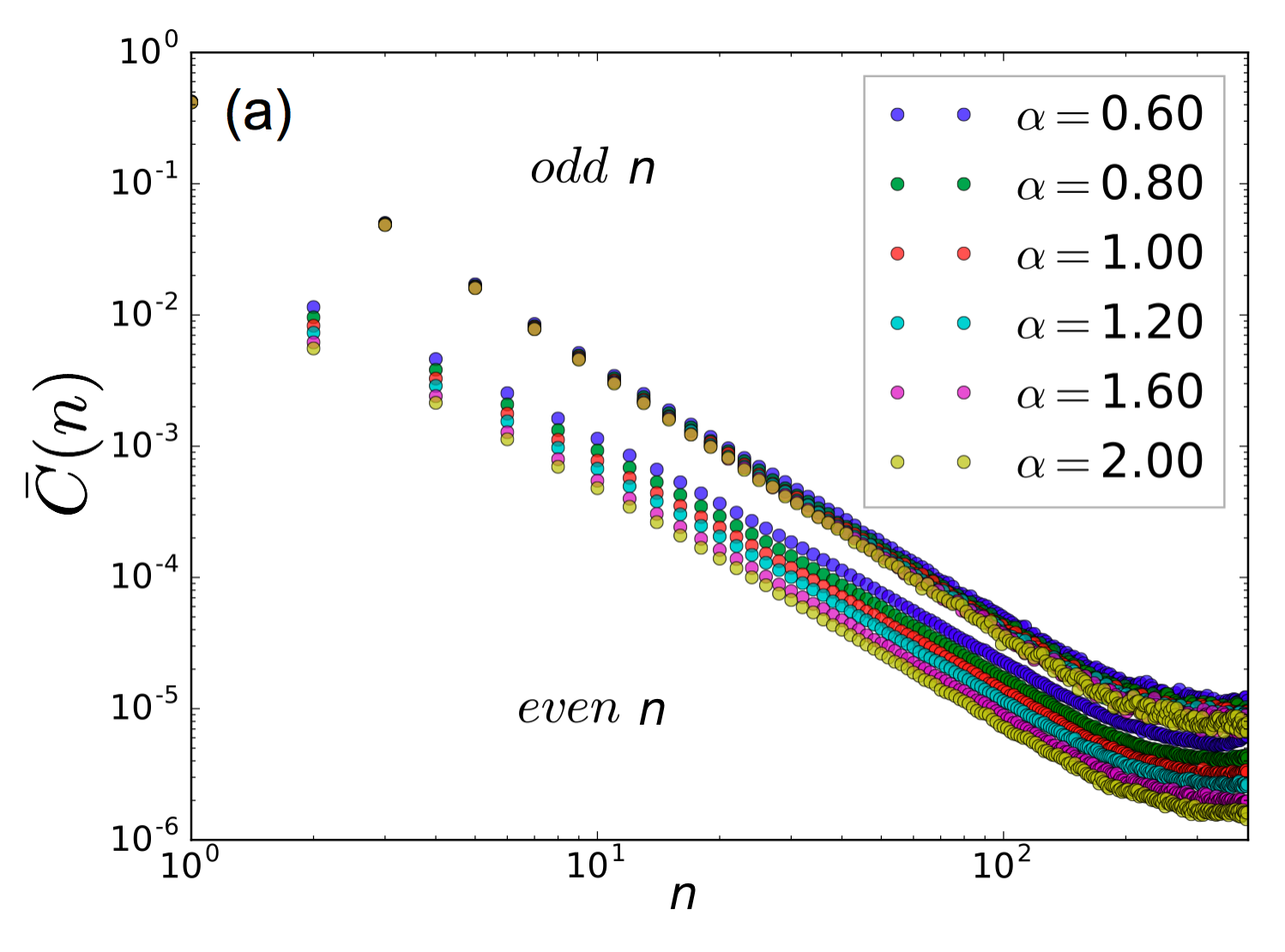}  \includegraphics[width=0.35\textwidth]{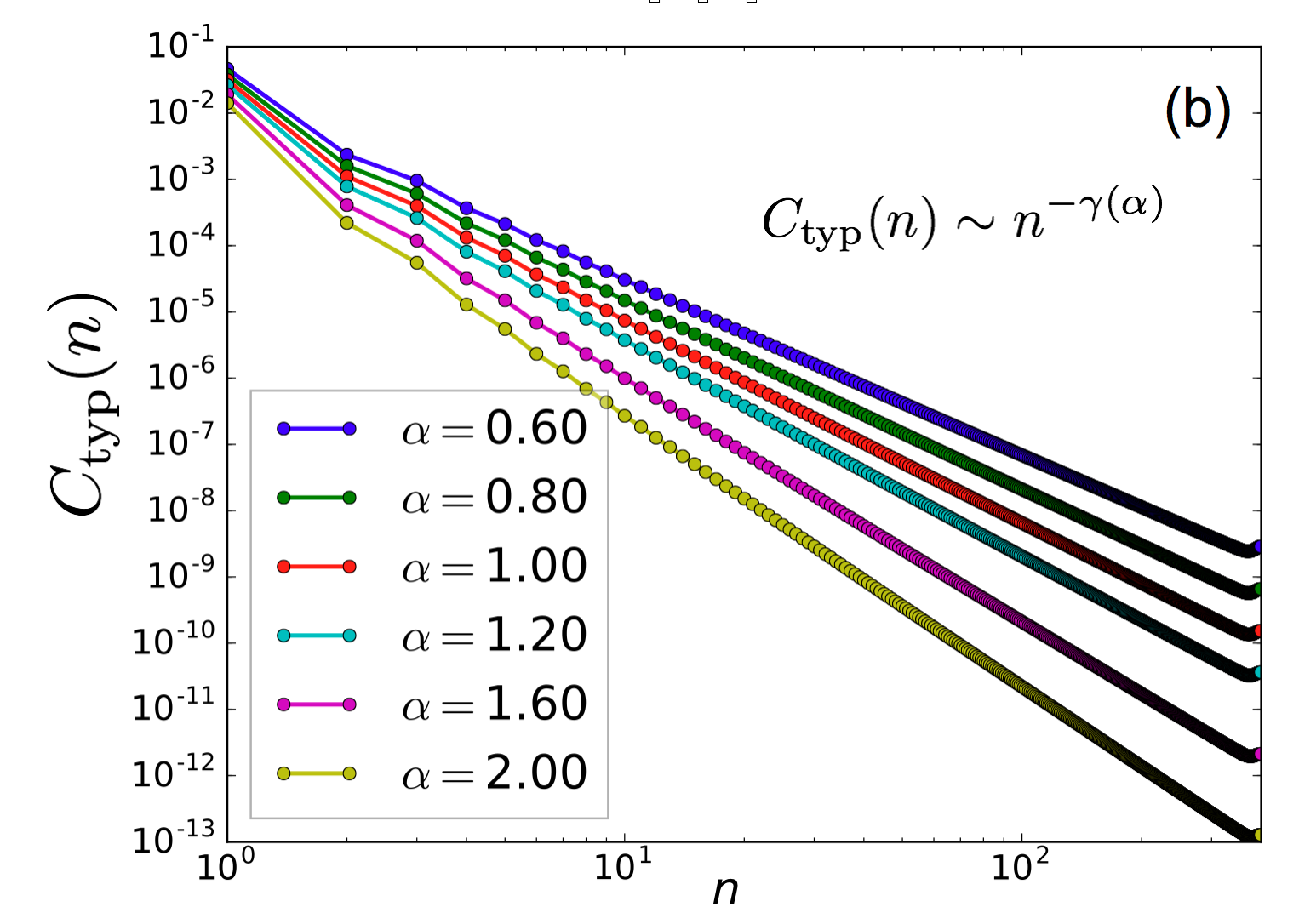}
 \includegraphics[width=0.35\textwidth]{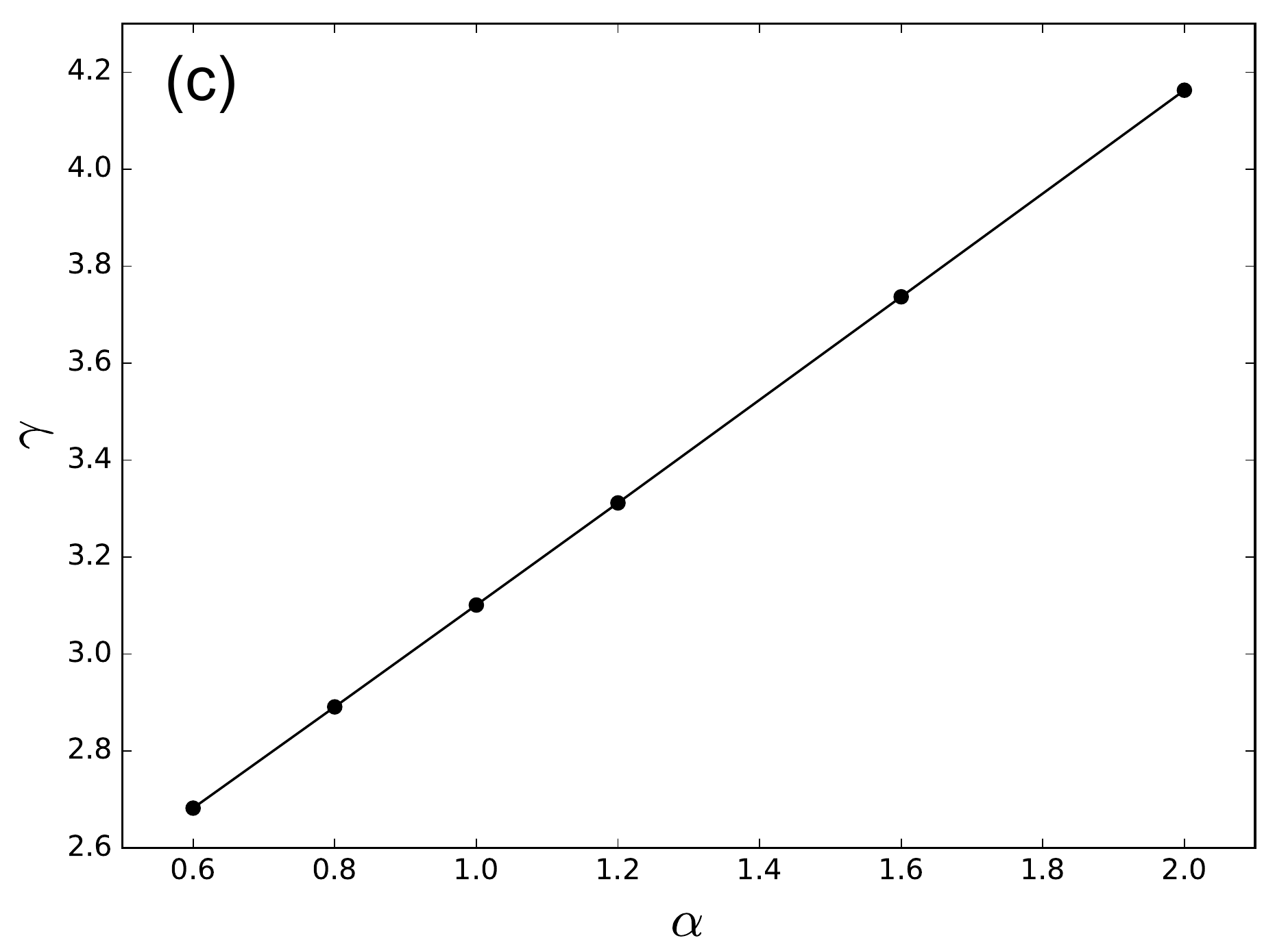}
\end{center}
\caption{(a) Mean concurrence of disordered  XX chains   with  power-law, long-ranged  couplings, averaged over $M=18960$  realizations, as function of index distance $n$. 
Power exponents of the coupling strengths (Eq. 2): $\alpha=0.6,...,2.0$,  $N=800$ spins placed randomly on one of 
 $L= 100 N$ sites. $n=|p-q|$ is the index distance. Top curves: odd  $n$, bottom curves: even  $n$. (b) typical concurrence as a function of index distance $n$. (c) Fits of linear regime in the typical concurrence to a   power-law decay with  power $\gamma$. The fitted exponents $\gamma$ are  plotted  as a function of $\alpha$.}\label{fig:Cmean&typ}
\end{figure}

In  Fig. \ref{fig:Cmean&typ}(a) we show numerical results for the mean concurrence as a function of index distance $n$ in the disordered long-range XX chain with $N= 800$
 spins randomly placed on $L = 80 000$ sites, as obtained 
by inserting the numerical results of the SDRG for the couplings into Eq. (\ref{CorrectedC}). First, we note that, in contrast with the short-range case, there is a finite concurrence for even values of  $n$. This is expected by looking at the form of Eq. (\ref{CorrectedC}) and recalling that $C(n)=0$ for even $n$ was due to the impossibility of crossing singlets in the RS state. However, there is still a clear difference between even values of $n$ (bottom) and odd ones (top), as indicated by the  clear separation of  two sets of  curves. Both sets of curves have a weak dependence on $\alpha$ and a regime in which it can be fitted with a power law,
\begin{equation}\label{Codd-even}
\bar{C}(n)\sim n^{-\gamma_{e,o}},
\end{equation}
where $\gamma_{e,o}$ are the decay powers for even and odd values of $l$, respectively. In fact, by using linear regression fits in the logarithmic scale  we find  $\gamma_e=1.75\pm 0.04$ and $\gamma_{o}=1.95\pm 0.04$ for both $\alpha=0.6$ and $\alpha=2.0$. 
Even with the corrections to the RS state the concurrence for odd values of $n$ is still dominated by rare events (singlets formed at long distances), as indicated by the small deviation from 
$\bar{C(n)} \sim n^{-2}$ decay for all values of $\alpha$. On the other hand, for even values of $n$, the decay is slower, the power-law regime is smaller, and the amplitudes have a stronger dependence on $\alpha$, as expected from Eq. (\ref{CorrectedC}). For both sets of curves, a saturation at large values of $n$ can be seen, which is a finite size effect.

The typical value of the concurrence is shown in Fig. \ref{fig:Cmean&typ}(b). A clear power-law behavior of the form
\begin{equation}\label{Ctyp}
C_{typ}(n) = \exp(\langle \ln C(n) \rangle)\sim n^{-\gamma(\alpha)},
\end{equation}
is found. Here, the power $\gamma(\alpha)$  has a strong dependence on $\alpha$, unlike the decay powers of the mean value. In fact, we find $\gamma(\alpha)$ to be linear in $\alpha$, with a  linear regression fit giving 
\begin{equation}\label{gamma-alpha}
\gamma(\alpha)=1.02\,\alpha+2.02,
\end{equation}
which indicates that the typical concurrence decays faster than the mean concurrence for all values of $\alpha$.
It is also worth noting that since the typical concurrence decays as a power law, it decays slower than in the IRFP case, where it has the extended exponential behavior stated in Eq. (\ref{typC_RSS}). 

\begin{figure}[hbt]
\begin{center}
\includegraphics[scale=0.63]{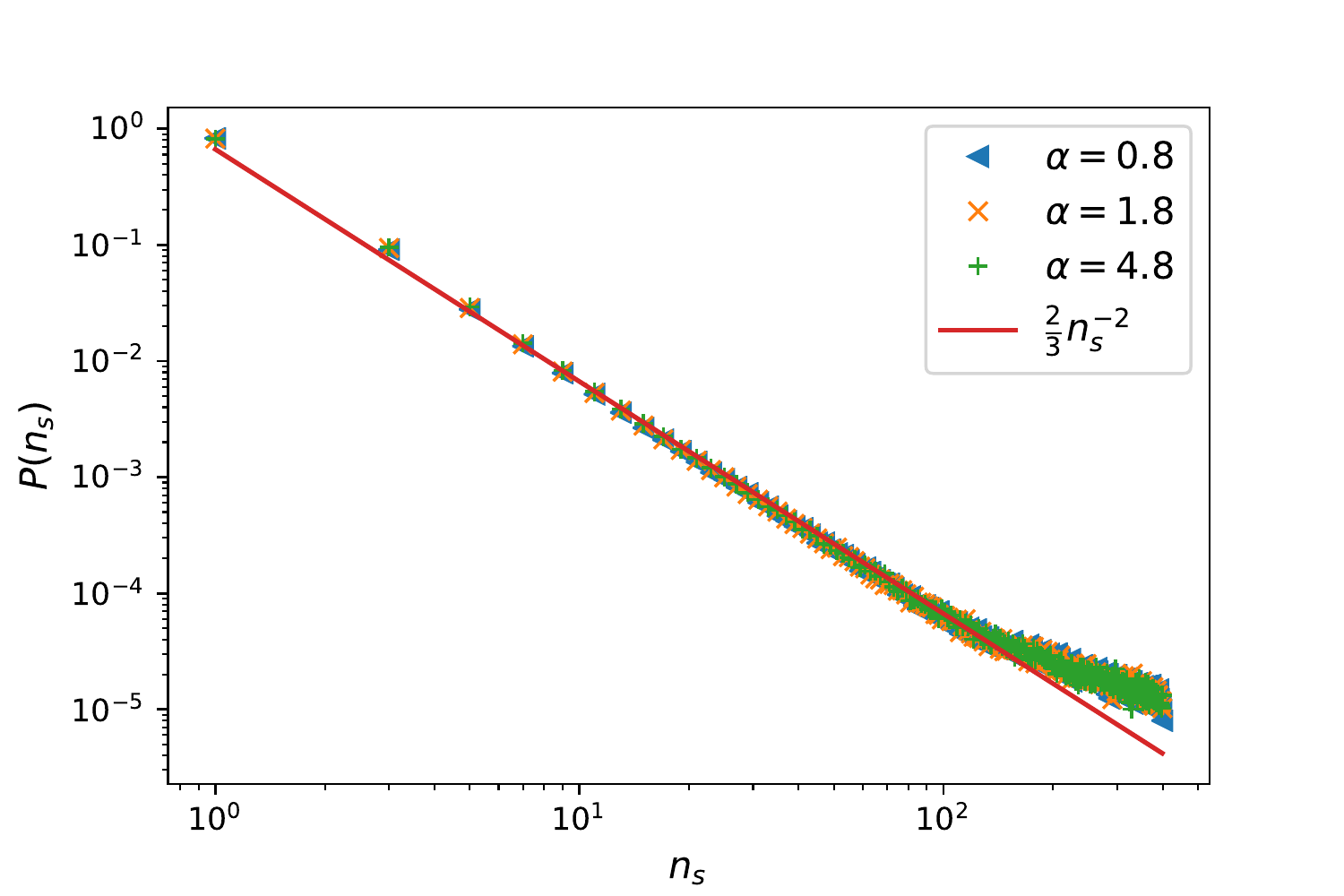}
\end{center}
\caption{Probability distribution of the singlet lengths,  $P(n_s)$, as a function of index distance, on a logarithmic scale, obtained via the SDRG for N=400 and a filling factor $\frac{N}{L}=0.1$. The data were obtained using 25000 realizations of the disorder for every  $\alpha$. The red dashed line corresponds to the function 
$\frac{2}{3} n_{s}^{-2}$, which fits 
the intermediate $n_s$  scaling regime. }
\label{indexSLPDF}
\end{figure}

 Now, let us see if we can use the corrections to the RS state,   Eq. (\ref{CorrectedC}) to find the scaling behavior of the typical concurrence analytically. The probability mass function as a function of the index distance $n$ 
 decays to leading order as $1/n^2$, in accordance
 with the distribution of lengths, Eq. (\ref{plsnum}). Thus, 
  noting that in the random singlet state only spins at odd index distance are paired we find 
\begin{equation}\label{pmfn}
P(n)=\left(c_2 \dfrac{1}{n^2} + O( \dfrac{1}{n^{5/2}})\right)\times\begin{cases} 1 &\text{if}\ n \ \text{is odd},\\
0 &\text{if}\ n \ \text{is even}.
	\end{cases}					  
\end{equation}
 Since $P(n)$ must be normalized, $\sum_{n=1}^{N-1} P(n) =1$, the 
  coefficient $c_2$ depends on the weight of the faster decaying terms. 
  In Ref. \cite{hoyos}, $c_2 = 2/3$ was derived for the IRFP of the short ranged disordered AFM spin chain. 
 In Fig. 
\ref{indexSLPDF} 
we show histograms of the index singlet length distribution  $P(n_s)$ on a logarithmic scale, obtained using the SDRG for N=400 with a filling factor $\frac{N}{L}=0.1$. The data were obtained from 25000 realizations of the disorder for each $\alpha$. The red dashed line corresponds to the function $\frac{2}{3} n_{s}^{-2}$, 
showing good agreement for intermediate and large values of $n_s$, in agreement with  Eq. (\ref{pmfn}) with $c_2 = 2/3$.
We also observe
deviations due to a faster decaying term at small $n$, in agreement with  Eq. (\ref{pmfn}). 
 The saturation at large values of $n$ has been checked to be a finite size effect.

 Now, 
  we can calculate the mean value of 
  the concurrence  as function of index distance $n$ via 
  \begin{equation}\label{mc}
\bar{C}(n) = \langle C_{pq}\rangle_{n=|p-q|} = P(n) C_{pq}^S + (1-P(n)) C_{pq}^{NS}, 
\end{equation}
and the typical value of the concurrence  via
 \begin{equation}\label{tc}
C_{typ}(n) = \exp \left( P(n) \ln   C_{pq}^S  + (1-P(n)) \ln C_{pq}^{NS}  \right).
\end{equation}

We note that in the random singlet state without corrections, Eq. (\ref{mc}) gives with Eq. (\ref{pmfn})
\begin{equation}\label{pmfnrss}
C^{RSS}(n)=(c_2 \dfrac{1}{n^2} + O(\dfrac{1}{n^{5/2}}) ) \times\begin{cases} 1 &\text{if}\ n \ \text{is odd},\\
0 &\text{if}\ n \ \text{is even},
	\end{cases}					  
\end{equation}
when  properly normalized, so that $\sum_{n=1}^{\infty}C^{RSS}(n) =1,$ since each spin can form 
 a singlet with only one other spin, in which case the concurrence
 is exactly equal to one, giving $c_2 = 2/3$.   Now, we can find the scaling of $C_{pq}^{NS}$ including the corrections to the RSS 
 by noting that the distance $l=|r_p-r_q|$
  is always larger than the distance between
  the spins which formed singlets in the RS state, see Fig. \ref{fig:pq_cases} top,
  $l_1=|r_p-r_s|$, $n_2=|r_q-r_r|$. Thus, we can Taylor expand  
Eq. (\ref{CorrectedC}) in $l_1/l$ and $l_2/l$. 
Thereby we
 find 
\begin{eqnarray}\label{cpqns}
 C_{pq}^{NS}   &=& c_{\psi}^2 \frac{l_1 l_2}{J_{l_1}+J_{l_2}} \partial_l^2 J_l
 \nonumber \\
   &=& c_{\psi}^2 \alpha (\alpha+1)\frac{l_1 l_2}{l_1^{-\alpha}+l_2^{-\alpha}} l^{-\alpha -2}.
\end{eqnarray}
 Noting that at the SDRG $J_l = J_0 (l)^{-\alpha}$ can be related to the index distance $n$
by assuming that $l$ scales with $n$ as dictated by the density of spins $n_0 = 1/l_0$,
 we can substitute
 $l \sim n l_0$.
As the bonding lengths $l_1$ and $l_2$ can take any value between $1$ and $l$, we can average over
  all possible values and find  as function of index distance 
  $C_{pq}^{NS}   = k^{NS} n^{-\alpha -2}$, with a constant $k^{NS}$.

Similarly, we can do an expansion   $C_{pq}^{S}$  
in the distance between the spins of the singlet states
in Fig. \ref{fig:pq_cases} bottom  $l = |l_p-l_q| $ and 
$l_2 =|l_k-l_m|$ to find
 \begin{equation}\label{cpqsscaling}
 C_{pq}^{S}   = 1 + c_{\psi}^2 \frac{\alpha }{4(\alpha+1)}  l^{1-\alpha}.
\end{equation}
 Thus, we can insert Eqs. (\ref{cpqns},\ref{cpqsscaling})
into Eqs. (\ref{mc},\ref{tc}) to get the scaling of the mean and typical concurrence with 
index distance $n$
\begin{eqnarray}\label{mcscaling}
\bar{C}(n) &=&  P(n) ( 1 + c_{\psi}^2 \frac{\alpha }{4(\alpha+1)}  (n l_0)^{1-\alpha}) 
\nonumber \\
&+& (1-P(n))  k^{NS} n^{-\alpha -2},
\end{eqnarray}
and the typical value of the concurrence  by 
 \begin{eqnarray}\label{tcscaling}
C_{typ}(n) &=& (k^{NS} (n l_0)^{-\alpha -2})^{1-P(n)} ( 1 + 
\nonumber \\&
 & \frac{c^2 \alpha }{4(\alpha+1)}  (n l_0)^{1-\alpha})^{P(n)}
 \sim  n^{-\alpha -2},
\end{eqnarray}
where we used  that $P(n) \ll 1$.
 For the typical value  we  thus find very good agreement of the
  power  $\gamma = 2+ \alpha$ with the  result obtained with numerical SDRG for system size,
 Eq. (\ref{gamma-alpha}), 
$\gamma(\alpha)=1.02\,\alpha+2.02$. 
 The mean value shows a more complicated behavior with different index distance regimes dominated by either $P(n)$ or the power law decay in Eq. (\ref{mcscaling}).

   When plotting the concurrence as function of physical distance $l$,
    we expect for small concentrations of spins $N/L \ll 1$ no even- odd effect.
     As we have numerically derived above $P(l) = c'_2 l^{-2} + O(l^{-5/2}) $, Eq. (\ref{plsnum})
     where it was found that $c_2'=1/3$.
    Thus, we get 
    the  average concurrence as function of real distance $l$,  
 \begin{eqnarray}\label{mcscalingl}
\bar{C}(l) &=&  P(l) ( 1 + c_{\psi}^2 \frac{\alpha }{4(\alpha+1)}  l^{1-\alpha}) 
\nonumber \\
&+& (1-P(l))  k^{NS} l^{-\alpha -2},
\end{eqnarray}
and the typical value of the concurrence  as function of the physical distance $l$ is 
 \begin{eqnarray}\label{tcscalingl}
C_{typ}(l) &=& (k^{NS} l^{-\alpha -2})^{1-P(l)} ( 1 + \frac{c^2 \alpha }{4(\alpha+1)}  l^{1-\alpha})^{P(l)}
\nonumber \\& \sim  & l^{-\alpha -2},
\end{eqnarray}
     
\begin{figure}[htb]
\begin{center}
\includegraphics[width=0.45\textwidth]{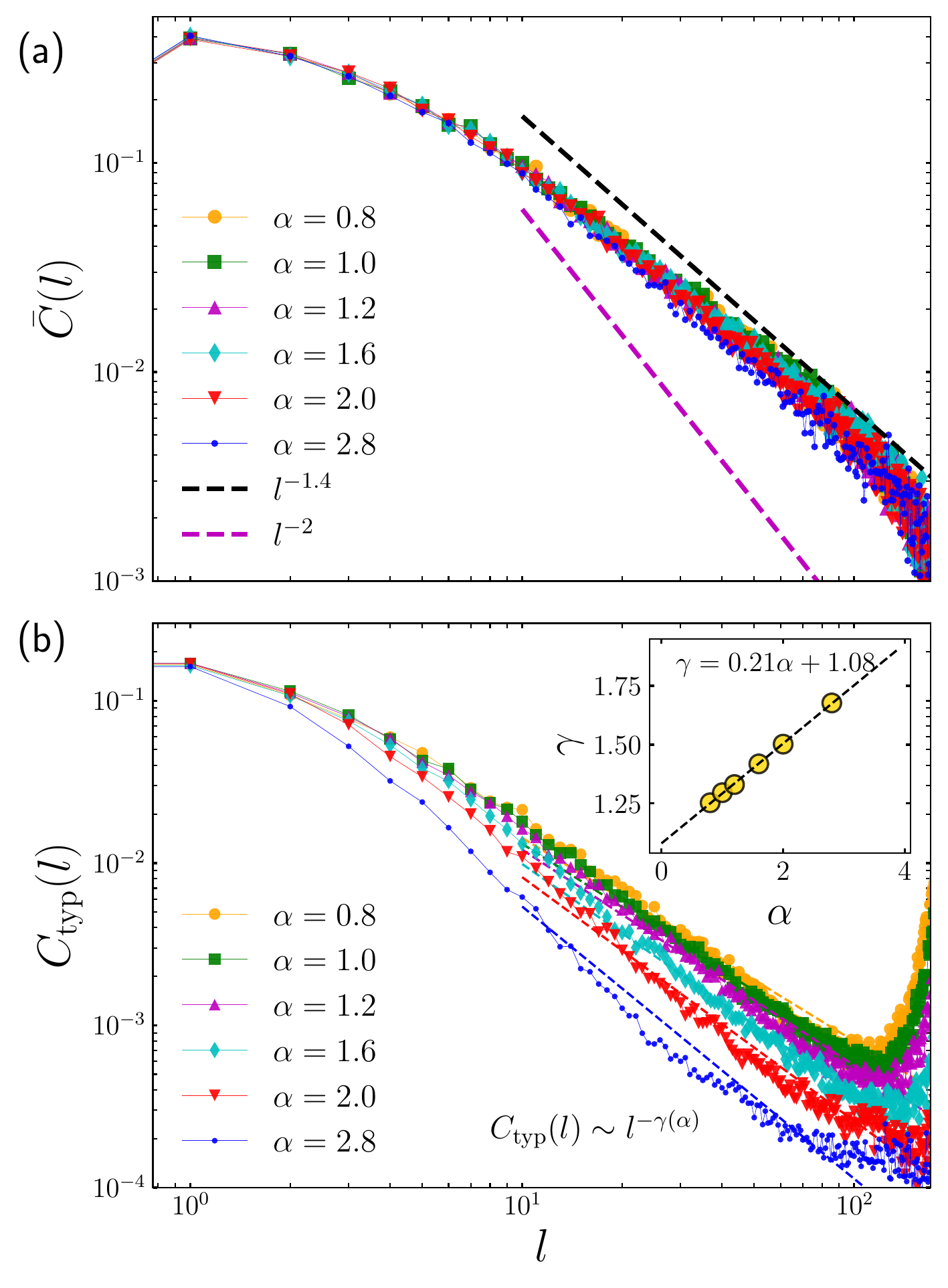}
\end{center}
\caption{Results from  exact numerical exact diagonalization. Panel (a):  Mean concurrence of disordered  XX chains   with  power-law, long-ranged  couplings, averaged over $M=2000$ number of realizations as function of the physical distance $l$ for various 
power exponents  $\alpha$,  $N=L/10 = 20$ spins placed randomly on one of 
 $L$ sites.  Panel (b):  typical concurrence as a function of the physical  distance $l$. Inset: Fits of linear regime in the typical concurrence to a  power-law decay with  power $\gamma$. The fitted exponents $\gamma$ are  plotted  as a function of $\alpha$.
 }\label{fig:Cmean&typED}
\end{figure}

  For comparison we show in Fig. \ref{fig:Cmean&typED} the results of numerical exact diagonalization for  spin chains with $N=20$ distributed randomly of length $L=200$,  as plotted as function of the physical distance $l$ and averaged over $M=2000$
  samples. 
  We find that  the average concurrence decays with a power which 
 increases slightly with increasing interaction power $\alpha$ but remains for 
 all $\alpha $
   smaller
   than the power $2$, obtained in SDRG. Also,  
  there is no 
  even odd effect, as expected when plotting the concurrence as function of 
  the physical distance $l$.  
  The typical concurrence is found to decay with a power law, 
   with exponent $\gamma = 0.21 \alpha + 1.09,$ as obtained by a fit of all results
    for $\alpha = 0.8,1.0,1.2,1.6,2.0,2.8.$, linearly increasing with $\alpha$
    as found from SDRG, but with a smaller slope. Note that the finite size effects seen in in SDRG, Fig. \ref{fig:Cmean&typ} are expected to be more dominant in the smaller size used in ED 
    as presented in Fig. \ref{fig:Cmean&typED}, which may explain the slower decay with smaller exponents observed with   ED.

\section{Entanglement Entropy}
The entanglement between two segments of a spin chain,  $A$ and $B$, can be quantified by the von Neumann entropy of the reduced density matrix, 
\begin{equation}
\label{entropy}
S=-\text{Tr}(\rho_A\ln\rho_A)=-\text{Tr}(\rho_B\ln\rho_B),
\end{equation}
where $\rho_A=\text{Tr}_B(|\psi\rangle\langle\psi|)$  and $\rho_B=\text{Tr}_A(|\psi\rangle\langle\psi|)$ is obtained by partially tracing the complete density matrix of the system over all degrees of freedom of subchain $B$ or $A$, respectively.\par 
This entanglement entropy  can be used to  characterize quantum phase transitions. For clean chains, it has been shown that at criticality, the entropy of a subchain $A$ of length $l$ scales  as \cite{vidal}
\begin{equation}
S(l)=\frac{b}{2}\dfrac{c+\overline{c}}{6}\ln(l/a)+k,
\label{S_CFT}
\end{equation}
where $c$ and $\overline{c}$ correspond to the central charges of the corresponding 1+1 conformal field theory. In the limit of infinite chains with finite partitions of length l, as well as for periodic chains with large length $L\gg l$,   $b=2$ since there are two boundaries of the partition, 
while $b=1$ for semi infinite chains, when the partition  of length $l$ is placed on one side of the chain. $k$ is a non-universal constant \cite{holzheyCFT}. This scaling behavior with a  logarithmic  dependence on the segment length $l$ is in contrast  to the area law expected for noncritical chains, where it does not depend on the length $l$ of the subchain for the one-dimensional case. The simple area law is recovered away from criticality, where it is found that the entropy saturates at large $l$ \cite{vidal,holzheyCFT}.\par 
In Ref. \cite{refael-entropy}, it was shown that Eq. (\ref{S_CFT}) also holds for the average entanglement entropy of antiferromagnetic  spin chains with random short-ranged interactions. In particular, using SDRG, they found that in the disordered transverse Ising Model, the effective central charges were given by $\tilde{c} = \overline{\tilde{c}}=\ln(2)/2$, whereas in the Heisenberg and XX model $\tilde{c}= \overline{\tilde{c}}=\ln(2)$.
Both cases correspond to a factor of $\ln(2)$ reduction  of the central charge
 and the entanglement entropy  of their corresponding pure systems\cite{vidal}, in accordance with a generalized $c$-theorem, 
  which states that if two critical points are connected by a relevant   RG flow, as here induced by the relevant disorder,  the  final  critical point has a lower conformal charge than the initial one\cite{refael-entropy}.

Eq. (\ref{S_CFT}) applies specifically to infinite systems. In Ref. \cite{calabrese}, Calabrese and Cardy derived a  formula valid for finite systems of length $L$,
\begin{equation}\label{S_b}
S_{b}(l)=b \dfrac{c}{6}\ln\left(\dfrac{L}{\pi a}\sin(\pi l/L)\right)+k',
\end{equation}
where $k'$ is a non-universal constant, and    $c=\overline{c}$ has been assumed
\cite{calabrese}. For periodic boundary conditions,  there is an additional factor $b=2$, since the subsystem is  then bounded by two boundaries, doubling the average number of singlets crossing one of the boundaries, while $b=1$ for 
open boundary conditions, when there are 2 partitions. 

\begin{figure}[hbt]
\begin{center}
\includegraphics[scale=0.4]{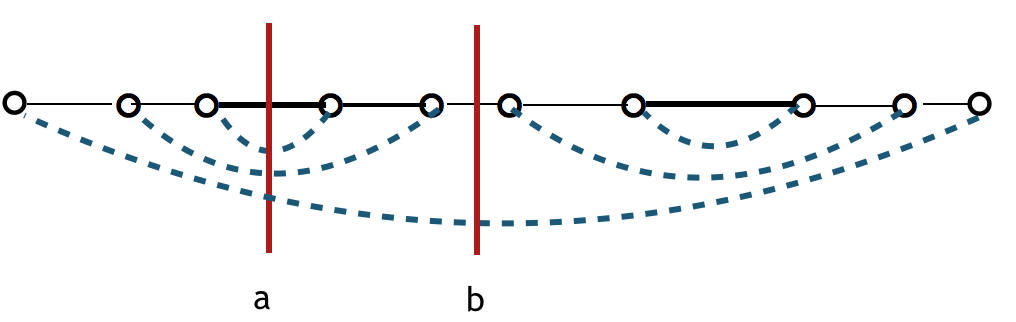}
\end{center}
\vspace*{-12pt}
\caption{Single realization of the random singlet state, illustrating the entanglement entropy calculation for two different boundaries $a$, $b$ (red lines) between subsystems.}\label{fig:S_RSS}
\end{figure}

Refael and Moore's method to calculate the average entanglement entropy in the presence of disorder is based on the assumption that the system has been drawn to the IRFP, and the random singlet state (RSS) is a correct representation of its ground state. Since the RSS corresponds to a product state of maximally entangled spin pairs, they note that the total entanglement entropy can be calculated by counting the number of singlets that cross the boundary between subsystems $A$ and $B$ and then multiplying this number with the entropy of a singlet $S_{0}=\ln(2)$ \cite{refael-entropy,hoyos}. A schematic representation of this method is shown in Fig. \ref{fig:S_RSS} for a specific realization. In this example, when the boundary between $A$ and $B$ is defined by line $a$, we obtain an entanglement entropy of $3\cdot S_0$, since three singlets cross over the boundary line. But if the boundary is defined by line $b$, the entanglement entropy is reduced to $1 \cdot S_{0}$.\par
Thus, for the   random singlet state $ | \psi_0 \rangle  $
 one finds $ S = M \ln 2$, where $M$ is the number 
  of singlets which cross the partition between 
   subsystems A and B. 
The average entanglement entropy of a random singlet state 
can therefore be
    related to the probability  to find a singlet of a certain 
     length $n_s$, in the spin chain of length $N$ \cite{refael-entropy,hoyos}, $P(n)$.
      As Refael notes in Ref. \cite{refael-entropy}, however, 
 one needs to take into account the correlation with the RG history. He found that the average distance between RG steps, where a decimated singlet may  cross the same partition, is exactly equal to $\langle m \rangle = 3$ for the disordered nearest neighbor AFM spin chain, resulting in  a  correlation factor $1/\langle m \rangle.$
     
     Another way to derive this  was outlined by  Hoyos et al. in Ref. \cite{hoyos,refael-entropy}. 
      The ratio of the number of singlets crossing the partition 
      is the probability that a spin on one side of the partition is 
    entangled with another one on the other side of the partition. 
     Thus, we can relate the entanglement entropy directly to 
      the  probability to have a singlet of  length $l$, $P(l)$, which we 
      derived in the previous chapter.  Moreover, as we consider
       partial filling of the lattice sites with spins with 
        density $n_0 = N/L <1,$ we need to distinguish the EE as a function of 
         the index distances $n= |i-j|$ between the spins 
         from the one plotted as function of 
         their physical distance $l = r_i-r_j$, where
         $r_i$ is the position of spin $i$.
   As a function of index distance $n$ we obtain   for open boundary conditions and 2 partitions, where one has the length $l$,  the 
     average entanglement entropy, 
      \begin{equation} \label{entropybycounting}
    \langle    S_n    \rangle  =  \ln 2 \sum_{n_s=1}^N \sum_i P_i(n_s)|_{\rm C. C.},
   \end{equation}
   where the crossing condition  (C.C.) ensures that only singlets are counted
    where spin 
    $i$ is on the  left  side of the partition (which contains  
    $n$ spins), while spin  $i+n_s$ is   on the right side of the partition.
    We first count the number of possible positions to place a singlet with index distance $n_s$ across the partition boundary, starting with the smallest distance $n_s=1$,
     and adding successively singlets of larger index length. 
     For $n_s < n$ there are, in principle, $n_s$ such possibilities,
     if the respective spins did not yet form a singlet with another spin,
     each with the probability to form a singlet of length $n_s$, $P(n_s)$. 
     
     In order to account  for  the correlation with existing singlets, Ref. \cite{hoyos} multiplied the probability $P(n_s)$ with a factor $1/2.$ This can be argued to be due to the fact that for every spin which may form a singlet with length $n_s$ across the boundary there is a second possibility to form a singlet, which is not crossing the boundary. 
     Thereby,  one arrives for the chain with OBC with a partition with $n$ spins and one partition boundary 
     at the following expression 
    \begin{eqnarray} \label{entropypbcnocorr}
    &&\langle    S_n    \rangle /  \ln 2  =  \frac{1}{2}
    \sum_{n_s=1}^n n_s P(n_s) 
    \nonumber \\
  &+&\frac{1}{2} n   \sum_{ n_s =n+1}^{N-n} P(n_s)  \nonumber \\ &+ & \frac{1}{2} \sum_{ n_s =N-n+1}^{N}
  (N-n_s) P(n_s).
   \end{eqnarray}
   This expression is equivalent to the one given in Ref.  \cite{hoyos} (with the difference that they considered an embedded partition with two boundaries). 
   By evaluating  Eq. (\ref{entropypbcnocorr})
   using the result Eq. (\ref{pmfnrss}) $P(n)= c_2/n^2 + O(1/n^{5/2})$ for odd $n$, $0$ for even $n$
   we
  find  in the limit of $N \gg 1$
 \begin{eqnarray} \label{entropypbc_polyg}
     \langle   S_n    \rangle  \approx \frac{1}{4} c_2  \ln 2   \ln n + k + O(1/n^{1/2}). 
   \end{eqnarray}
 We 
thus  recover by comparison with 
Eq. (\ref{S_CFT})  the conformal charge given by  
$
\tilde{c}=\overline{\tilde{c}}= (6/4) c_2 \ln 2 $.
 Using $c_2 = 2/3$, as derived in Ref. \cite{hoyos} and 
 confirmed numerically above, we thus find
 $
\tilde{c}= ln 2$ in agreement with \cite{refael-entropy,hoyos}.

For finite $N$  we can do the summations, 
    written 
   in terms of Polygamma functions as function of  n :
        \begin{eqnarray} \label{entropypbc_polyg2}
    && \langle   S_n    \rangle  =   \frac{2 \ln 2}{<m>\pi^2} 
     (2 \gamma - \frac{4}{3 - n + N} \nonumber  + \ln 16 \\ &+& 
   2 {\rm PG}[0, 3/2 + (n-1)/2]  - 2 {\rm PG}[0, 3/2 + (N-1)/2]  \nonumber \\
  &+& 2 {\rm PG}[0, 5/2 - (n-1)/2 + (N-1)/2] 
  \nonumber \\&+& n  ~ {\rm PG}[1, 
     3/2 + (n-1)/2] - N  ~
     {\rm PG}[1, 3/2 + (N-1)/2] \nonumber \\ &+& 
  (N-n)  {\rm PG}[1, 3/2 - (n-1)/2 + (N-1)/2])
   \end{eqnarray}
where $\gamma = {\rm EulerGamma}=0.577$, and  
${\rm PG}[x,y] = {\rm PolyGamma}[x,y]$ is the PolyGamma function.

 In an attempt to  take into account the correlation with the location of other singlets in the RSS state more rigourously,  we  could argue that we need to 
      multiply each term with  the probability that the two spins did not yet form a singlet 
     of other length with another spin, that is 
     $\prod_{n'_s \neq n_s}(1- P(n'_s))$. 
    Thereby, one finds  for a random singlet state, using  Eq. (\ref{pmfnrss})
    \begin{eqnarray} \label{entropypbc}
    &&\langle    S_n    \rangle /  \ln 2  =   \sum_{n_s=1}^n n_s P(n_s) \prod_{n'_s }
    \left(1- P(n'_s) \right)  
    \nonumber \\
  &+&n   \sum_{ n_s =n+1}^{N-n} P(n_s) \prod_{n'_s } \left(1- P(n'_s) \right)  \nonumber \\ &+ & \sum_{ n_s =N-n+1}^{N}
  (N-n_s) P(n_s) \prod_{n'_s } \left(1- P(n'_s) \right).
   \end{eqnarray}

By evaluating  Eq. (\ref{entropypbc}) we
  find  in the limit of $N \gg 1$
 \begin{eqnarray} \label{entropypbc_corr}
     \langle   S_n    \rangle  \approx \frac{1}{6} c_2  \ln 2   \ln n + k + O(1/n^{1/2}). 
   \end{eqnarray}
 We 
thus  recover 
Eq. (\ref{S_CFT}) with the conformal charge given by  
$
\tilde{c}=\overline{\tilde{c}}=  c_2 \ln 2  <1$.
 Here, for pure $P(n) = c_2/n^2$ for odd $n$, $P(n) = 0$ for even $n$ we have  $c_2 = 8/\pi^2 =0.81$.
If there are faster decaying correction terms $O(1/n^{5/2})$, Eq. (\ref{pmfnrss}), as is confirmed by 
  our numerical results we get  $c= c_2 \ln 2  = 2/3 \ln 2$, smaller by a factor $2/3$ than found previously.

  We can also derive the EE as function of the physical distance $l$. For small 
   filling, $N\ll L$, the even odd effect as function of  the physical distance $l$ is negligible and we get for intermediate range $a \ll l_s \ll L$, Eq. (\ref{plsnum}), as confirmed numerically. 
  We
thus  recover 
Eq. (\ref{S_CFT})   with  the  central charge given by  
$\tilde{c}=\overline{\tilde{c}}=  c' \ln 2  < 1.$
Here, using  Eq. (\ref{plsnum}) we get 
 $c' = 1$.

\begin{figure}[hbt]
\begin{center}
\includegraphics[scale=0.65]{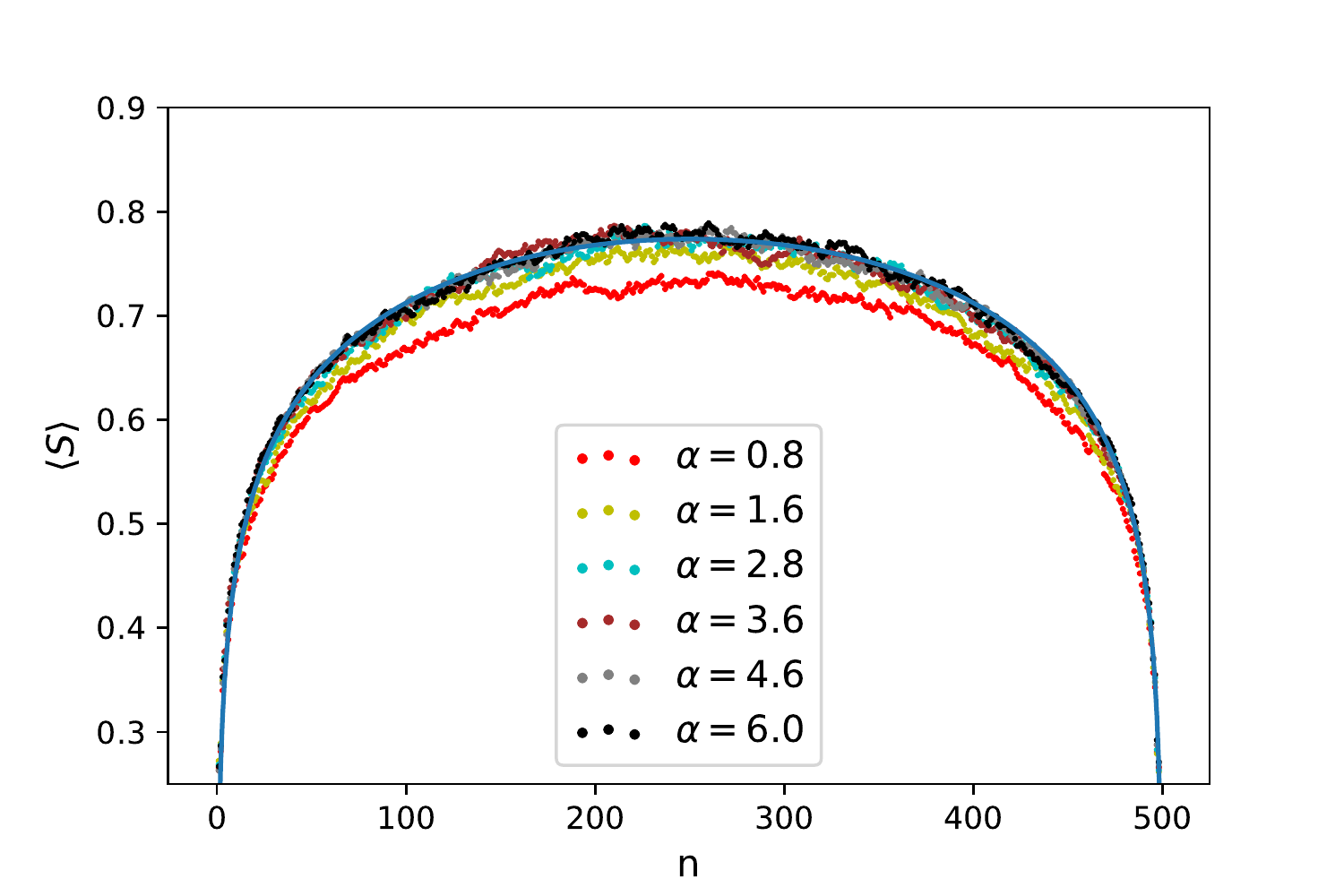}
\end{center}
\vspace{-0.4cm}
\caption{Average block entanglement entropy
as obtained with the numerical SDRG
for the long ranged XX-chain with $N=500$ spins, open boundary conditions, and various values of $\alpha$ as function of 
index distance $n$. The average was evaluated over $2 \times 10^4$ realizations for each $\alpha$, The blue  line is Eq. (\ref{S_b}) with a central charge $c=\ln(2)$
$b=1$ and $k'=0.18$.}\label{fig:entropy0}
\end{figure}

    Let us next implement the SDRG numerically. 
    Fig. \ref{fig:entropy0} shows  results, numerically  calculating  the mean block entanglement entropy using Refael and Moore's prescription illustrated in Fig. \ref{fig:S_RSS},
     which is plotted as  function of the index distance size  $n$  of the partition,  that counts how many spins are inside a partition. The XX chain with open boundary conditions has $N=500$ spins with long-range power-law interactions Eq. (\ref{jcutoff}).  As mentioned above, such a prescription implies that the RSS is a good approximation of the actual ground state of the chain. 
 
  The result for the average entanglement entropy as function of 
   index distance $n$ is shown  in Fig. \ref{fig:entropy0} for  various values of $\alpha$ with open boundary conditions. The blue  line is Eq. (\ref{S_b}) with a central charge $c=\ln(2)$,
$b=1$ and $k'=0.18$.

We thereby  find that the EE is in good agreement with 
Eq. (\ref{S_b}), and  we observe only a weak dependence on  $\alpha$, decaying by only a few percent as $\alpha$ is changed from $\alpha=6$ to $\alpha=0.8$.

\begin{figure}[hbt]
\begin{center}
\includegraphics[scale=0.65]{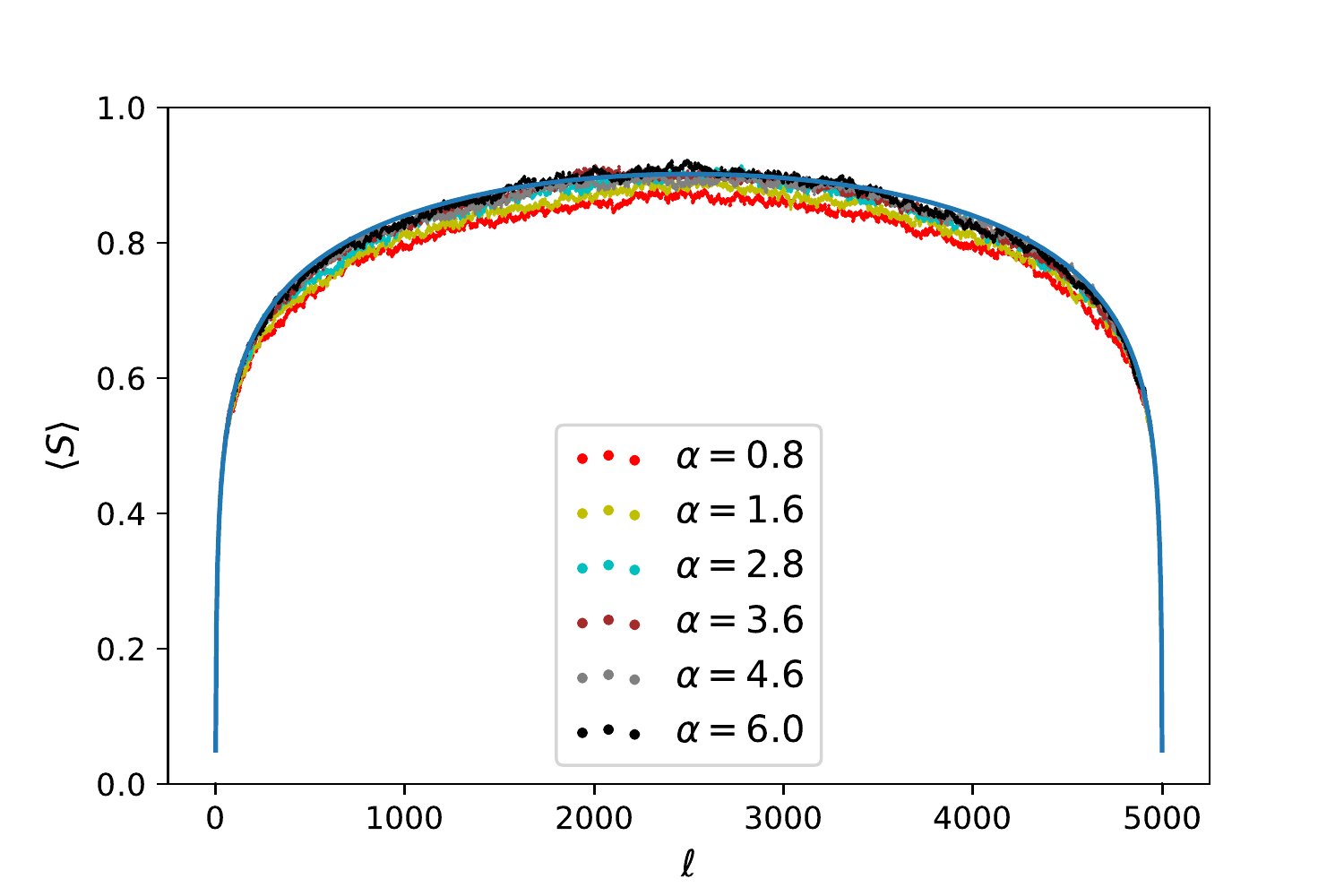}
\end{center}
\vspace{-0.4cm}
\caption{Average block entanglement entropy,  obtained from the numerical SDRG for the long ranged XX-chain of length $L=5000$, open boundary conditions for  various values of $\alpha$, and for a filling factor $\frac{N}{L}=0.1$ as function of the physical distance $l$. The average was evaluated over $2 \times 10^4$ realizations for each $\alpha$.
The blue  line corresponds to the Cardy law Eq. (\ref{S_b}) with a central charge $c=\ln(2)$, $b=1$ and $k'=0.05$.}\label{fig:entropy real length}
\end{figure}

\begin{figure}[ht]
\centering
\includegraphics[width=.45\textwidth]{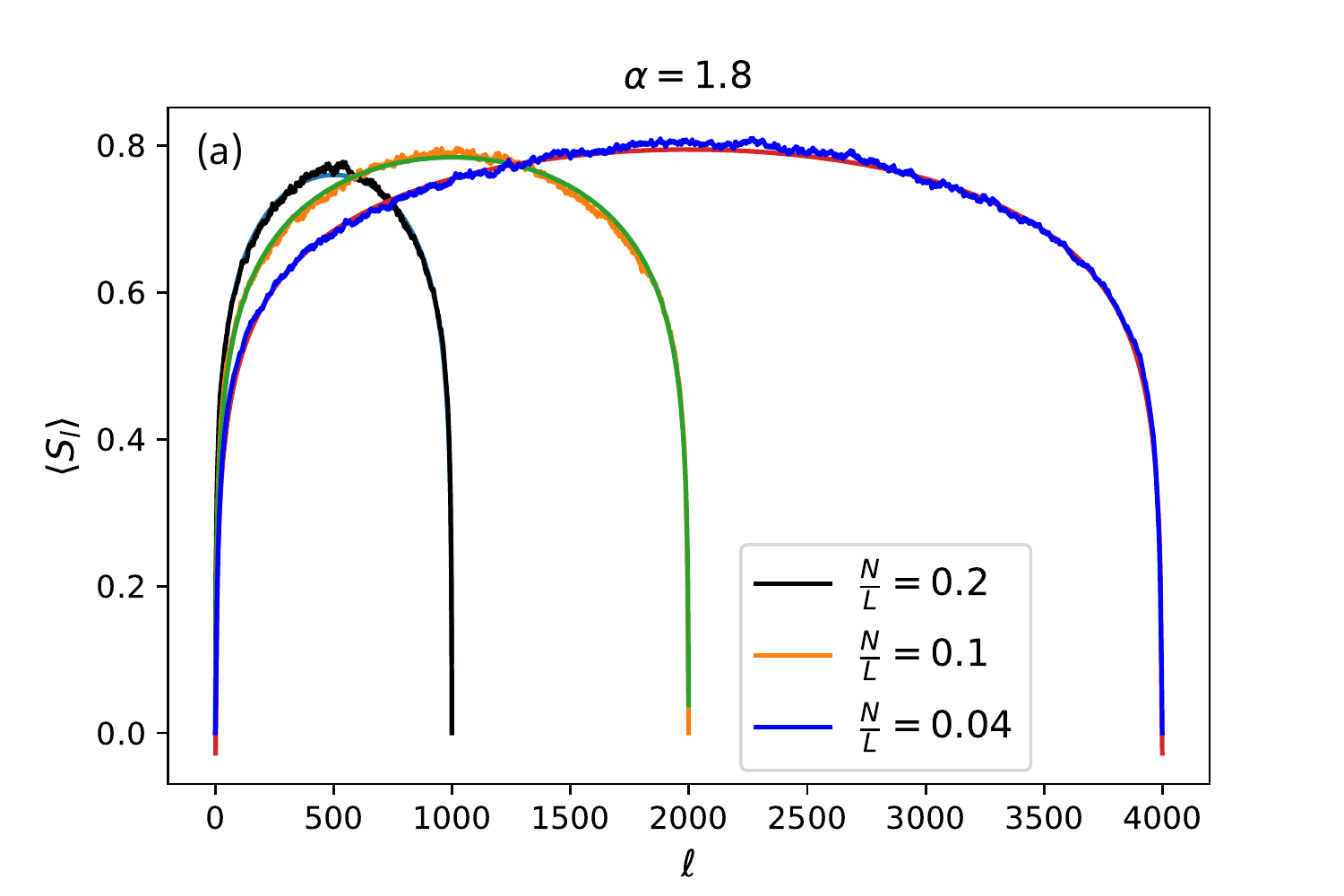}
\includegraphics[width=.45\textwidth]{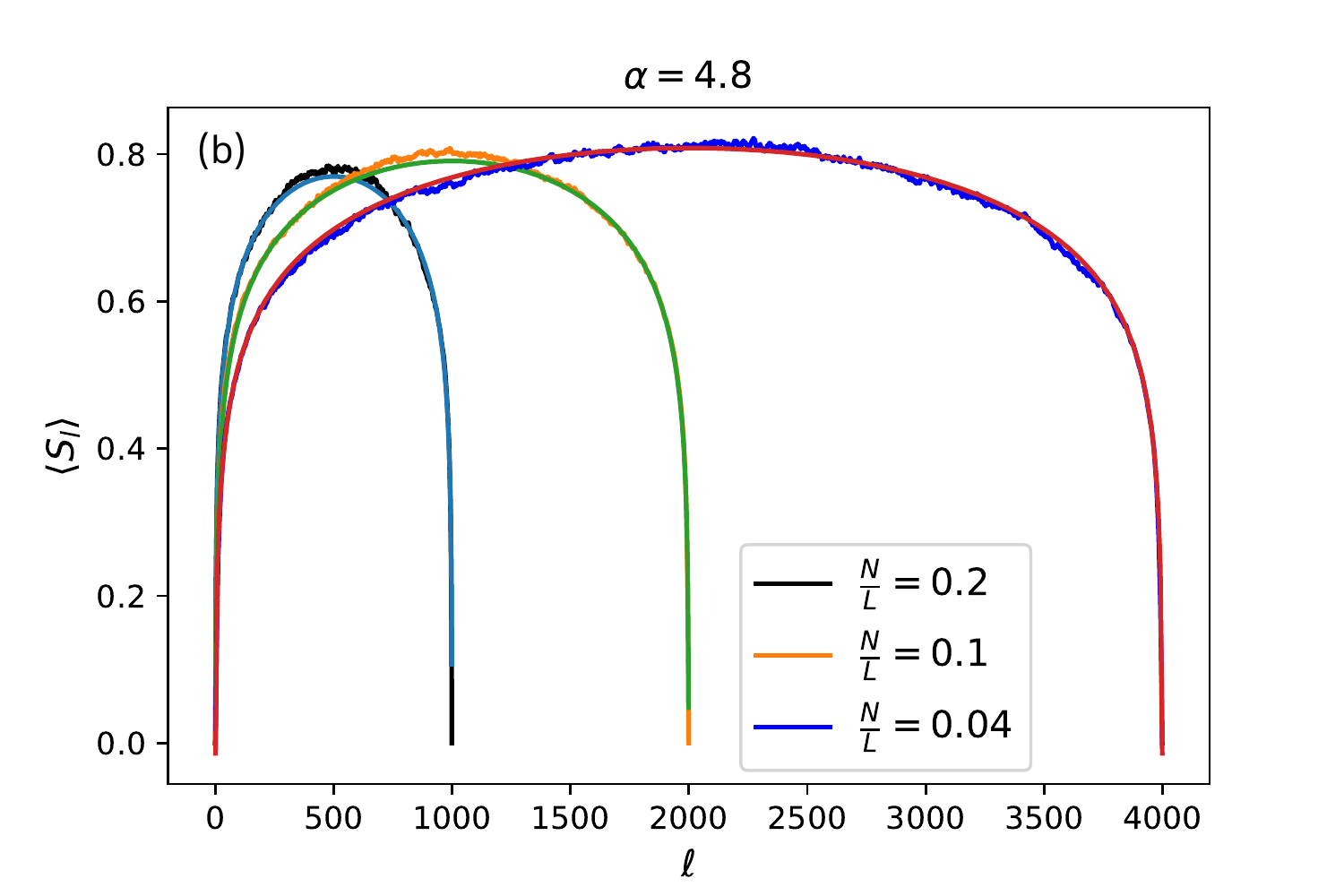}
\caption{Average block entanglement entropy as a function of the partition length $l$ (physical distance), obtained from numerical SDRG for the long ranged XX-chain with open boundary conditions  for $N=200$ spins. Various filling factors $\frac{N}{L}=0.2, 0.1, 0.04$ were considered for both $\alpha=1.8$ (a)
and  $\alpha= 4.8$ (b). The average was evaluated over $M=100000$ realizations for each $\alpha$.
The full  lines correspond to the Cardy law Eq. (\ref{S_b}) with a central charge $\tilde{c}=\ln(2)$, $b=1$.}
\label{realEEvdens}
\end{figure}

 In Fig. \ref{fig:entropy real length} we plot the 
 average block entanglement entropy as function of the physical distance $l$ for the long ranged XX-chain  with $N=500$ spins, open boundary conditions for  various values of $\alpha$, and for a filling factor $\frac{N}{L}=0.1$. The average was evaluated over $20000$ realizations for each $\alpha$.
The blue dashed line corresponds to the Cardy law Eq. (\ref{S_b}) with  $b=1$, $k'=0.05$ and a central charge $c= \ln(2)$, in very good agreement with the analytical result $c= \ln(2)$.
    In Fig. \ref{realEEvdens} the  average block entanglement entropy as function of partition length $l$  is shown, 
    as obtained with numerical SDRG for the long ranged XX-chain  with $N=200$ spins, and for various filling factors $\frac{N}{L}=0.2, 0.1, 0.04$, with open boundary conditions and for  two values of $\alpha$ . The average was evaluated over $M=100000$ realizations for each $\alpha$.
The dashed  lines correspond to the Cardy law Eq. (\ref{S_b}) with a central charge $\tilde{c}=\ln(2)$, $b=1$, in good agreement with the analytical result.

\begin{figure}[hbt]
\begin{center}
\includegraphics[scale=0.4]{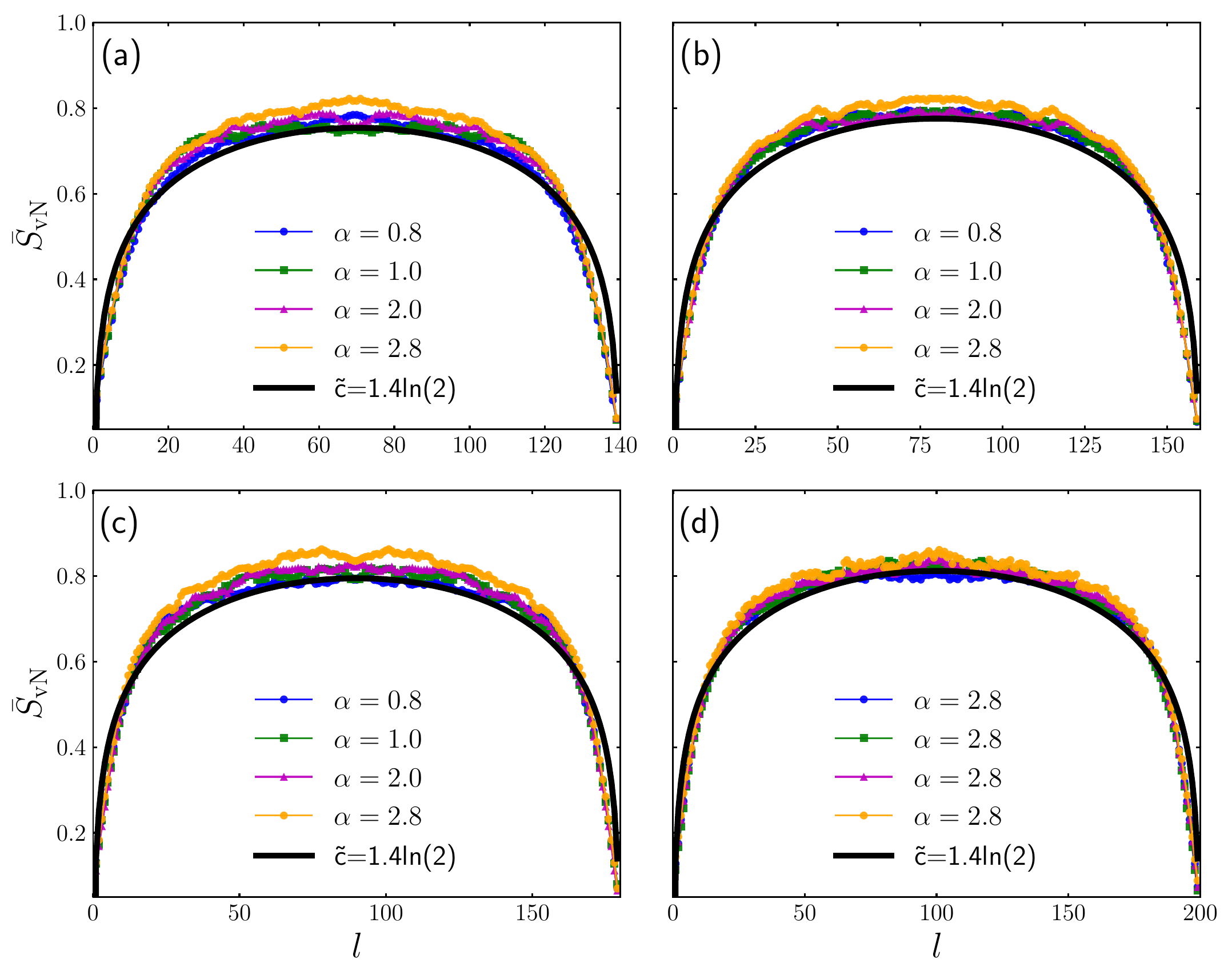}
\end{center}
\caption{Average block entanglement entropy, obtained from 
numerical exact diagonalization for the long ranged XX-chain with  a filling factor $\frac{N}{L}=0.1$ 
for sizes $L=140,160,180,200$, open boundary conditions for  various values of $\alpha=0.8,1.0,2.0,2.8$ as a function of the partition length $l$ (physical distance). The average was evaluated over $M=2000$ realizations for each $\alpha$.
The full lines correspond to the Cardy law Eq. (\ref{S_b}) with a central charge $\tilde{c}=1.4 \ln(2)$, $b=1$ and $k'=0.13$.}
\label{fig:entropyreallength1}
\end{figure}

\begin{figure}[ht]
\begin{center}
\includegraphics[scale=0.6]{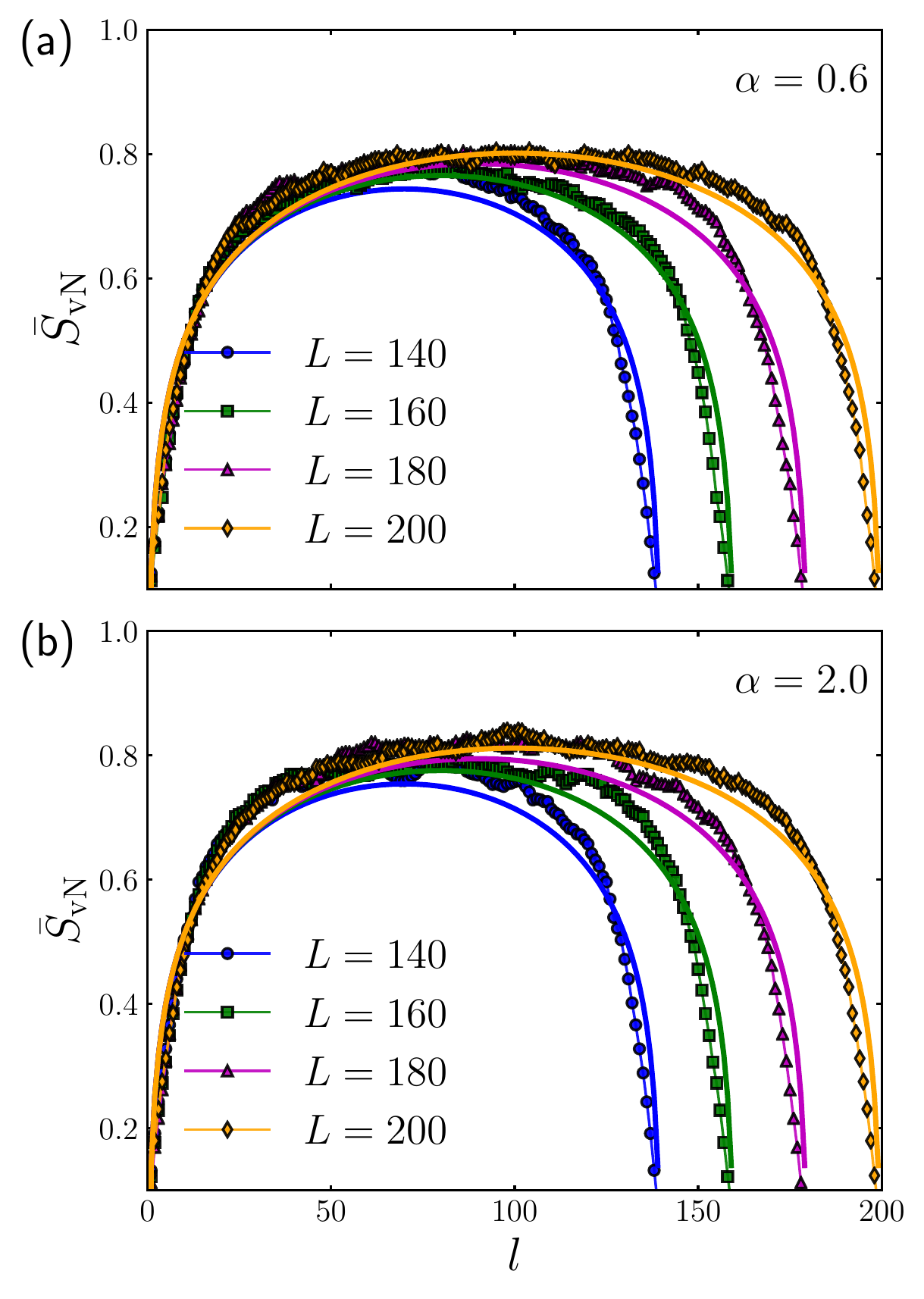}
\end{center}
\caption{Average block entanglement entropy as obtained by 
numerical exact diagonalization for the long ranged XX-chain with a filling factor $\frac{N}{L}=0.1$ 
for sizes $L=140,160,180,200$, for  two values of (a) $\alpha=0.6$ and (b) $\alpha=2.0$ as a function of the partition length $l$ (physical distance). The average was evaluated over $M=2000$ realizations for each $\alpha$.
The yellow  line corresponds to the Cardy law Eq. (\ref{S_b})with a central charge $\tilde{c}=1.4\ln(2)$, $b=1$ and $k'=0.13$.}\label{fig:entropyreallength2}
\end{figure}

\begin{figure}[ht]
\begin{center}
\includegraphics[scale=0.5]{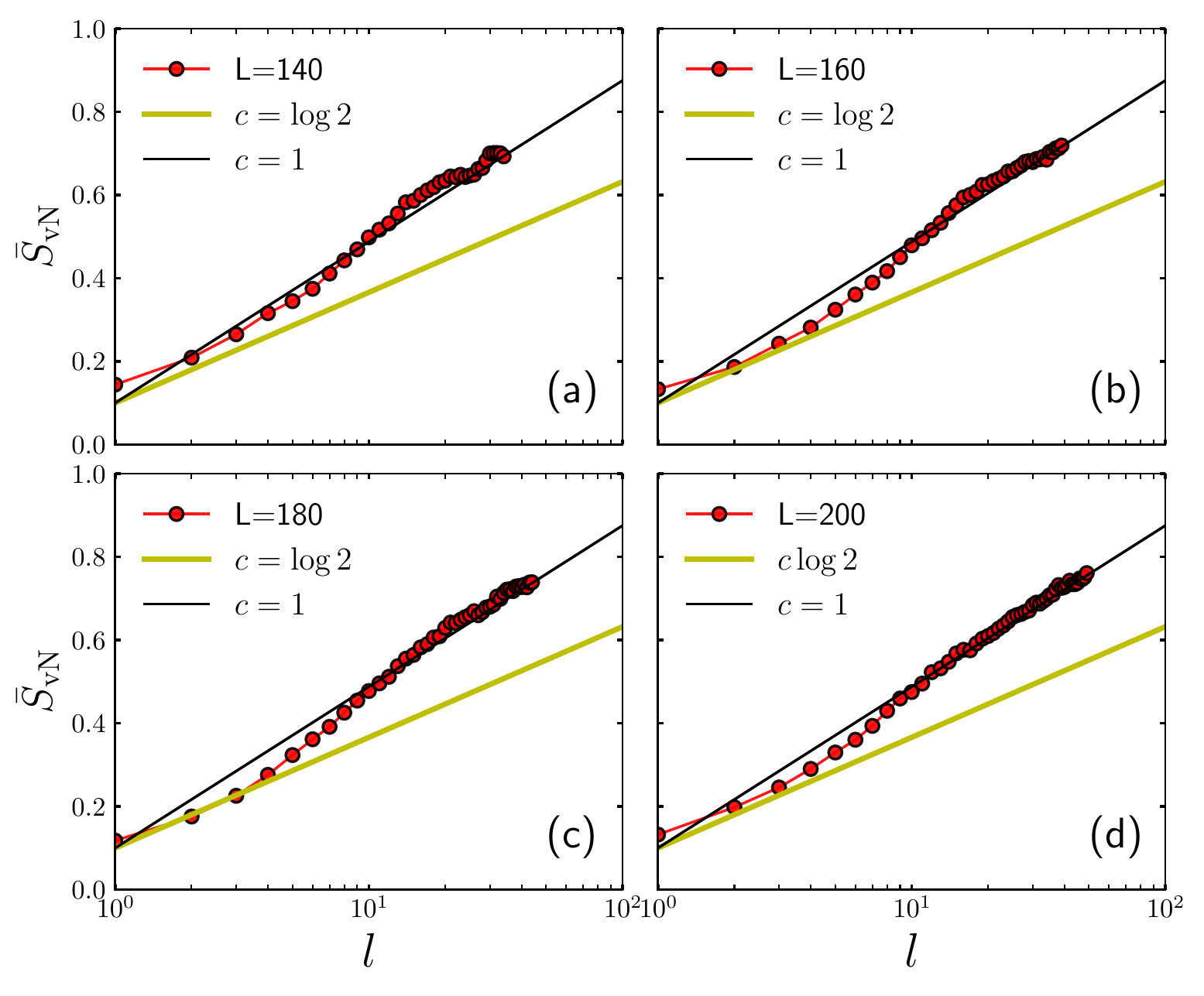}
\end{center}
\caption{ Same as Fig. \ref{fig:entropyreallength2} (b) 
plotted only for $1< l < L/4$. 
The black  line is  the critical entanglement entropy  Eq. (\ref{S_CFT}), corresponding to the   Cardy law Eq. (\ref{S_b}) for $l\ll L$ with a central charge $\tilde{c}=1$, $b=1$, the yellow line is Eq. (\ref{S_CFT}) with  $\tilde{c}=\ln 2$.}\label{fig:entropyfit}
\end{figure}

  As  both analytical and numerical results based on SDRG  could be artefacts of the assumption  that the ground state remains a   random singlet state,
  let us  next calculate the entanglement entropy  using 
the numerical  exact diagonalization method. 
In Figs. \ref{fig:entropyreallength1} 
and \ref{fig:entropyreallength2} 
we show the results 
for  the 
average block entanglement entropy, obtained by
numerical exact diagonalization for the long ranged XX-chain  for a filling factor $\frac{N}{L}=0.1$ with open boundary conditions,
in Fig. \ref{fig:entropyreallength1} for $L=140,160,180,200$ and various
values of $\alpha$, and in Fig. \ref{fig:entropyreallength2} 
for various sizes $L=140,160,180,200$, for  $\alpha =0.6,2$, as averaged over $M=500$ random samples as a function of the real subsystem size $l$. Here, $\langle\cdots\rangle_{\rm ens}$ denotes the ensemble average.  The dashed  lines correspond to  Cardy's law, Eq. (\ref{S_b}), with a central charge $\tilde{c}=1.4 \ln(2)$, $b=1$  and $k'=0.13$, which 
 confirms the weak dependence on $\alpha$.
 The central charge is  found  by exact diagonalization to be larger than the one for the short ranged disordered AFM spin chain, and  by $1.4$ larger than obtained with the numerical implementation of the SDRG above.
 In Fig. \ref{fig:entropyfit} we show the same results for  average block 
 entanglement entropy as obtained from  numerical exact diagonalization   for   $\alpha=2.0$ as a function of the partition length $l$ (physical distance), for $1< l < L/4$, where the 
 critical entanglement entropy  Eq. (\ref{S_CFT}) is plotted as 
the black  line  corresponding to the approximation of the   Cardy law Eq. (\ref{S_b}) for $l\ll L$ with a central charge $\tilde{c}=1$, $b=1$.  The yellow line is Eq. (\ref{S_CFT}) with  $\tilde{c}=\ln 2$, corresponding to the result obtained with SDRG. 
 We see that while  for a wide range of $l$, the central charge
 seems to fit to the one of a clean critical spin chain $c=1$,
 at small $l$ there are strong deviations tending rather to the SDRG result. 
 
As the averaging may diminish the dependence on $\alpha$, let us next also consider the full  distribution of the entanglement entropy for (a) $\alpha=0.1$, (b) $\alpha=0.8$ and (c) $\alpha=2.8$, where 500 random samples have been taken, and the system size $N=22$ is used, and $S_{\rm vN}^0=\log2$. We note that also the distribution shows only a weak dependence on $\alpha$. It is remarkable that the distribution is peaked at integer multiples of $\ln 2$,
 which means that even when averaged over many ensembles, an   integer number of singlets crossing the partition is most likely. 

\begin{figure}[ht]
	\includegraphics[width=0.46\textwidth]{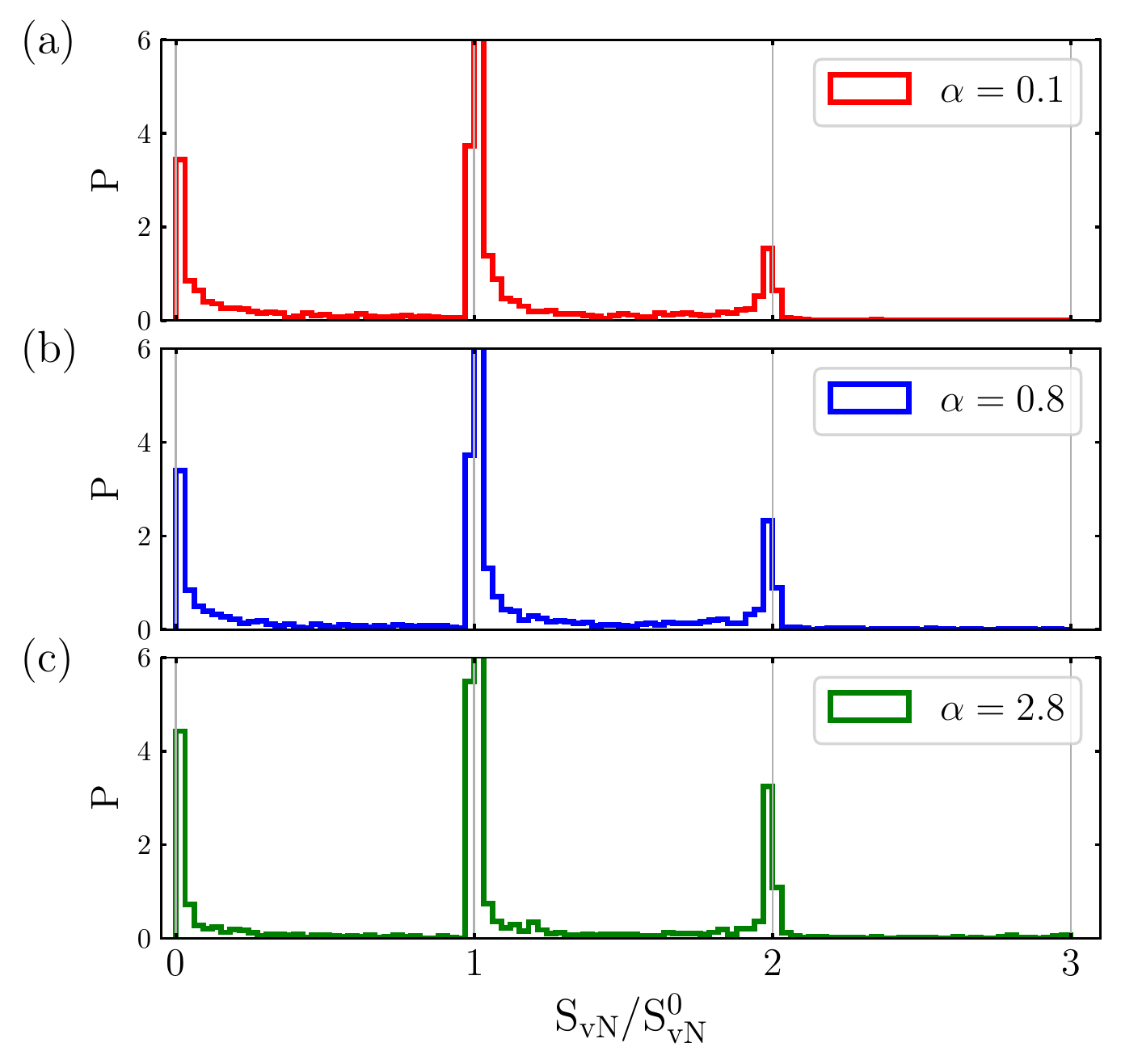}
	\caption{Distributions of the entanglement entropy for (a) $\alpha=0.1$, (b) $\alpha=0.8$ and (c) $\alpha=2.8$, where 500 random samples and the system size $N=22$ are used, and $S_{\rm vN}^0=\log2$.}
\label{fig:ED2}
\end{figure}

\section{Entanglement Measures obtained by  Density Matrix Renormalization Group}

We consider again the  XX
model in which $N$ spins are randomly distributed 
on a finite lattice with length $L$ and 
interact with each other with long-range interactions, but 
consider also an external 
 an external Zeeman magnetic field ${\bf B}$,
\begin{align}
	H_{xx}\!=\! \sum_{i,j<i} J_{ij}(\alpha,\eta_0) 
	\left( S_i^x S_j^x + S_i^y S_j^y \right) \!-\! {\bf B} \sum_i {\bf S}_i,
	\label{eq:hamiltonian}
\end{align}
where  the coupling function is long-range
\begin{align}
	J_{ij}(\alpha,\xi) = |{\bf r}_i-{\bf r}_j|^{-\alpha}\eta_0^{|{\bf r}_i-{\bf r}_j|},
	\label{eq:coupling_ftn}
\end{align}
and we introduced an exponential cutoff $\eta_0=\exp(-1/\xi)$
with correlation length $\xi$.
Here, the position ${\bf r}_i$ is randomly distributed 
on the chain of length $L$ with lattice spacing $a$.

\subsection{DMRG applied to Spin Chains with Power Law Interactions}
To find the ground states\,(GS) of Eq.\,\eqref{eq:hamiltonian} for different random realizations, we use the density matrix renormalization group\,(DMRG) method\,\cite{white92,white93,schollwock05} and the noise algorithm\,\cite{white05} to avoid converging to a local minimum in DMRG calculations. As shown in Ref.\,\onlinecite{verstraete07}, DMRG can be regarded as a method for optimizing variational wave functions known as matrix product states\,(MPS)\,\cite{verstraete07, schollwock11}:

\begin{eqnarray}
	|\psi\rangle = \sum_{\{p_i\}} \sum_{\{ \alpha_i \}} A_{\alpha_1}^{p_1} A_{\alpha_1 \alpha_2}^{p_2} \cdots	A_{\alpha_{N-1}}^{p_N} |p_1,\cdots,p_N\rangle	
\end{eqnarray}
which is a representative one-dimensional tensor network state\,\cite{orus14}. In MPS representation, one can rewrite quantum operators in a similar tensor network, so-called the matrix product operator\,(MPO), 
\begin{align}
	\hat{O} \!=\! \sum_{\{p_ip_i'\}}\!\sum_{\{\alpha_i\}} O_{\alpha_1}^{p_1p_1'} O_{\alpha_1 \alpha_2}^{p_2p_2'} \cdots O_{\alpha_{N-1}}^{p_Np_N'} |p_1,\cdots \rangle\langle p_1',\cdots|.
\end{align}
Here, $\{\alpha_i\}$ are the virtual indices which are traced out, and their dimensions are called the bond dimension of MPO. The success of DMRG for one-dimensional systems is due to the existence of an exact MPO representation with {\it finite} bond dimensions for a Hamiltonian with short-range or exponentially decaying interactions. For example, the Hamiltonian in Eq.\,\eqref{eq:hamiltonian} with $\alpha=0$ and $\eta_0 > 0 $ can be exactly written in MPO representation with the following tensor

\begin{align}
	O_{\alpha \beta}^{pp'} = 
	\begin{bmatrix}
		(\mathbb{I})^{pp'} & 0 & 0 & 0 & 0\\
		(S_x)^{p p'} & \eta_0 (\mathbb{I})^{pp'} & 0  & 0 & 0\\
		(S_y)^{p p'} & 0  & \eta_0 (\mathbb{I})^{pp'} & 0 & 0\\
		(S_z)^{p p'} & 0 & 0 & \eta_0 (\mathbb{I})^{pp'} & 0 \\
		B(S_z)^{p p'} & (S_x)^{p p'} & (S_y)^{p p'} & \zeta(S_z)^{p p'}  & (\mathbb{I})^{pp'} \\
	\end{bmatrix}_{\alpha\beta},
\end{align}
where $\mathbb{I}$ is the $2\times2$ identity matrix.
It becomes the nearest neighbor XX-model by setting $\eta_0=0$.
Unfortunately, an {\it exact} MPO expression of a Hamiltonian with the {\it power-law} decaying interaction is not allowed with finite bond dimension regardless of its exponent $\alpha$. However, one can decompose the power-law functions into several exponential functions as follows 
\begin{align}
	|{\bf r}_j- {\bf r}_i|^{-\alpha} \simeq \sum_{l=1}^m \lambda_l \eta_l^{|{\bf r}_j- {\bf r}_i|}, 
	\label{eq:exp_decomp}
\end{align}
where $m$ depends on the distance $|{\bf r}_j- {\bf r}_i|$ and $\alpha$. Finding proper $\{\lambda_l\}$ and $\{\eta_l\}$ is not a trivial problem, but Pirvu {\it et al.} in Ref.\,\onlinecite{verstraete10} have found a systematic and elegant way to find them. Employing the fitting procedure in Ref.\,\onlinecite{verstraete10}, we have decomposed the power-law function in Eq.\,\eqref{eq:coupling_ftn} into 17 different exponential functions, and they are in excellent agreement with the original power-law function in the range of $\alpha \in [0.2,2.8]$ on a lattice with $L=800$\,($\alpha \in [0.2,2.0]$ on $L=1200$). 
Considering the random distances between neighboring spins, one can find the following MPO tensor encoding both the long-range interaction and randomness:

\begin{widetext}
\begin{align}
	O_{\alpha \beta}^{p_ip_i'} = 
	\begin{bmatrix}
		(\mathbb{I}_2)^{p_i p_i'} & 0 & 0 & 0 & 0\\
		(\vec{\Gamma}_i\otimes S_x)^{p_i p_i'} & (\vec{\Gamma}_i \otimes \mathbb{I}_2)^{p_i p_i'} & 0 & 0 & 0\\
		(\vec{\Gamma}_i \otimes S_y)^{p_i p_i'} & 0 & (\vec{\Gamma}_i \otimes \mathbb{I}_2)^{p_i p_i'} & 0 & 0\\
		(\vec{\Gamma}_i \otimes S_z)^{p_i p_i'} & 0 & 0 & (\vec{\Gamma}_i \otimes \mathbb{I}_2)^{p_i p_i'} & 0\\
		B (S_z)^{p_i p_i'} & (\vec{\lambda}^T \otimes S_x)^{p_i p_i'}  & (\vec{\lambda}^T \otimes S_y)^{p_i p_i'} & \zeta (\vec{\lambda}^T \otimes S_z)^{p_i p_i'} & (\mathbb{I}_2)^{p_i p_i'}
	\end{bmatrix}_{\alpha\beta},
	\label{eq:mpo_tensor}
\end{align}
\end{widetext}
where $T$ denotes the transpose, $\vec{\lambda} = [\lambda_1,\lambda_2,\cdots\lambda_m]^T$ and $\Gamma_i$ is a diagonal matrix with diagonal elements
\begin{align}
	&\vec{\Gamma}_i = [ (\eta_0 \eta_1)^{r_i}, \,\,\,(\eta_0 \eta_2)^{r_i}, \cdots ,  \,\,\,(\eta_0 \eta_m)^{r_i} ]^T,
\end{align}
and $r_i = |{\bf r}_{i+1}-{\bf r}_{i}|$.
Performing DMRG with MPO tensor in Eq.\,\eqref{eq:mpo_tensor}, one can find the ground state of Eq.\,\eqref{eq:hamiltonian} for a given $\alpha$, $\xi$ and random sample. 
\subsection{Results}
\begin{figure}[!ht]
	\includegraphics[width=0.5\textwidth]{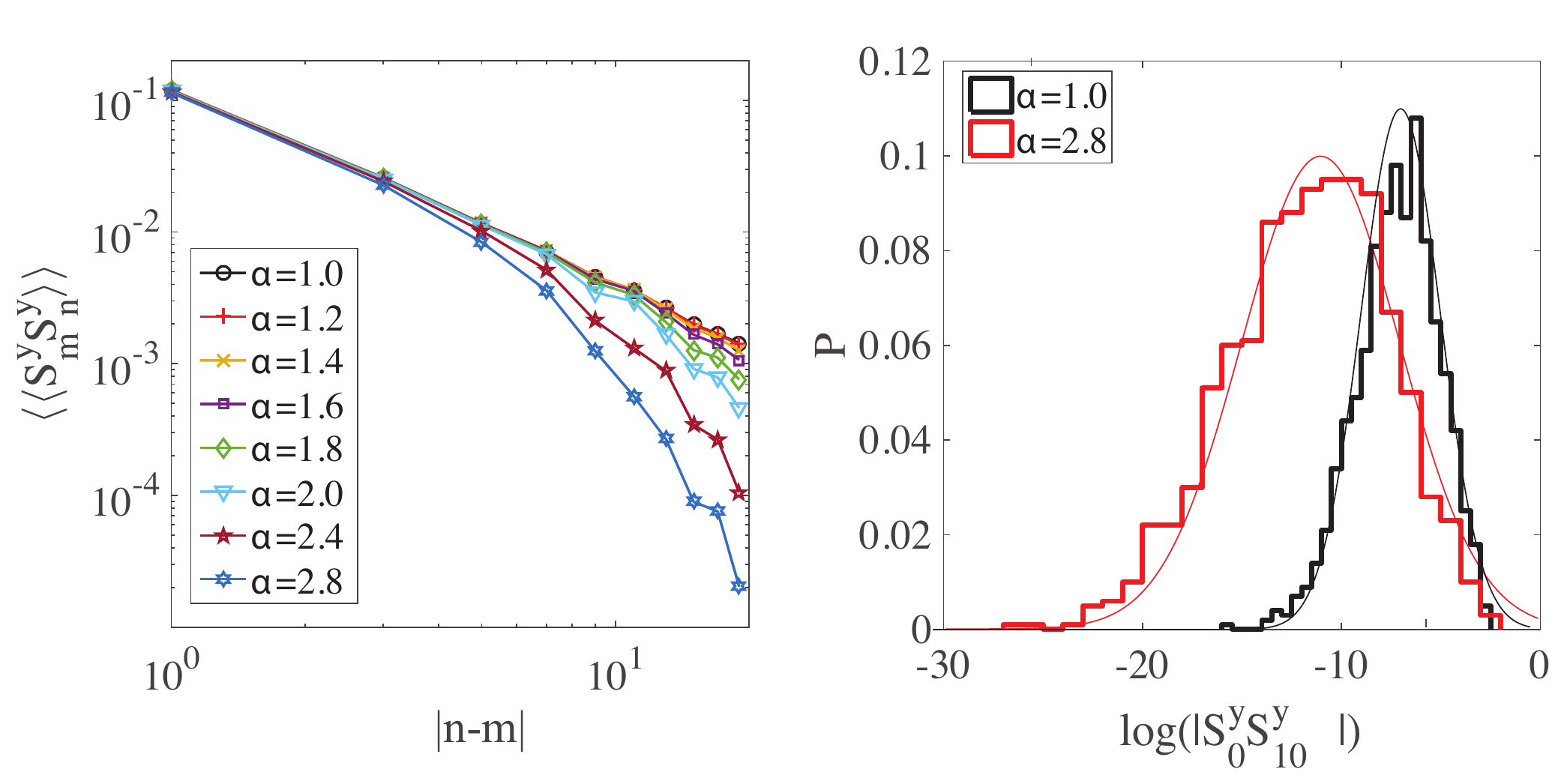}
	\caption{(Left)Concurrence  averaged over 1000 random samples as a function of index distance $n_{nm}=|n-m|$. (Right) Distribution of the concurrence at $n_{nm}==10$. 
	Filling factor is fixed to $N/L =0.1$}
	\label{fig:corr_con}
\end{figure}
\begin{figure}[!ht]
	\includegraphics[width=0.4\textwidth]{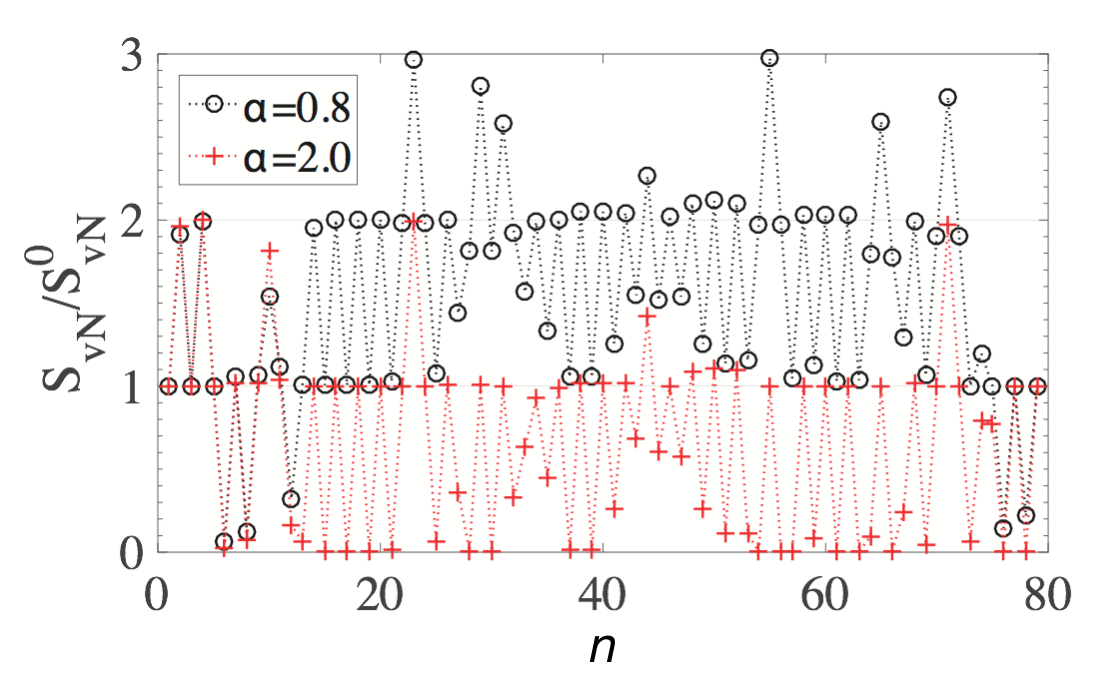}
	\caption{Entanglement entropy as a function of the subsystem size
	$n$ for a given random sample, where $S_{\rm vN}^0 = \log 2$. 
 }
	\label{fig:ee_sample}
\end{figure}
Now, let us discuss the results obtained that way with DMRG for a chain with  open boundary condition and  filling factor 
 $N/L=1/10$.

{\it Concurrence}- First, let us begin with the concurrence results presented in Fig.\,\ref{fig:corr_con}. The left panel shows the averaged concurrence  as a function of the index distance\,($n_{nm}=|n-m|$) between two spins at sites $m$ and $n$. The distribution of concurrence at a given index distance $n_{nm}=10$ with $\alpha=1$ and $\alpha=2.8$ is shown on the right.

The averaged concurrence  $\bar{C}(n)$
 shows for some range a power law decay, followed by an exponential decay at larger distance $l$.
 We observe onset of the exponential decay at smaller lengths $l$ for larger 
 $\alpha$.  Near $\alpha=1$ we find only a power law decay.
 This is consistent with the appearance of a delocalized critical state. 
 We extract the power of the concurrence function at $\alpha=1$ and obtain  $\gamma = 1.56$, i.e. $\langle C_n \rangle \sim n^{-1.56}$, which is different from the result obtained at the IRFP where  $\langle C_n \rangle \sim n^{-2}$, as reviewed above. Note that the horizontal axis of distribution on the right panel is on a log-scale. As expected, the concurrence is widely distributed. It  follows
 a log-normal distribution 
 ,   its center is shifted by changing $\alpha$, and its width decreases with decreasing $\alpha$.
\begin{figure}[h!]
	\includegraphics[width=0.45\textwidth]{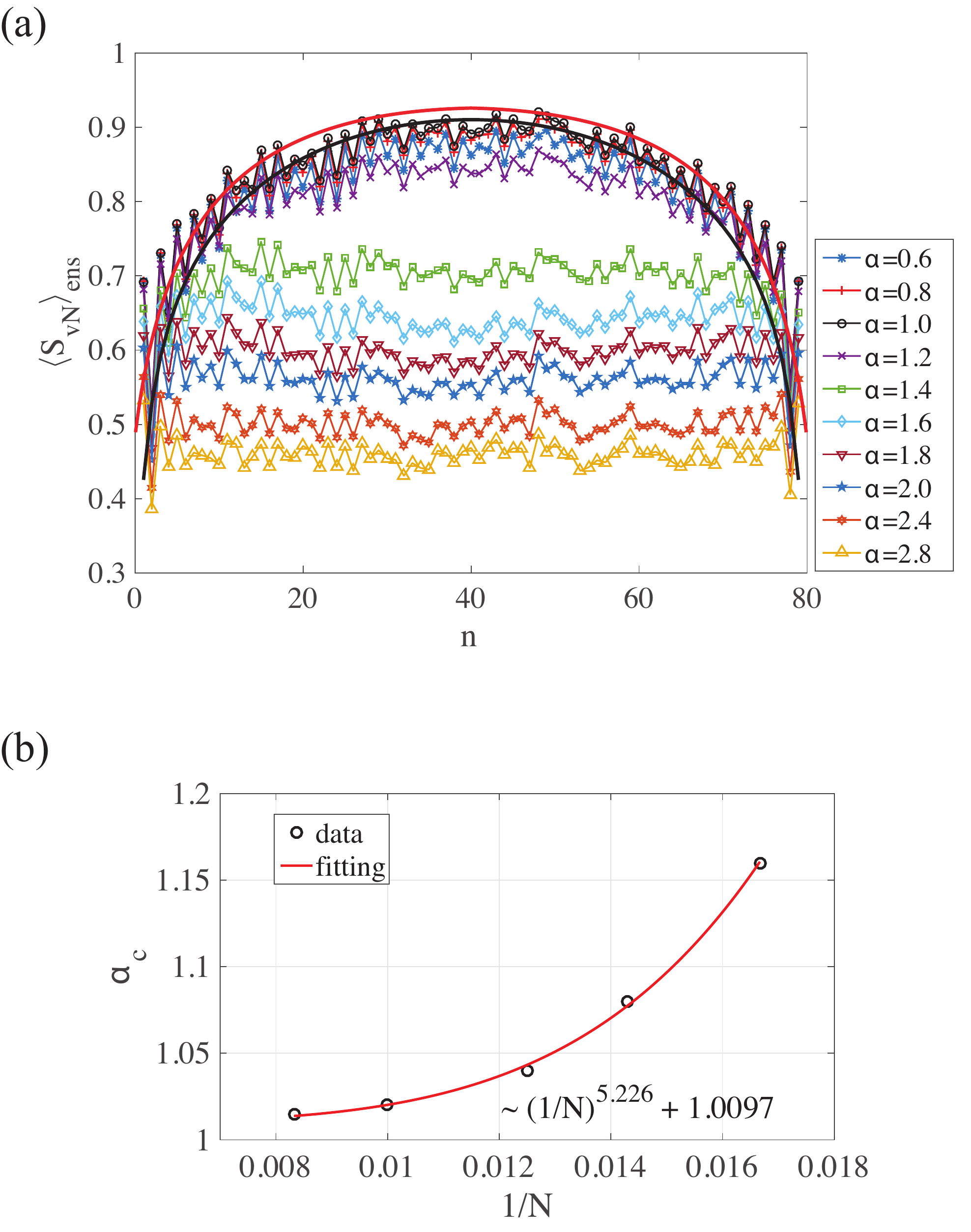}
	\caption{(a) Entanglement entropy  averaged over 1000 random samples as a function of the subsystem size $n$. Here, $\langle\cdots\rangle_{\rm ens}$ denotes the ensemble average. The red and black solid lines are fitting curves to Eqs.\,\eqref{entropypbc_polyg} and \eqref{S_b}, respectively. 
	(b) Finite size scaling of $\alpha_c$ at which the ground state is most entangled for a given $N$. The density of spins is fixed at $N/L=0.1$.  }
	\label{fig:ee_avg_wrt_L}
\end{figure}
%

{\it Entanglement entropy}-  Let us next 
 consider the entanglement entropy\,Eq.\,(\ref{entropy}), which one can 
 obtain directly with DMRG  from the entanglement spectrum. In Fig.\,\ref{fig:ee_sample}, an exemplary result is presented. We see a distinct structure of entanglement of the ground states for a given random sample at $\alpha = 0.8$ and $2$. Here, the horizontal axis $n$ stands for the length of the subsystem. At $\alpha=2$, the entanglement entropy alternates between $0$ and $S_{\rm vN}^0(=\log2)$ throughout the entire lattice, which indicates that the ground state is to a good approximation 
a random singlet state, a product of local singlets, whose length is narrowly distributed  and thus localized, as expected for the SDFP RS state. On the other hand, at $\alpha=0.8$, the $S_{\rm vN}$ is non-zero everywhere. It implies that the ground state has a finite degree of  entanglement\, throughout the system. This is consistent with a delocalization transition at $\alpha=\alpha_c$. The averaged entanglement entropy\,[$\langle S_{\rm vN} \rangle_{\rm ens}$] over $1000$ random samples with $N=80$ is presented in Fig.\,\ref{fig:ee_avg_wrt_L}\,(a) as a function of the subsystem size $n$. 
For $1.2<\alpha\leq 2.8$  we  find that the average entropy is independent of subsystems size $n$, i.e.  area law scaling.
At smaller $\alpha$ the  entanglement entropy increases  and we find good  agreement near $\alpha=1$  with  Cardy's formula  Eq. (\ref{S_b}) for a finite system with open boundary conditions.

Fitting curves are displayed in Fig.\,\ref{fig:ee_avg_wrt_L}\,(a) as red\,[Eq.\eqref{entropypbc_polyg}] and black\,[Eq.\eqref{S_b}] solid lines. Here, the extracted central charge  is $c_{\rm fit} = 0.93$, with a constant $k = 0.39$. After the entanglement entropy reaches a maximum at $\alpha \approx 1$, it decreases
 again with decreasing $\alpha$\,, as seen for $\alpha = 0.6,0.8$ in Fig.\,\ref{fig:ee_avg_wrt_L}\,(a).
 We  find that the $\alpha_c$ with maximum entanglement entropy depends on the system size $N$. In analyzing its system size dependence, Fig.\,\ref{fig:ee_avg_wrt_L}, we find  the critical exponent to be $\alpha_c = 1.0097$ in the thermodynamic limit $1/N \rightarrow 0$. This is in  very good agreement with the SDRG result in Ref.\,\onlinecite{ourPRB}, where $\alpha_c$ is estimated to be $1.066 \pm 0.002$. 
\begin{figure}[ht]
	\includegraphics[width=0.5\textwidth]{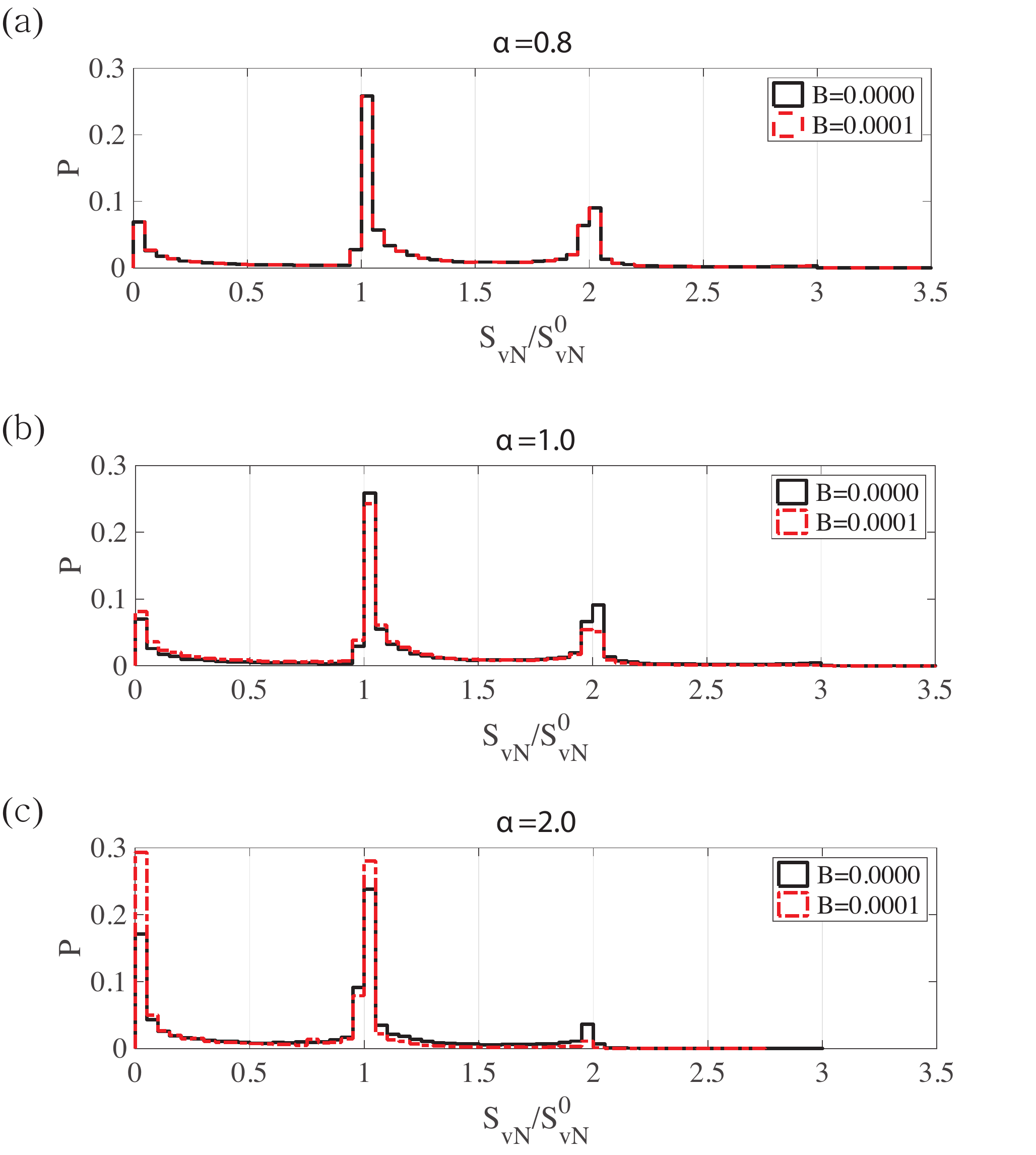}
	\caption{Distributions of the entanglement entropy for (a) $\alpha=0.8$, (b) $\alpha=1.0$ and (c) $\alpha=2.0$, where 1000 random samples and the system size $N=L/10=80$ are used, and $S_{\rm vN}^0=\log2$.}
	\label{fig:ee_dist}
\end{figure}

 Fig.\,\ref{fig:ee_dist} presents the distribution of the  entanglement entropy obtained from  $M=1000$ random samples. 
 There are  peaks  at integer multiples of $S_{\rm vN}^0=\log2$, both with (red lines) and without magnetic field (black lines). Note that the effect of the magnetic field on the GS  depends  strongly on $\alpha$. The distributions of EE are  for $\alpha<1$\, Fig.\,\ref{fig:ee_dist}\,(a),  hardly influenced by a weak field, which indicates that  spins are  are correlated, and there are no free spins. At $\alpha=1$, \,[Fig.\,\ref{fig:ee_dist},\,(b)], the distribution
 without magnetic field is almost identical to  the one at $(\alpha,B)=(0.8,0)$. However,  it is affected by the weak magnetic field, such that peaks at $S_{\rm vN}^0$ and $2S_{\rm vN}^0$ are lowered while the one at $0$ is enhanced by the field. This indicates   the emergence of free spins with no entanglement entropy. We see that the enhancement of probability at $S_{\rm vN}=0$, the density of free spins increases further with $\alpha$\, see Fig.\,\ref{fig:ee_dist}\,(c)].
\begin{figure}[ht]
	\includegraphics[width=0.4\textwidth]{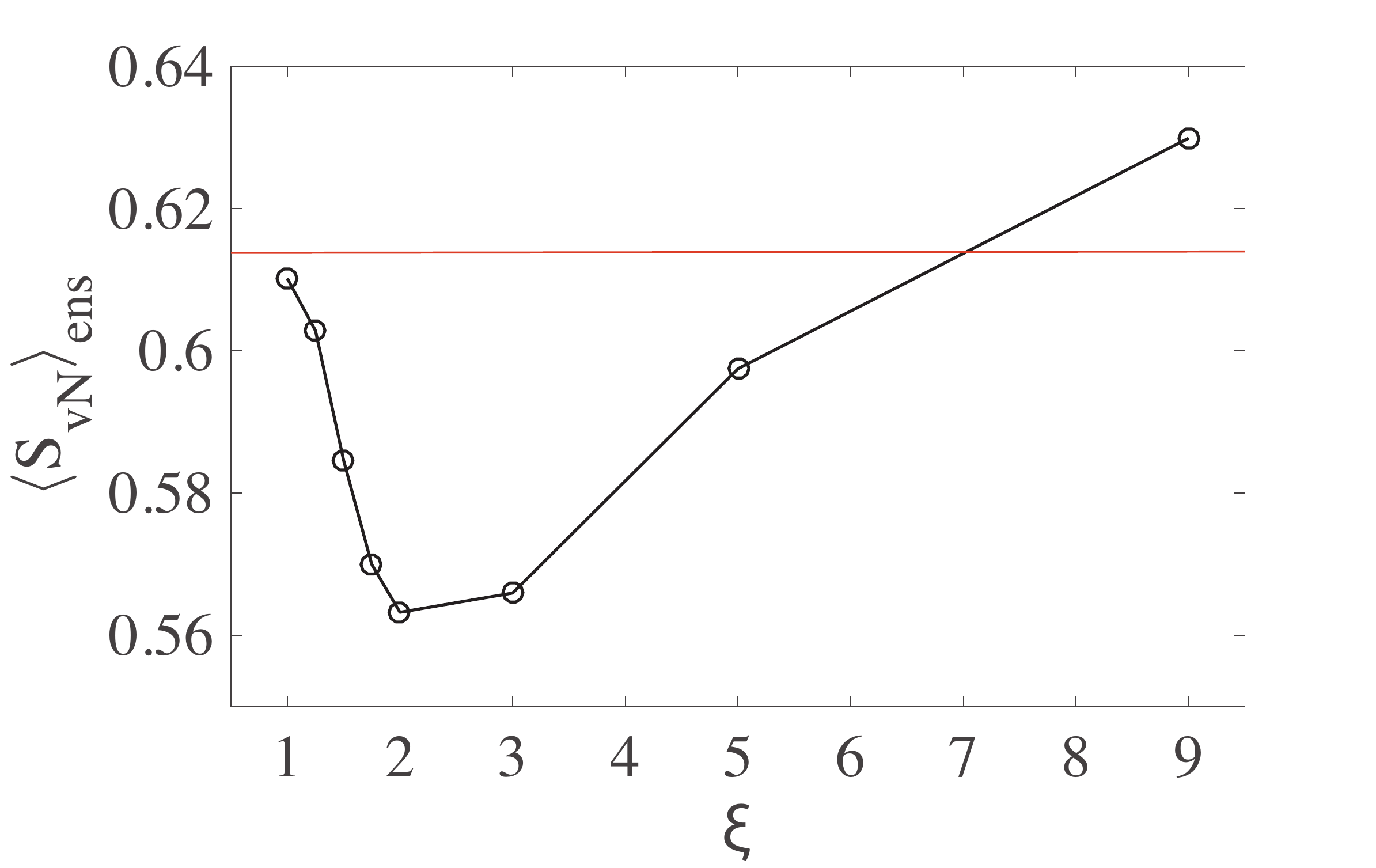}
	\caption{Averaged entanglement entropy at $l=40$ as a function of the correlation length in units of the lattice spacing. The red solid line denotes the exact EE $S_{\rm vN} = \frac{\ln 2}{6} \log_2 40 \simeq 0.6148 $ at IRFP\cite{refael-entropy}.}
	\label{fig:ee_avg_wrt_alpha}
\end{figure}
\begin{figure}[t]
	\includegraphics[width=0.4\textwidth]{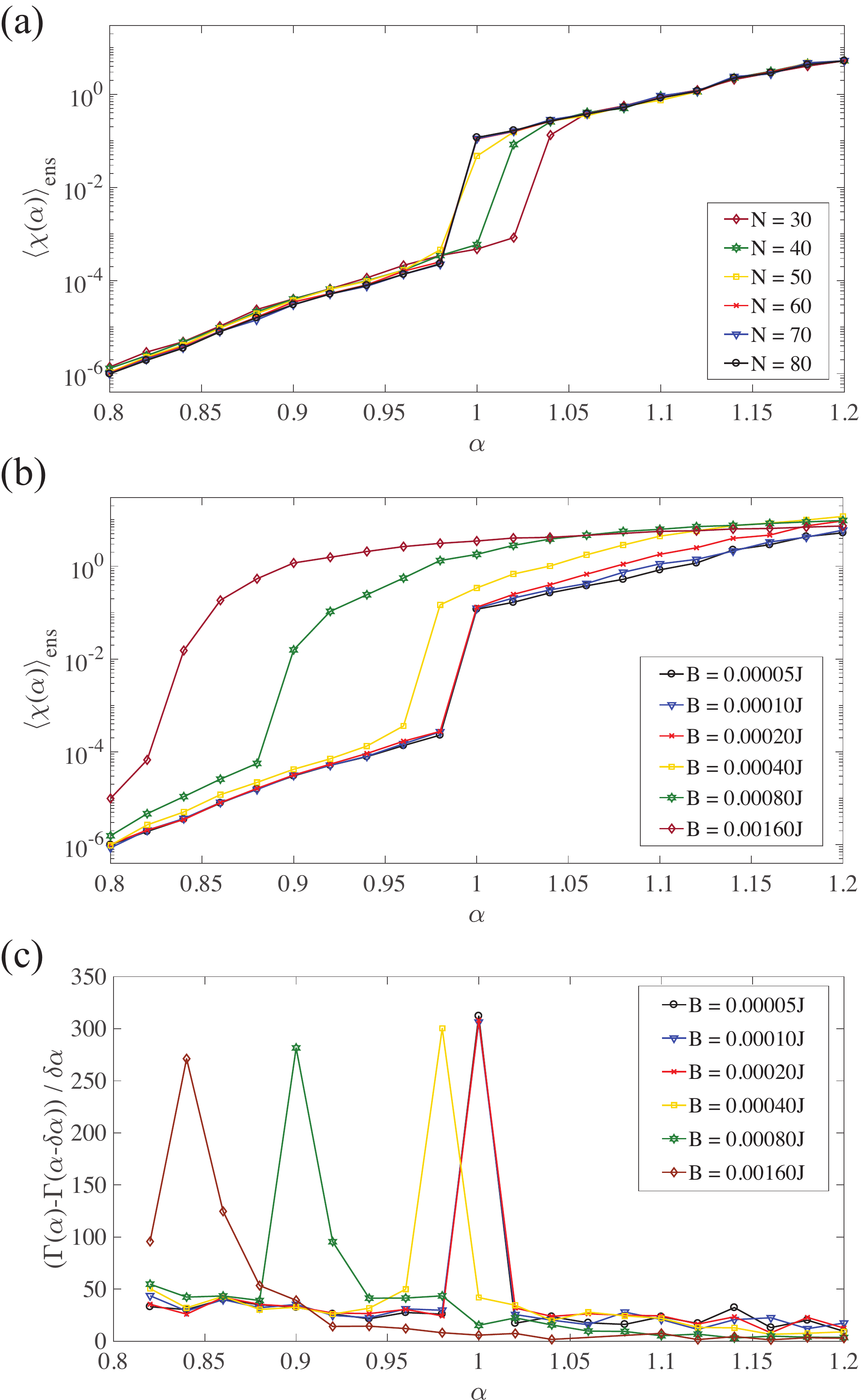}
	\caption{Averaged susceptibility of the ground state as a function of $\alpha$ for (a) various system sizes $N$ with fixed density $N=L/10$ and fixed magnetic field  $B=0.0001J$. b) for various magnetic fields $B$ for $N=80$. (c) The  derivative of the logarithm of the averaged  magnetic susceptibility with respect to $\alpha$,  where $\Gamma(\alpha) = \log[\langle\chi(\alpha)\rangle_{\rm ens}]$ . }
	\label{fig:chi}
\end{figure}
  Thus, the DMRG shows that the entanglement entropy
   decreaes as $\alpha$ is increased beyond the critical value $\alpha \approx 1$ and approaches a constant 
   value independent of the subsystem size, as expected in the noncritical regime. 

     While we cannot extend the DMRG to sufficiently large $\alpha$ to detect any increase with $\alpha$, we can check if we can see the increase of the entanglement entropy when the interaction is cutoff exponentially, and the IRFP with 
     critical entanglement entropy is recovered. 
     In Fig. \ref{fig:ee_avg_wrt_alpha} the averaged entanglement entropy as a function of the correlation length is shown as obtained with DMRG. Indeed, while initially a decrease with decreasing correlation length $\xi$is observed at a $\xi$ of the order of twice the lattice spacing, we observe an increase of the EE again, in agreement with the approach to the IRFP. 

{\it Susceptibility.}- Let us consider next the magnetic  susceptibility of  the ground state under a weak magnetic field $B$: $\langle\chi(\alpha)\rangle_{\rm ens}=\langle\Delta M(\alpha)/\Delta B\rangle_{\rm ens}$, where $M=(1/N)\sum_{i=1}^N \langle S_i^z \rangle$. Figure\,\ref{fig:chi}\,(a) shows the result for that  susceptibility as function of the system size $N$. Note that it converges to a single curve as increasing the system size, and $N=80$ is large enough to see the thermodynamic limit. The susceptibility shows a very sharp increase in $0.98< \alpha \leq 1.0$ and then increases monotonously with $\alpha$. We believe that the emergence of {\it free} spins leads to the significant increase of $\chi$. We also investigate the field dependence of susceptibility with fixed $N=80$, and results are shown in Fig.\,\ref{fig:chi}\,(b) and (c). These results strongly suggest $\alpha^*$ to be close to 1 which is  consistent with the result obtained above from the concurrence and the result obtained  by SDRG in Ref.\,\onlinecite{ourPRB}.

\section{Comparison between Exact Diagonalization and DMRG}

\begin{figure}[ht]
	\includegraphics[width=0.48\textwidth]{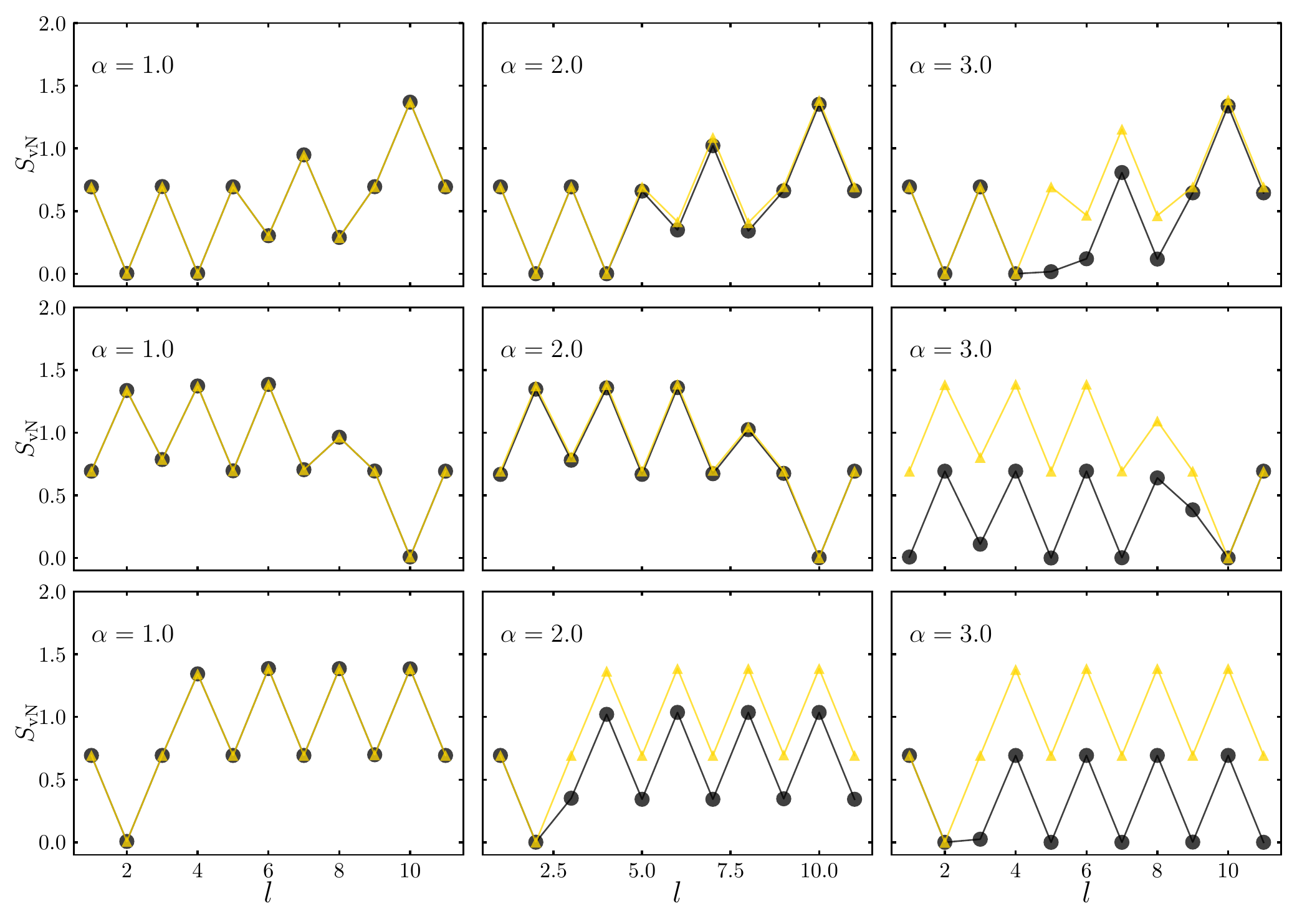}\\
	\includegraphics[width=0.48\textwidth]{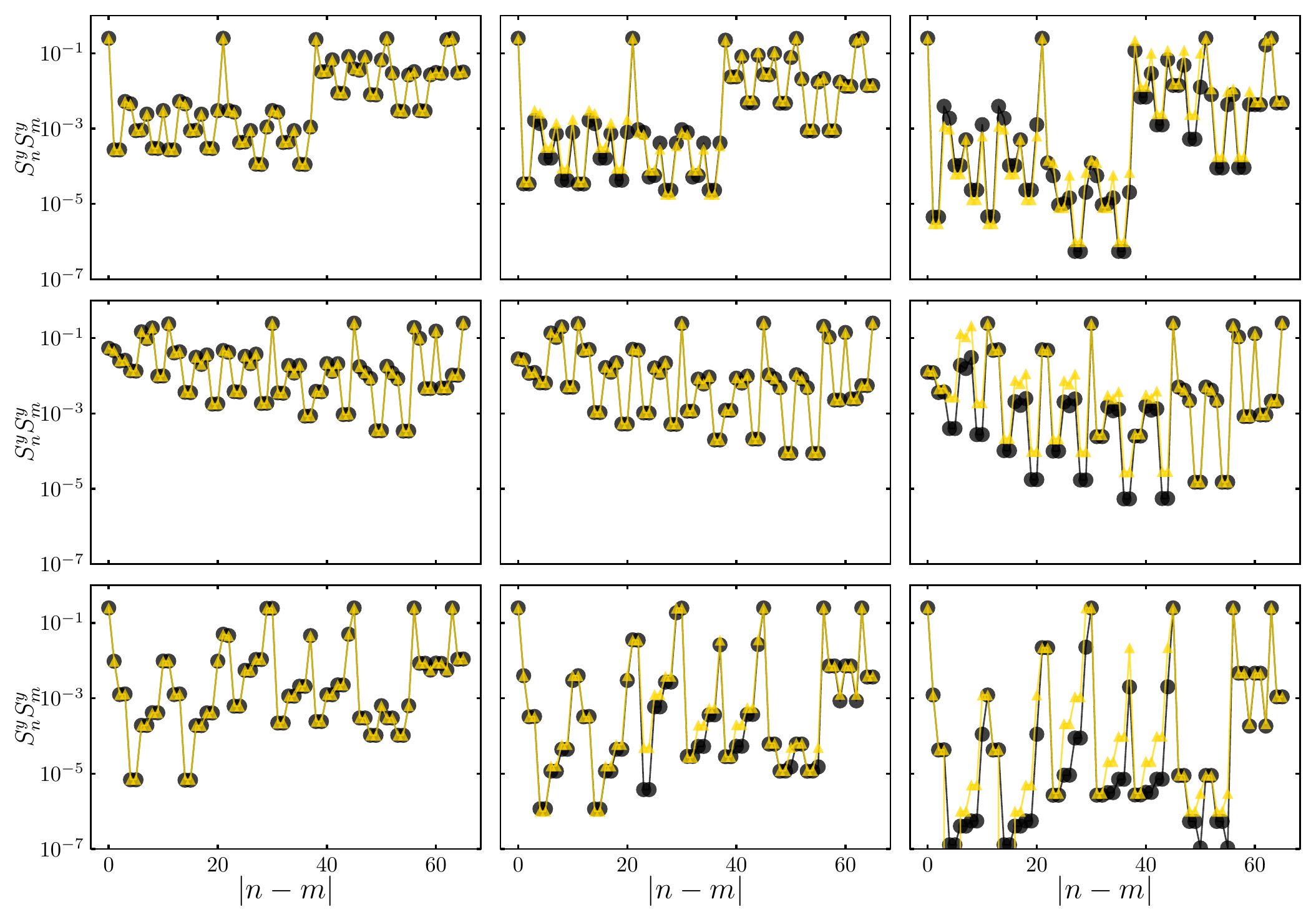}\\
	\includegraphics[width=0.48\textwidth]{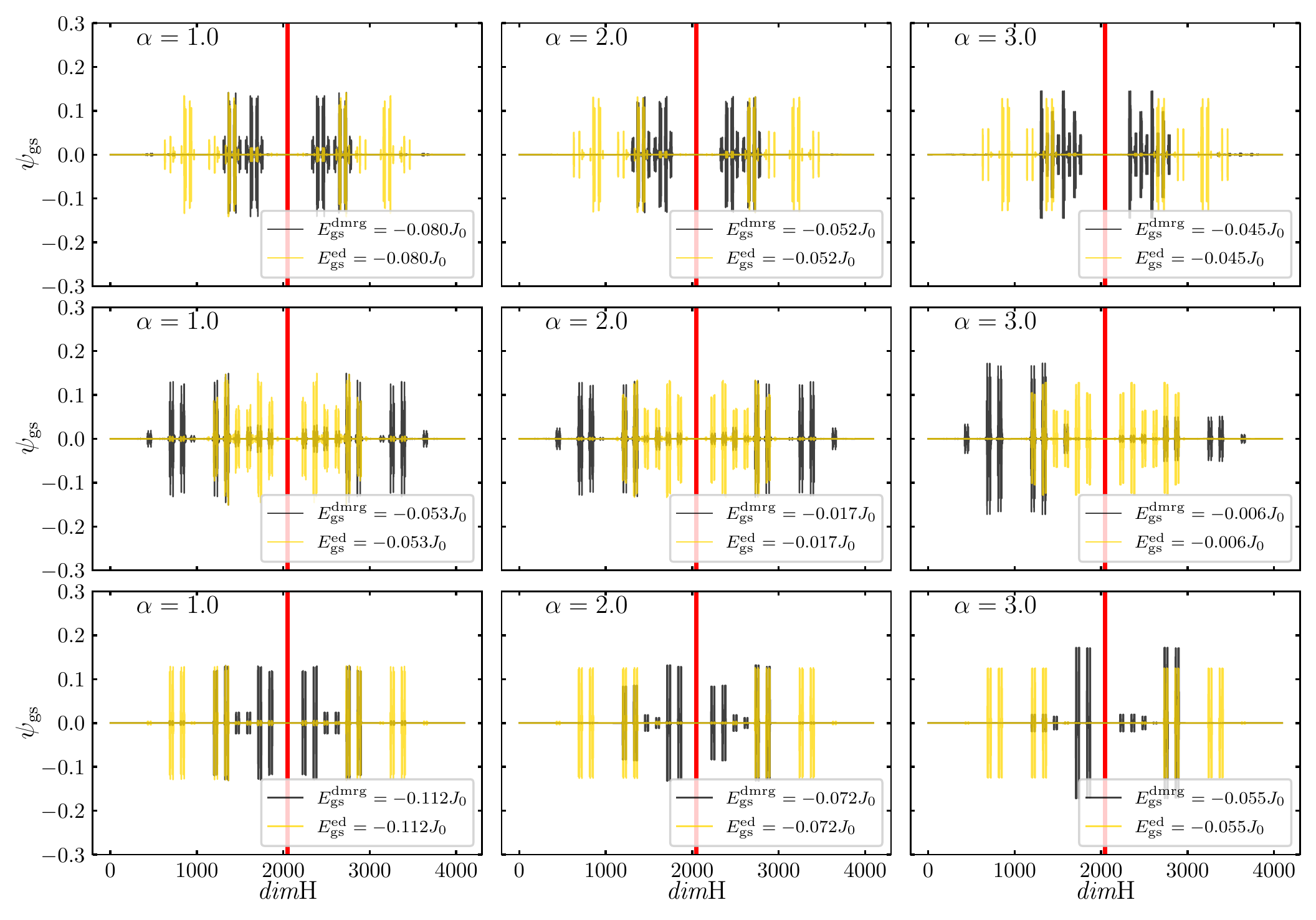}
	\caption{Comparison between ED  and DMRG  for $N=12$ spins  distributed randomly  over length $L=120$ with  open boundary condition: From  the $M= 1000$ realizations, 
	we  show    the entanglement entropy, the  spin-spin correlation, and the ground-state wave function of three samples In the lower figure  the vertical red-line is plotted at the particle-hole point, as a  guide  to the eyes.}
\label{fig:ED3}
\end{figure}

We have seen in the previous section that the tensor network extension of the DMRG yields results 
 for entanglement measures which are, at least for $\alpha >1$ not in good agreement with the results obtained 
  with the strong disorder renormalization group method nor the exact diagonalzation (ED) presented above. 
  Therefore, let us compare the DMRG results for the ground state properties
 directly with the ED results. In order to compare with ED, we consider a small chain with $N=12$ spins, distributed randomly in  a chain of length $L=120$.
 We have used ED and DMRG to calculate the ground-state wave function, entanglement entropy, and spin-spin correlations among all pairs. Results are illustrated in  Fig.~\ref{fig:ED3} for  open boundary condition.
 Calculations with matrix product state (MPS) optimization have been performed using the ITensor C++ library\cite{iten}. We run enough sweeps for the entropy to converge to at least $10^{-10}$, and a large number of states, up to $1000$, was kept so that the truncation error is less than $10^{-12}$. Regarding the implementation of  long-range interaction, as the system is small,  we used the \emph{AutoMPO} method available in ITensor and input all of the terms connecting sites $i$ and $j$, so here we did not consider further approximation like fitting interactions to a sum of exponentials as done in the previous section~\cite{verstraete10} or more recent SVD compression approaches~\cite{Stoudenmire17}. The ED results were obtained  with the
 standard ARPACK diagonalization routine as implemented in SciPy\cite{scipy}.

 For  small power exponent $\alpha=1$  both methods are in  agreement  for all random realizations, both  in the entanglement entropy and 
 the spin-spin correlations measurement,
  as seen  in Fig.~\ref{fig:ED3} where each row corresponds to a specific sample. 
  When increasing $\alpha$, however,  one can see that the results of DMRG gradually deviate from ED. Particularly, focusing on the entanglement entropy, it is clear that DMRG converges to a state with lower entanglement. As seen from the amplitudes of 
   the many body wave function in the lower  Fig. \ref{fig:ED3} (black) this is  accompanied by a breaking of the  particle-hole symmetry. 
  In fact, it is well known that the matrix product state 
  Ansatz of DMRG 
  tends to prefer states of lower entanglement, when states are close in energy. 
   Indeed, although the states obtained with ED shown   in the lower  Fig. \ref{fig:ED3} (yellow)  are found to have the same energy,  
    the ED ground state is particle-hole symmetric and more strongly entangled. 
   Thus, this is evidence that the DMRG
   omits some of   the  singlets  formed at long distances, which therefore  tends to underestimate the  entanglement  while changing the energy only by an amount smaller than the numerical accuracy.  
   It has been reported that  extensions of  tree tensor networks (TTN)  can capture entanglement properties  in disordered systems better\cite{TTN}.

\section{Conclusions}
We find that the strong disorder fixed point, characterized by a  fixed point distribution  of the couplings with a finite dynamical exponent, describes disordered quantum systems of long-range coupled antiferromagnetic  spin chains  consistently.
However, the lowest-order SDRG, with its RS state, is found  not to be  sufficient to obtain the typical value of the concurrence. 
We therefore proposed and implemented a correction scheme to the RS state, allowing us to obtain the leading order corrections. These 
corrections  yield a power law distance scaling of the typical value of the   concurrence, which we demonstrate both by a numerical implementation of these corrections and by an analytical derivation. They are found to be in agreement with each other.

The   entanglement entropy (EE) is calculated 
using SDRG numerically and analytically and found  to be logarithmically enhanced for all $\alpha$, whereas the effective charge is found not to depend on $\alpha$ and to be $c = \ln 2$, in  agreement with an  analytical derivation. However, the analytical derivation uses assumptions on the correlation between singlets, and in a first attempt to 
 include these  correlations, we arrived at a smaller central charge. Therefore, a more rigorous derivation is called for, which we leave for future research. 

While we confirm with numerical exact diagonalization (ED)  the 
logarithmic enhancement of EE and a weak dependence on $\alpha$,
it fits in a wide range of distances $l$ a critical behavior with a central charge 
close to $c=1$,  reminiscent of the clean Haldane-Shastry model with power law decaying interaction with $\alpha =2$. 
 Indeed, also the concurrence, derived with numerical  ED is found to decay with a power law, whose exponent is smaller than 
  the one found by SDRG, $\gamma =2$ and closer to the one known for the Haldane-Shastry model, $\gamma =1\cite{Haldane,HaldaneCFT}$.
  However, at small distances $l\ll L$ we find strong deviations, 
  which may indicate that the  central charge converges to the SDRG value  $c = \ln 2 <1$  for large sizes. 
 Therefore in future research the 
 exact diagonalization should be extended  to larger systems 
  to check for which ranges of   $\alpha$ disorder is relevant 
  so that the system converges to  the SDRG fixed point.

We also present results obtained with DMRG and  find agreement with ED for sufficiently small $\alpha < 2$, while for larger $\alpha$ DMRG is found to tend to underestimate the   entanglement entropy and finds a faster decaying concurrence. 
  As it is known from previous studies that 
DMRG underestimates Entanglement,  extensions like the  tree tensor network have been suggested, which also might allow to study larger system sizes.
We note that it has been previously suggested that a 
delocalization  occur at a critical value of $\alpha_c$\cite{ours}.
 As we find a logarithmic length dependence for all $\alpha$, as expected at a critical point, we cannot discern 
the delocalization transition at a specific $\alpha_c$ in  the  entanglement properties within this approach.

\section{Acknowledgements}
R.N.B. acknowledges support from DOE BES Grant No. DE-SC0002140.
S.H. acknowledges support from DOE under Grant No.
DE-FG02-05ER46240. Computation for the
work described in this paper was supported by the University of Southern California’s
Center for High-Performance Computing (hpc.usc.edu). S. K. acknowledges support from DFG KE-807/22. H.-Y.L. was supported by a Korea University Grant and National Research Foundation of Korea (NRF- 2020R1I1A3074769).

\appendix{}

\section{Solution of the  Master Equation at the Fixed Point}
In this appendix we show how to derive the solution of the  master equation at the fixed point. We  denote $\mu=\frac{1}{\Gamma(\Omega)}$.\newline
First, we multiply by $\int z^J dJ$ the two sides of Eq. (\ref{Meq}), and then plug in the Ansatz for $P(J,\Omega)$, Eq. (\ref{Ansatz}), \small
\begin{align}
\int_{0}^{\Omega}dJ z^J \Omega^{\mu} J^{\mu-1} (\mu^2(\Omega)-\Omega\mu^{'}\mu -\Omega\mu^{'} \mu  \ln\left(\frac{J}{\Omega}\right))\nonumber\\=
\mu^6 \Omega^{-5\mu-1} \int_{0}^{\Omega} dJ_5 J_5^{\mu-1} z^{J_5} \int_{0}^{\Omega} dJ_3dJ_4 (J_3J_4)^{\mu-1}\nonumber\\\times\left[\int_{0}^{\Omega} dJ_1 J_1^{\mu-1} J_1^{\frac{J_1(J_4-J_3)}{\Omega}}\int_{0}^{\Omega} dJ_2 J_2^{\mu-1} J_2^{\frac{J_1(J_3-J_4)}{\Omega}}\right]
\label{caracfct}
\end{align}


The left hand side of Eq. (\ref{caracfct}) can be  integrated and expressed in terms of hypergeometric functions. One can also integrate over $J_5$,$J_1$ and $J_2$ on the L.H.S., yielding  \small
\begin{eqnarray}
\label{intlhs}
&& M(\mu, \mu+1,\Omega\ln(z))\left(\frac{\mu}{\Omega} -  \frac{\mu'}{\mu } \right)
\nonumber \\
&+&\frac{\mu'}{\mu } {}_2 F_2(\mu,\mu,\mu+1,\mu+1,\Omega\ln(z))
\nonumber \\ &=&  \mu^6\Omega^{-3\mu-1}(-\ln(z))^{-3\mu} \gamma(\mu,-\Omega\ln(z))\int_{0}^{\Omega} dJ_3dJ_4 (J_3J_4)^{\mu-1} \nonumber  \\ && \left[(J_4-J_3)^{-2\mu}\gamma(\mu,(J_3-J_4)\ln(z))\gamma(\mu,(J_4-J_3)\ln(z))\right]
\end{eqnarray}
Here $\gamma$, ${}_2F_2$ and $M$, are respectively the lower incomplete gamma function, the generalized hypergeometric function, and the confluent hypergeometric function,  defined by 
\begin{equation}
\label{defi}
\begin{cases}    
\gamma(s,x)=\int\limits_{0}^{x}t^{s-1} e^{-t} dt\\
\\
M(\mu,\mu+1,x)=\sum\limits_{n=0}^{\infty}  \frac{\mu x^n}{n!(\mu+n)}\\ \\
{}_2 F_2(\mu,\mu,\mu+1,\mu+1,x)=\sum\limits_{n=0}^{\infty} \frac{\mu^2  x^n}{n!(\mu+n)^2}
\end{cases}
\end{equation}
Using the identity
$\gamma (a,-z) = -(z^a/ a) M(a,a+ 1,z)$
and Eqs. (\ref{defi}) one can rewrite Eq. (\ref{caracfct}) as  
\begin{eqnarray}
&& M(\mu, \mu+1,\Omega\ln(z))\left(\frac{\mu}{\Omega}- \frac{\mu'}{\mu } \right)
\nonumber \\ 
&+ &\frac{\mu'}{\mu } {}_2 F_2(\mu,\mu,\mu+1,\mu+1,\Omega\ln(z))
\nonumber 
\\ &=& \mu^6 \Omega^{-3\mu-1}(-\ln(z))^{-\mu}\gamma(\mu,-\Omega\ln(z))\sum_{k=0}^{\infty}\frac{ B(\mu,2k+1)(\ln(z))^{2k}}{(2k)!(k+\mu)} \nonumber \\ &&
\left[\int_{0}^{\Omega} dJ_3dJ_4(J_3 J_4)^{\mu-1}(J_4-J_3)^{2k}\right]
\end{eqnarray}
with $B(\mu,2k+1)$ being the standard Beta function.\newline

Finally we evaluate the last double integral,
\begin{equation}\label{integral}
\int_{0}^{\Omega}\int_{0}^{\Omega} dJ_3dJ_4(J_3 J_4)^{\mu-1}(J_4-J_3)^{2k}= \frac{\Omega^{2\mu} \Omega^{2k}}{k+\mu} \frac{\Gamma(\mu)\Gamma(2k+1)}{\Gamma(\mu+2k+1)}
\end{equation}
Allowing us to get the integrated form of the constraint equation,
\begin{eqnarray}
&&M(\mu, \mu+1,\Omega\ln(z))\left(\frac{\mu}{\Omega}- \frac{\mu'}{\mu } \right)+\frac{\mu'}{\mu }  {}_2 F_2(\mu,\mu,\mu+1,\mu+1,\Omega\ln(z)) \nonumber \\ &=&\frac{\mu^5}{\Omega} M(\mu,\mu+1,\Omega\ln(z))\sum_{k=0}^{\infty}\frac{\Gamma^2(\mu)\Gamma(2k+1)(\Omega\ln(z))^{2k}}{(k+\mu)^2\Gamma^2(\mu+2k+1)}
\end{eqnarray}
Which can be rewritten using Eq. (\ref{defi}) in the form 
\begin{eqnarray}
-\mu'(\Omega)\sum_{n=1}^{\infty}\frac{(\Omega t)^{n}}{(n-1)!(\mu+n)^2}=\frac{\mu^6(\Omega)}{\Omega}
\nonumber 
\\ \times\sum_{k=1,k'=0}^{\infty}\frac{\Gamma^2(\mu)(2k)!(\Omega t)^{2k+k'}}{(k+\mu)^2\Gamma^2(\mu+2k+1)k'!(\mu+k')}
\end{eqnarray}
Here we defined $t=\ln(z)$.   

\section{Benchmark Model Results}
\subsection{Haldane-Shastry model}
\begin{figure}[!h]
	\includegraphics[width=0.45\textwidth]{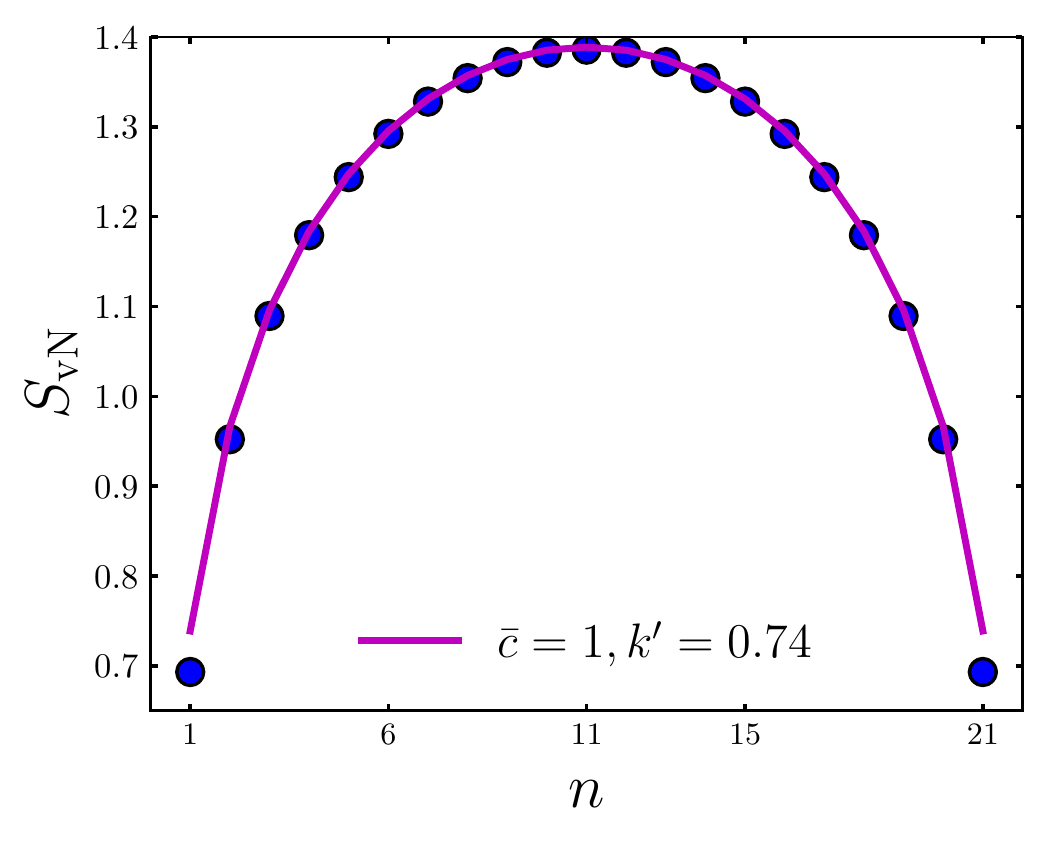}
	\caption{Entanglement entropy obtained via exact diagonalization  for the Haldane-Shastry model with $N=22$ sites with periodic boundary conditions. The solid  line corresponds to the Cardy law Eq. (\ref{S_b}) with a central charge $\tilde{c}=1.0$, $b=2$  and $k'=0.74$. }
\label{fig:ED6}
\end{figure}

Among the spin models in one dimension, the Haldane-Shastry  chain \cite{Haldane} is interesting for several reasons. It is an antiferromagnet with $1/r^2$
exchange interactions, and it possesses a Yangian symmetry which makes it  integrable, therefore, exactly solvable. This model is defined by  
\begin{equation} \label{hs}
    H= J \frac{\pi^2}{N} \sum_{j<i} \frac{S_i.S_j}{d(z_i,z_j)^2},
\end{equation}
where $d(z_i,z_j)$ is the distance between two arbitrary sites on a ring, as given by
\begin{equation}
    d(z_i, z_j)= \mid z_i-z_j\mid= 2 \mid \sin(\frac{\pi(i-j)}{N}) \mid.
\end{equation} 
The HS spin chain is known to be critical and indeed connected to the WZW conformal field theories in the long wave length limit. More precisely, the  critical theory  of the model in Eq. (\ref{hs}) is  the WZW model $SU(2)_k$ at level $k = 1$, with a central charge $c=1$.
In this section we show, as a benchmark, the result of our implementation of exact diagonalization for the  clean Haldane Shastry model, for $N=22$ spins, recovering the $c=1$ analytical result, as shown in Fig. \ref{fig:ED6}.

\subsection{Random Heisenberg XX-model}
\begin{figure}[!h]
	\includegraphics[width=0.48\textwidth]{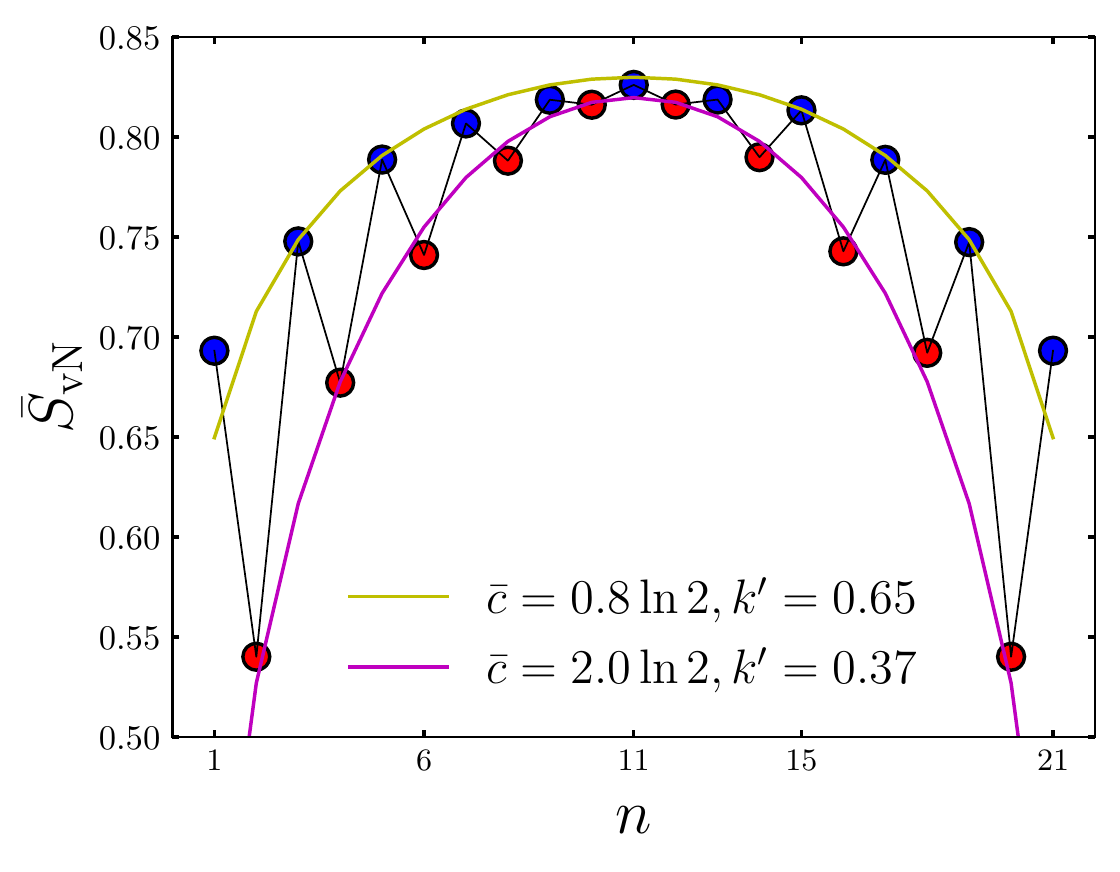}
	\caption{ED results for a sample with $N=22$ spins with open boundary conditions  for the entanglement entropy averaged over 200 random samples as function of partition size $n$.
	The yellow line is the Cardy law Eq. (\ref{S_b}) with $\tilde{c}=0.8 \ln 2$, $b=1$  and $k'=0.68$.
	The pink line is  Eq. (\ref{S_b}) with  $\tilde{c}=2 \ln 2$, $b=1$  and $k'=0.37$.}
\label{fig:ED4}
\end{figure}

In this section, we implement the exact diagonalization procedure for the random  short ranged AFM 
XX-Heisenberg model, defined by its Hamiltonian :
\begin{equation}
\label{HamNN}   
H=\sum_{i=1}^{N-1} J_i\left(S_i^{x} S_{i}^{x} +S_{i}^{y} S_{i+1}^{y}\right)
\end{equation}
Where $\{J_i\}_{i=1}^{N-1} $are uncorrelated positive random variables, drawn from a
distribution $P(J)$.
In Ref. \onlinecite{refael-entropy}, it was shown with SDRG method that, given an interval of length
$l$ embedded
in the infinite line, the average entanglement entropy of this interval, with the rest of the chain
scales for large $l$ as Eq. (\ref{S_b}) with $b=2$,
corresponding to the entanglement entropy of a critical system Eq (\ref{S_CFT}), with an effective  central charge $\tilde{c}= c\times\ln(2)$, where $c=1$ is the central charge of the pure XX-Heisenberg model. 
Fig. (\ref{fig:ED4}) shows
ED results for a sample with $N=22$ spins with open boundary conditions  for the entanglement entropy averaged over 200 random samples as function of partition size $n$. A strong even-odd effect is seen. 
	The yellow line is the Cardy law Eq. (\ref{S_b}) with $\tilde{c}=0.8 \ln 2$, $b=1$  and $k'=0.68$, which is in good agreement with the result obtained for a RS state Eq. (\ref{entropypbc_polyg}), when $c_2 =0.8$.
	The pink line is  Eq. (\ref{S_b}) with  $\tilde{c}=2
	\ln 2$, $b=1$  and $k'=0.37$.

\begin{figure}[!h]
	\includegraphics[width=0.45\textwidth]{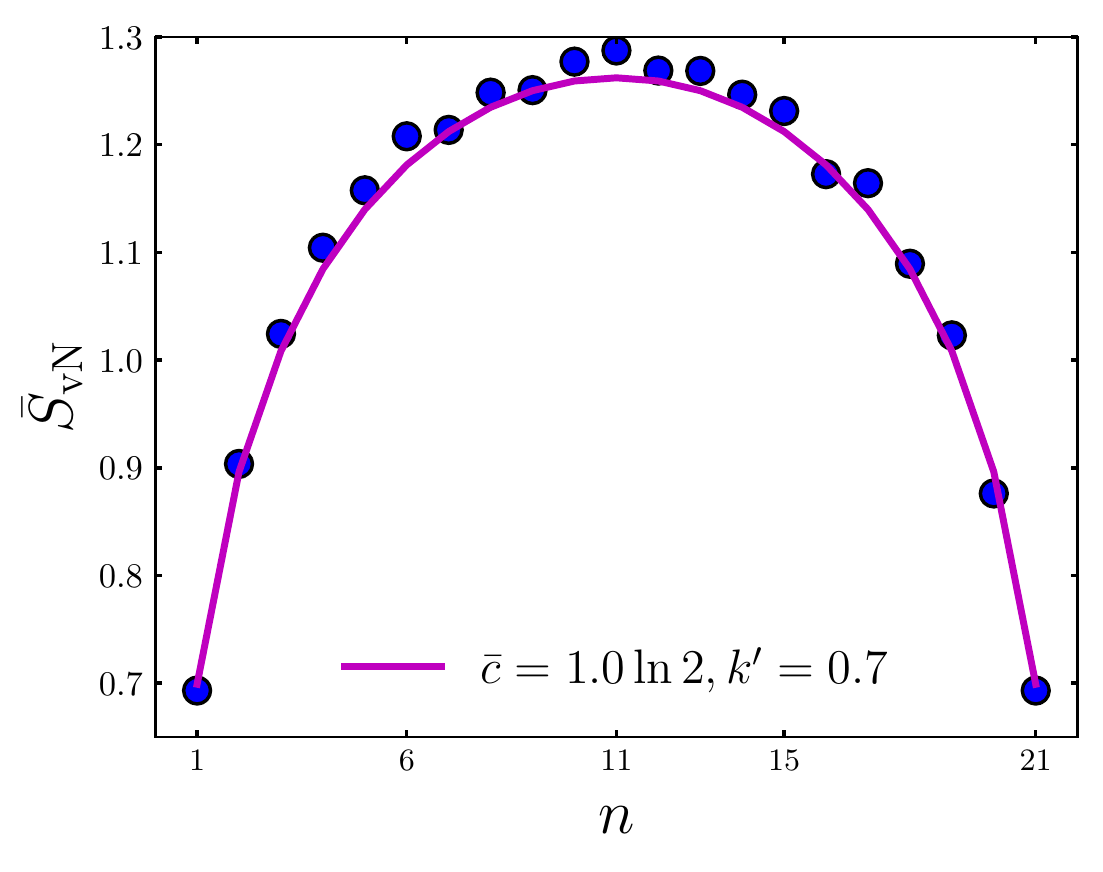}\\
	\caption{The ED results for the random nearest-neighbor model with $N=22$ sites on periodic chain. Data  were averaged over 1000 random sample.  The solid  line corresponds to the Cardy law Eq. (\ref{S_b}) with a central charge $\tilde{c}=1.0 \ln(2)$, $b=2$  and $k'=0.7$.} 
\label{fig:ED5}
\end{figure}

Fig. (\ref{fig:ED5}) shows the result of exact diagonalization for this model, considering a system of $N=22$ spins with periodic boundary conditions.  
ED  reproduces the results obtained analytically in Ref. \onlinecite{refael-entropy}, and derived in this article,  Eq. (\ref{entropypbc_polyg})
for $b=2$, when $c_2 =1$.

\section{  Entanglement Entropy With Corrections to the  Random Singlet State.}

 Having the RS state with corrections $|\psi\rangle$, Eq. (\ref{Psi_corrected})
we can write the density matrix  to calculate the entanglement entropy
beyond the RSS using directly the definition of the von Neumann entropy Eq. (\ref{entropy}). 
Given that the RS state with corrections $|\psi\rangle$ is not a product state, there are no 
simple combinatorical arguments, such as the counting of crossing singlets, since the entropy of a superposition state is not the sum of the individual entropies. Moreover, a closed formula using the definition  Eq. (\ref{entropy}) is not feasible due to the dependence of the sums on the specific realization of the RSS, which makes taking the partial trace inconceivable without considering every single possible scenario, i.e., there are as many outcomes for the partial trace as there are possible random singlet states on the chain ($\sim N$).

A possible solution to this problem is to start by    taking into account
 in Eq. (\ref{Psi_corrected})
the term with the largest coefficient in the sum corresponding to the corrections to two singlet states, only.
 Once this is achieved one possibly can then close the argument recursively to take into account all corrections. 
Keeping only the largest coefficient in  Eq. (\ref{Psi_corrected})
we get
\begin{eqnarray}\label{Psi_approx}
&|\psi\rangle &\approx c|\psi_{RS}\rangle +
\nonumber \\
& c \delta J & (|+_{nl}\rangle|-_{mk}\rangle+|-_{nl}\rangle|+_{mk}\rangle)\bigotimes_{
\{ij\}\neq \{nl\} \neq\{mk\} } |0_{ij}\rangle,
\end{eqnarray} 
where
\begin{equation} 
\delta J=\frac{J_{nm}+ J_{lk} -J_{nk} -J_{ml}}{J_{nl} + J_{mk}},
\end{equation}
is the maximum coefficient appearing in Eq. (\ref{Psi_corrected}), and the coefficient $c_{\psi}$ needs to be redefined as

\begin{eqnarray}\label{capp}
c_{\psi}=\frac{1}{\sqrt{1+2 \delta J^2}},
\end{eqnarray}
in order to keep the approximated state properly normalized. 
Now, let us consider a situation, where the RSS state is such that 
 $k$ singlets cross the partition boundary, giving 
  the EE
\begin{equation} 
S(k)=k\ln 2.
\label{S0}
\end{equation}
With the corrected state Eq. (\ref{Psi_approx})
we find after a lengthy but straightforward calculation,  that it is possible to arrive at a conditional closed form for the entanglement entropy that depends where the two converted singlet pairs $\{nl\}$ and $\{mk\}$ are located, relative to the partition boundary.
There are three distinguishing cases that give rise to different expressions for the entanglement entropy as a function of the number of crossing singlets and triplets $k$ ( Here we set $c= c_{\psi}$ in Eq. (\ref{capp}), 
which in the limit of no corrections ($\delta J\rightarrow 0, c\rightarrow 1$),  simplify  to  Eq. (\ref{S0}).

\textbf{Case 1:} Each of the two converted singlets $\{nl\}$ and $\{mk\}$ are at opposite sides of the cut and none of them cross the boundary: 
\begin{equation}
S(k)=-c^2\ln\left(\dfrac{c^2}{2^k}\right)-(1-c^2)\ln\left(\dfrac{1-c^2}{2^{k+1}}\right).
\label{S_1}
\end{equation} 

\textbf{Case 2:} Both converted singlets cross the boundary between subsystems, for $k\geq 2$:
\begin{equation}\label{S_2}
\begin{split}
S(k)=-\dfrac{c^2}{2}\ln&\left(\dfrac{c^2}{2^k}\right)-\dfrac{c^2}{4}(2\,\delta J+1)^2\ln\left(\dfrac{c^2}{2^k}(2\,\delta J+1)^2\right)\\-&\dfrac{c^2}{4}(2\,\delta J-1)^2\ln\left(\dfrac{c^2}{2^k}(2\,\delta J-1)^2\right).
\end{split}
\end{equation}

 \textbf{Case 3:} Any other relationship between the converted pairs and the boundary, \textit{e.g.}, both pairs are part of the same subsystem or only one of them crosses the boundary. In this case, the approximated state brings no correction to the entropy, giving the same value obtained at the IRFP \cite{refael-entropy}, Eq. (\ref{S0}).

\begin{figure}[h]
\begin{center}
\includegraphics[scale=0.2]{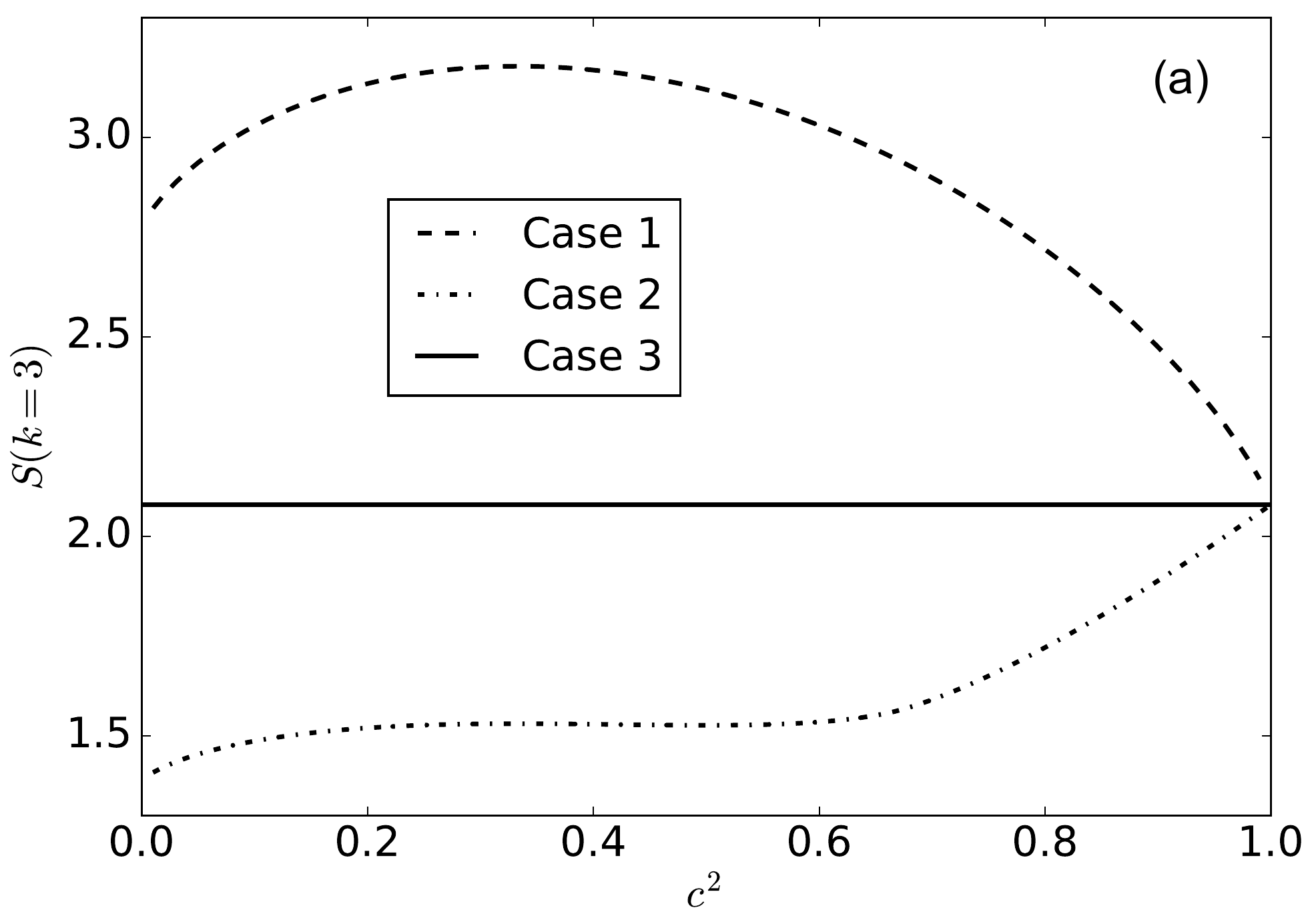}
\includegraphics[width=0.25\textwidth]{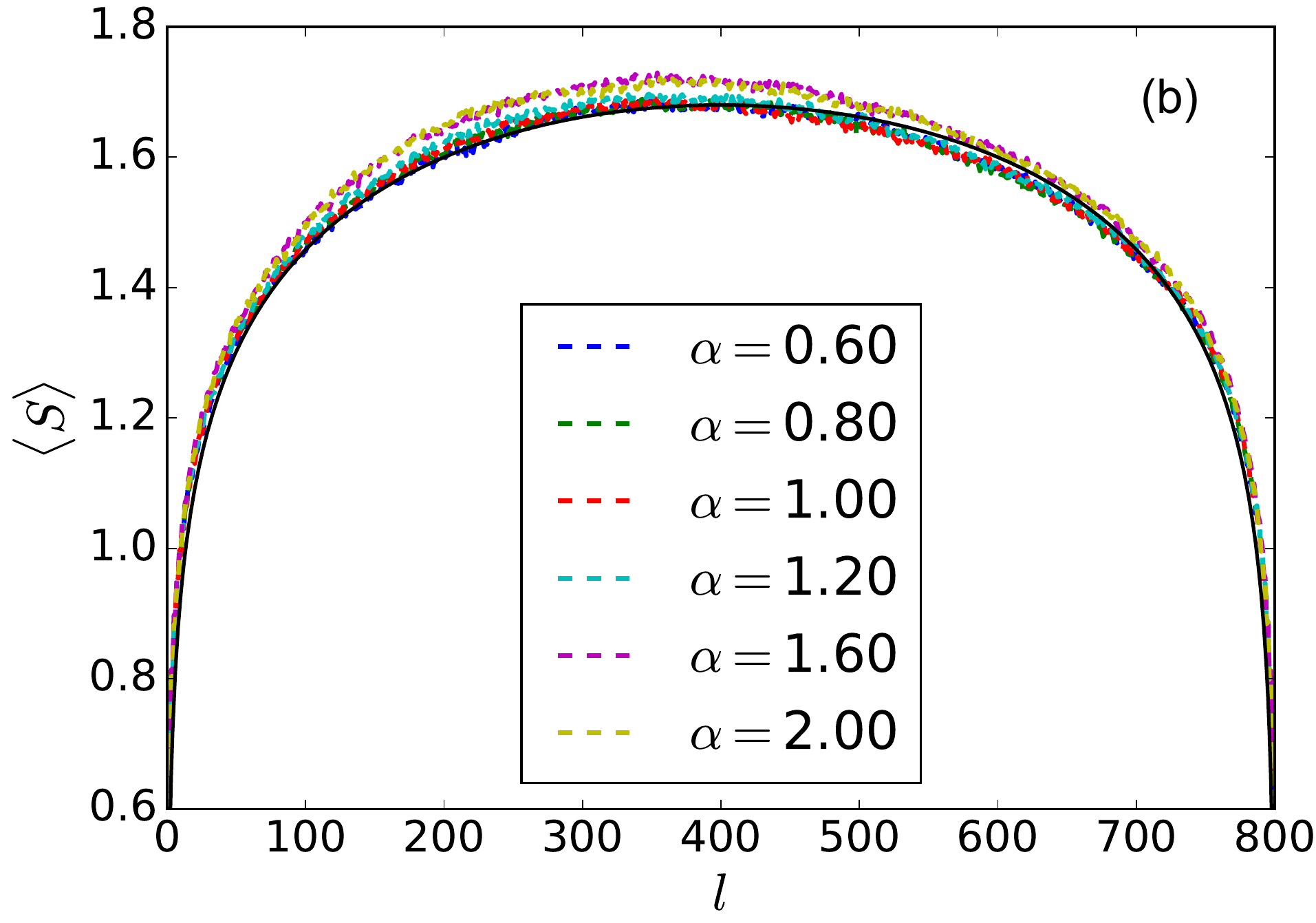} \hspace*{6pt} 
\end{center}
\caption{(a) Entanglement entropy as  function of $c^2$
 for the three instances that occur after approximating the corrected state to Eq. (\ref{Psi_approx}). We note that for $0<c^2\lesssim 0.5$, perturbation theory is no longer valid and a different behavior in this regime might occur.
(b)Average block entanglement entropy calculated with the approximated state in Eq. (\ref{Psi_approx}) for a chain with $N=800$ spins. The power $\alpha$ is varied from $0.60$ to $2.00$. Calabrese and Cardy's formula (Eq. (\ref{S_b})) is plotted along for reference as a black continuous line. 
}
\label{fig:Sk3}
\end{figure}
 Moreover, as seen in Fig. \ref{fig:Sk3}(a) for the specific instance $k=3$,  case 1 (Eq. (\ref{S_1}), dashed line)  gives a  higher entropy  than the one at the IRFP (Eq. (\ref{S0}), continuous line). This is expected since the corrected state is a superposition of states that differ only in spin pairs $\{nl\}$ and $\{mk\}$, which live on opposite sides of the subsystem boundary, and therefore results in an enhancement of the quantum correlations between subchains. On the other hand, case 2 (Eq. (\ref{S_2}), dashed-dotted line) gives a lower entropy than that of Eq. (\ref{S0}), also for all values of $c\neq 1$. Again, this is expected due to the fact that the extra correction terms are destroying the RSS, which in this case is the maximally entangled state, given that both pairs cross the boundary. It is worth noting, that in order to plot the entropy in Eq. (\ref{S_2}), Eq. (\ref{capp}) was inverted in order to obtain $\delta J(c^2)$, and the positive root was chosen. However, since Eq. (\ref{S_2}) is an even function of $\delta J$, this choice becomes trivial. 

We observe that, even though the plot is only shown for the specific case  of  $k=3$ crossing singlets, the above statements  remain true for all values of $k$,  as can be inspected via Eqs. (\ref{S_1}) and (\ref{S_2}).

As seen in Fig. \ref{fig:Sk3}(b) and (c) for a chain of length $N=800$, the approximation  in Eq. (\ref{Psi_approx}) that gives rise to Eqs. (\ref{S_1},\ref{S_2},\ref{S0}) does not give a significant dependence on the power $\alpha$, and the entropy remains close to Cardy's result.  By looking at the difference of the entropy calculated with corrections and the one calculated solely with the RSS, we find that they are about two orders of magnitude smaller than the respective entropy values, which is not surprising, since the corrections to only two singlets are taken into account so far.

Therefore, we can conclude that sases 1 and 2, the two cases in which corrections appear, are not frequent enough throughout realizations to notably affect the average entropy. 

In conclusion, even though the corrected state in Eq. (\ref{Psi_corrected}) is useful to calculate the typical  concurrence, it does not give a sizable correction to  the average entanglement entropy  governed by the RSS.
 Next,  we would have to find a way to include the corrections to the EE from  all singlet-triplet excitations by 
 taking recursively weaker and weaker corrections into account, which we leave for future research.
 \clearpage
 \clearpage
 \clearpage
\bibliography{bibliography}

\end{document}